\documentclass[journal]{IEEEtran}
\usepackage{amsfonts}
\usepackage{amssymb}   
\usepackage[cmex10]{amsmath}
\usepackage{graphicx}
\usepackage{subfigure}
\usepackage{verbatim}
\usepackage{cite}
\usepackage{color}
\usepackage{booktabs}\usepackage[ruled]{algorithm2e}
\usepackage{multirow}
\usepackage{diagbox}
\usepackage{bm}
\usepackage{array}
\usepackage{amsthm}
\usepackage{enumerate}

\newcommand{\bff}{{\mathbf{f}}}
\newcommand{\bY}{{\mathbf Y}}
\newcommand{\bW}{{\mathbf W}}

\newcommand{\bX}{{\mathbf X}}
\newcommand{\bx}{{\mathbf x}}

\newcommand{\bpi}{{\boldsymbol\pi}}

\newtheorem{Def}{Definition}
\newtheorem{Pro}{Proposition}

\newtheorem{Rem}{Remark}
\allowdisplaybreaks
\newcommand{\proproof}[1]{\noindent\textbf{Proof. } See Appendix #1.}
\IEEEoverridecommandlockouts
\begin{document}

\title{Multi-object Tracking  for Generic Observation Model Using Labeled Random Finite Sets}
\author{
Suqi Li, Wei Yi, Reza Hoseinnezhad, Bailu Wang and Lingjiang Kong\\
\IEEEauthorblockA{{University of Electronic Science and Technology of China, School of Electronic Engineering, Chengdu City, China} \\
{Email: kussoyi@gmail.com}}}

\author{
Suqi Li, Wei Yi$^*$, Reza Hoseinnezhad, Bailu Wang, Lingjiang Kong
\thanks{This work was supported
by the Chang Jiang Scholars Program, the National Natural Science Foundation of China under Grant 61771110, the Fundamental Research Funds of Central Universities under Grants ZYGX2016J031, the Chinese Postdoctoral Science Foundation under Grant 2014M550465 and Special Grant 2016T90845,  and the Australian Research Council (ARC) through the Linkage Project Grant LP160101081. (\textit{Corresponding author: Wei Yi.})

S. Li, B. Wang, W. Yi,  and L. Kong  are with the School of  Electronic Engineering, University of Electronic Science and Technology of China, Chengdu 611731, China (Email: qi\_qi\_zhu1210@163.com;  kussoyi@gmail.com; w\_b\_l3020@163.com; lingjiang.kong@gmail.com).

R. Hoseinnezhad is with the School of Aerospace, Mechanical and Manufacturing Engineering, RMIT University, Victoria 3083, Australia (Email: reza.hoseinnezhad@rmit.edu.au). 


}
}

\maketitle

\begin{abstract}
This paper presents  an exact Bayesian filtering solution for the multi-object tracking problem with the generic observation model. The proposed solution is designed in the labeled random finite set framework, using the product styled representation of labeled multi-object densities, with the standard multi-object transition kernel and no particular simplifying assumptions on the multi-object likelihood. 
Computationally tractable solutions are also devised by applying a principled approximation involving the replacement of the full multi-object density with a labeled multi-Bernoulli density that minimizes the Kullback-Leibler divergence and preserves the first-order moment. To achieve the fast performance, a dynamic grouping procedure based implementation is presented with a step-by-step algorithm. The performance of the proposed filter and its tractable implementations are verified and compared with the state-of-the-art in numerical experiments.
\end{abstract}
\IEEEpeerreviewmaketitle
\section{Introduction}
Finite set statistics (FISST)~\cite{refr:Mahler_book} has become a hot spot  in multi-object inference for the random finite set (RFS) framework can perfectly accommodate relatively accurate models for the behavior of multi-object dynamic systems, especially in terms of its ability to capture the randomness of both the number of, and the values of object states, as well as their statistical correlations. FISST has attracted
substantial interest from academia as well as the commercial
sector with applications spanning many areas such as, biology~\cite{refr:biology}, physics~\cite{refr:physics}, computer vision~\cite{refr:vedio-tracking}, 
 multi-object
tracking~\cite{refr:tracking-1,refr:tracking-2,refr:GCI-MB}, and robotics~\cite{refr:robotics}. At the core of multi-object tracking is Bayes filter which is usually intractable due to suffering from the curse of dimensionality with computing set integrals and the combinatorial growth of computations involved with increasing number of objects.  In order to solve these problems, several tractable approximations of multi-object Bayes filter have been proposed successively, namely the  probability hypothesis density (PHD) filter~\cite{refr:tracking-1,refr:PHD}, the  cardinalized PHD filter~\cite{refr:CPHD,refr:CPHD-2}, and the multi-Bernoulli filter~\cite{refr:Mahler_book,refr:MeMber_filter1,refr:MeMber_filter}.

With the recent development of labeled set  filters~\cite{refr:label_1,refr:label_2,refr:label_3,refr:label_4, refr:label_5, refr:label_6, refr:label_7,Vo-Vo-JMS,robust-distributed-fusion} and their enhanced performance  compared to previous unlabeled versions, 
the study on the FISST-based multi-object tracking has recently turned its focus on the labeled random set filters. Vo \textit{et al.}~\cite{refr:label_1} proposed  a class of generalized labeled multi-Bernoulli (GLMB) densities\footnote{GLMB distribution was also termed as Vo-Vo distribution by Mahler in his book~\cite{refr:tracking-2}.}  and the relevant tracking filter, the GLMB filter.  The advantages of GLMB RFS family are that they are conjugate priors with standard multi-object likelihood, and are closed under the multi-object Chapman-Kolmogorov equation with respect to the standard multi-object transition kernel. 
 Nevertheless, the $\delta$-GLMB filter involves exponential growth in the number of posterior components with the number of objects and therefore, tractable techniques for truncating the posterior and prediction densities are also proposed in~\cite{refr:label_2}. Later, to further  decrease the computational costs, principled approximations of the  GLMB filter were proposed, including the labeled multi-Bernoulli (LMB) filter~\cite{refr:label_5} and the marginalized $\delta$-GLMB filter~\cite{refr:label_3}.  These two filters are not only computationally cheaper, but also preserve the key statistical properties of the full multi-object posterior density.

All of the aforementioned labeled set filters are originally designed  for the standard observation model, and are not necessarily suitable for the generic observation model (GOM) which involves no simplifying assumptions made on the multi-object likelihood. 
 In many applications, there might be sensor observations that cannot be accurately modelled by the standard multi-object likelihood. Examples include
 the   track-before-detect (TBD) problem~\cite{refr:MeMber_filter,refr:tbd-2,refr:tbd-3,refr:tbd-4,refr:tbd-5,haichao1,haichao2}, superpositional sensors~\cite{refr:superpositional-1, refr:superpositional-2, refr:superpositional-3, refr:label_7}, merged observations~\cite{refr:label_4}, and extended objects~\cite{refr:extended-1}. Sensors providing such non-standard observations are widely used in  applications such as 
  vehicle tracking using automotive radars, person tracking using laser sensors, acoustic amplitude sensors~\cite{refr:amplitude},  
 and video tracking~\cite{refr:computer-vision, refr:vedio-tracking}. Consequently, there is a substantial demand for devising  multi-object tracking algorithms that work with the GOM.

There is no specified class of labeled RFSs that can be closed under the Bays' rule with respect to the GOM. In an independent work from this paper, Papi \textit{et al.}~\cite{refr:label_6} proposed  a decomposition of the general case of the  labeled multi-object (LMO) density, as the product of the joint existence probability of the label set and the joint probability density of states conditional on their corresponding labels. This decomposition  provides an explicit expression for the  LMO density, and is  fundamental  in the labeled multi-object filtering context especially with  the GOM.
Papi~\textit{et al.}~\cite{refr:label_6} also proposed an extension of the $\delta$-GLMB filter that  works with the GOM,  by replacing the multi-object posterior with a principled $\delta$-GLMB density approximation that  minimizes the  Kullback-Leibler divergence (KLD), and preserves the cardinality distribution and the first-order moment.
To distinguish it from the conventional $\delta$-GLMB filter, it is referred to as the $\delta$-GLMB-GOM filter in this paper.

Unlike the $\delta$-GLMB-GOM filter which is an approximate solution for the  multi-object tracking problem with the GOM, the novel solution presented in this paper, the  LMO-GOM filter\footnote{Preliminary results have been published in~\cite{refr:conference-GOM}. This paper provides a complete and detailed picture with extended results, proofs, and experiments.}, is an exact solution for the same problem.
The prediction equations of the LMO-GOM filter are exact under the standard multi-object transition kernel which embeds  the basic assumptions commonly made with multi-object tracking solutions, such as Markovian dynamics for object states, and the independence of the  birth process from other object states.  The update equations of the LMO-GOM filter are not  based on any approximations or simplifying assumptions with the multi-object likelihood model. Essentially, the $\delta$-GLMB-GOM filter~ is an  approximation of the LMO-GOM filter with the multi-object posterior  approximated as a principled $\delta$-GLMB density. 


Another major contribution is a generalization of the LMB filter, called the  LMB-GOM filter, that works with generic multi-object likelihoods. The LMB-GOM filter is devised by approximating the original multi-object posterior  with the closest LMB density in terms of its KLD. The approximate LMB density also matches the  first-order moment of the original multi-object posterior.  
Our analysis shows that    the computational cost of the LMB-GOM filter  is  less than the $\delta$-GLMB-GOM filter.

A third major contribution of is  this paper is a variant of the proposed LMB filter, called the grouping based LMB-GOM (G-LMB-GOM) filter which is essentially an efficient implementation  of the  LMB-GOM filter.  The G-LMB-GOM filter is based on a dynamic grouping procedure which enables parallelization.
This parallel implementation significantly reduces both the number and the dimension of integrals,    leading to a substantial improvement in computational costs  as well as the numerical accuracy  when the computing and memory  resources are  limited.  In some cases, the resulting improvements in the numerical accuracy are well beyond the extent of inaccuracies stemmed from the grouping procedure.


The performance of the proposed algorithms including the LMO-GOM and LMB-GOM/G-LMB-GOM filters, implemented via Sequential Monte Carlo (SMC) method, are presented and  demonstrated in numerical experiments.

The rest of the paper is organized as follows. A background on  notations,  labeled RFSs and the formal statement of the labeled multi-object tracking problem is provided in Section \ref{chp:2}. Section \ref{chp:4} proposes the LMO-GOM filter and Section \ref{chp:5} presents the ``best'' LMB  approximation for the general LMO density and the resulting LMB-GOM filter. Section \ref{chp:6} provides a comparative summary for different labeled multi-object tracking algorithms with the GOM.  Section \ref{chp:7} demonstrates the performance of the proposed algorithms via numerical experiments. Conclusions are remarked in Section \ref{chp:8}.

\section{Background}\label{chp:2}
\subsection{Notations}
We adhere to the convention that single-object states are
represented by lowercase letters, e.g., $\bx$, $x$, while multi-object
states are represented by uppercase letters, e.g., $\bX$, $X$.  To distinguish labeled states and distributions from the
unlabeled ones, bold-type letters are adopted for the labeled
ones, e.g., $\bx$, $\bX$, $\bpi$. Moreover, blackboard bold letters represent spaces, e.g., the
state space is represented by $\mathbb{X}$, the label space by $\mathbb{L}$. The collection of all finite subsets of $\mathbb{X}$
is denoted by $\mathcal{F}(\mathbb{X})$.

The labeled single-object state $\bx$ is constructed by augmenting
a state $x\in\mathbb{X}$ with a label $\ell\in\mathbb{L}$. The labels are usually
drawn from a discrete label space, $\mathbb{L}=\{\alpha_i,i\in\mathbb{N}\}$, where
all $\alpha_i$s are distinct and  $\mathbb{N}$ is the set of positive
integers.

We use the multi-object exponential notation
\begin{equation}\label{multi-object exponential notation }
  h^{X}\triangleq{\prod}_{x\in X}h(x)
\end{equation}
for real-valued function $h$, with $h^\emptyset=1$ by convention.

To admit arbitrary arguments like sets, vectors and integers, the generalized Kronecker delta function and the inclusion function are repectively given by
\begin{equation}\label{delta}
  \delta_Y(X)\triangleq\left\{\begin{array}{l}
\!\!1, \,\,\mbox{if}\,\,\,X=Y\\
\!\!0, \,\,\mbox{otherwise},
\end{array}\right.
 1_Y(X)\triangleq\left\{\begin{array}{l}
\!\!1, \,\,\mbox{if}\,\,\,X\subseteq Y\\
\!\!0, \,\,\mbox{otherwise}.
\end{array}\right.
\end{equation}
If $X$ is a singleton, i.e., $X=\{x\}$, the notation $1_Y(x)$ is used instead of $1_Y(\{x\})$. For functions $a(\bx)$ and $b(\bx)$ defined on $\mathbb{X}\times\mathbb{L}$, the inner product is denoted by
$\big<a,b\big>=\int a(\bx)b(\bx)d\bx$.

\subsection{Labeled RFS}
The notion of labeled RFSs was firstly proposed in \cite{refr:label_1} to address the uniqueness of tracks.   A \textit{labeled RFS} ~\cite{refr:label_1,refr:label_2} with (kinematic) state space $\mathbb{X}$ and (discrete) label space
 $\mathbb{L}$ is an RFS on $\mathbb{X}\times\mathbb{L}$ such that each realization $\mathbf{X}$ has distinct labels. Let $\mathcal{L}: \mathbb{X}\times\mathbb{L}\!\rightarrow \!\mathbb{L}$ be the projection $\mathcal{L}((x,\ell))\!\!=\!\!\ell$, and hence $\mathcal{L}(\bX)\!\!=\!\!\{\mathcal{L}(\bx),\bx\!\!\in\!\!\bX\}$ is the set of labels of $\bX$. A labeled RFS and the set of its labels have the same cardinality, namely, $|\mathcal{L}(\mathbf{X})|\!\!=\!\!|\mathbf{X}|$. The function $\Delta(\bX)\!\!=\!\!\delta_{|\bX|}(\mathcal{L}(\bX))$ is called the distinct label indicator.
\subsubsection{Decomposition of LMO Density}
For an arbitrary labeled RFS, its multi-object density can be decomposed as the product of the joint existence probability of the label set and the  joint probability density of states conditional on their corresponding labels~\cite{refr:label_6}. 
The definitions of necessary quantities  and the decomposition of the LMO density are briefly reviewed by providing a more rigorous  definition.

The set of labels $\mathcal{L}(\bX)$ of a labeled RFS $\bX$ (distributed according to $\bpi$) is distributed according to the marginal
\begin{equation}
\begin{split}\label{joint existence probability}
 &\omega(\{\ell_1,\!\cdots\!,\ell_n\})\triangleq\\&
  \!\!\!\!\!\!\left\{\begin{array}{l}
\!\!\! \int \bpi(\{(x_1,\ell_1),\!\cdots\!,(x_n,\ell_n)\})d(x_1,\!\cdots\!,x_n), 
\,\,\mbox{if}\,\,\,n>0\\
\,\,\,\,\,\,\,\,\,\,\,\,\,\,\,\,\,\,\,\,\,\,\,\,\,\,\,\,\,\,\,\,\,\,\,\,\,\,\bpi(\emptyset),\,\,\,\,\,\,\,\,\,\,\,\,\,\,\,\,\,\,\,\,\,\,\,\,\,\,\,\,\,\,\,\,\,\,\,\,\,\,\,\,\,\,\,\,\,\,\,\,\,\,\,\mbox{if}\,\,\,n=0.
\end{array}\right.
\end{split}
\end{equation}
The quantity $\omega(\{\ell_{1},\cdots,\ell_{n}\})$ is referred to as the joint existence probability of the label set $\{\ell_{1},\cdots,\ell_{n}\}$ in this paper.
\begin{Def}\label{definition:1}
Given an LMO density $\bpi$ on $\mathcal{F}(\mathbb{X}\times\mathbb{L})$, we define a function $P(\bX)$ on $\mathcal{F}(\mathbb{X}\times\mathbb{L})$ as (\ref{joint probability density}).
\begin{figure*}
\begin{equation}
\label{joint probability density}
 P(\{(x_1,\ell_1),\cdots,(x_n,\ell_n)\})\triangleq
  \left\{
 \begin{array}{cl}
\ \frac{\bpi(\{(x_1,\ell_1),\cdots,(x_n,\ell_n)\})}{\omega(\{\ell_1,\cdots,\ell_n\})},  &\mbox{if}\,n>0\,\,\mbox{and}\,\, w(\{\ell_{1},\cdots,\ell_{n}\})>0\\
\ 1,                                                                                                                       & \mbox{if}\,\,n>0\,\, \mbox{and}\,\,w(\{\ell_{1},\cdots,\ell_{n}\})=0\\
\ 1,                                                                                                                       &\mbox{if}\,n=0.
\end{array}
\right.
\end{equation}
\hrulefill
\end{figure*}

\end{Def}
\begin{Rem}\label{remark:1}
Given a certain  set of distinct labels $\{\alpha_1,\cdots,\alpha_n\}$, if $n>0$ and  the weight $\omega(\{\alpha_1,\cdots,\alpha_n\})>0$, $P(\{(x_1,\alpha_1),\cdots,(x_n,\alpha_n)\})$ is essentially a joint probability density on $\mathbb{X}^{n}$ conditional on their corresponding labels $\alpha_{1},\cdots,\alpha_{n}$. Indeed, from Definition \ref{definition:1}, the LMO density $\bpi$ can be decomposed as 
 \begin{equation}\label{P-LMO}
  \bpi(\mathbf{X})=\omega(\mathcal{L}(\mathbf{X}))P(\mathbf{X}).
\end{equation}
\end{Rem}

\subsubsection{Common Labeled RFSs}
The most commonly used labeled RFSs in  existing labeled multi-object filtering algorithms belong to  the GLMB RFS family \cite{refr:label_1}  \cite{refr:label_2}. 
They are distributed according to 
\begin{equation}\label{GLMB}
\bpi(\bX)=\Delta(\bX){\sum}_{c\in\mathbb{C}}\,\,\omega^{(c)}(\mathcal{L}(\bX)){[p^{(c)}]}^\bX
\end{equation}
where $\mathbb{C}$ is a discrete space, each $p^{(c)}(\cdot,\ell)$ is a probability density, and each $w^{(c)}(I)$ is non-negative with $\sum_{(I,c)\in\mathcal{F}(\mathbb{L})\times\mathbb{C}} w^{(c)}(I)=1$.

The class of LMB RFSs is a  subclass of the GLMB RFS family. An  LMB RFS with state space $\mathbb{X}$ and label space $\mathbb{L}$ 
 is distributed according to \cite{refr:label_1,refr:label_2,refr:label_5}
\begin{equation}\label{LMB}
\begin{split}
\bpi(\mathbf{X})=\Delta(\bX)\omega(\mathcal{L}(\bX))p^\bX
\end{split}
\end{equation}
where
\begin{align}\label{weight-LMB}
\omega(L)=&{\prod}_{i\in\mathbb{L}}(1-r^{(i)}){\prod}_{\ell\in L }\frac{1_{\mathbb{L}}(\ell)r^{(\ell)}}{1-r^{(\ell)}}
\end{align}
and $r^{(\ell)}$ represents the existence probability of track $\ell$, and 
$p(\cdot,\ell)$
 is the probability density of the kinematic state of track $\ell$ given its existence. 

From  (\ref{LMB}) and (\ref{weight-LMB}), an LMB RFS is completely determined by the parameters $r^{(\ell)}$ for each $\ell\in\mathbb{L}$ and a function $p(x,\ell)$ defined on $\mathbb{X}\times\mathbb{L}$. 
Also, an LMB RFS can be completely characterized by its
LMB
parameters, i.e.,
\begin{equation}\label{LMB-parameter}
\bpi=\{(r^{(\alpha)},p^{(\alpha)}(x,\ell)):\alpha\in\mathbb{L}\},\end{equation}
with 
\begin{equation}\label{p-alpha}
p^{(\alpha)}(x,\ell)\triangleq \delta_{\alpha}(\ell)p(x,\ell)
\end{equation}
Note that the definition in (\ref{p-alpha}) is applied for all the LMB RFSs throughout the paper.
\begin{Def}\label{definition:2}
Given the 
LMB
parameters $\{(r^{(\alpha)},p^{(\alpha
)}(\cdot))\}_{\alpha\in\mathbb{L}}$, the labeled Bernoulli component $(r^{(\alpha)},p^{(\alpha)}(\cdot))$ is referred to as  track $\alpha$, with $p^{(\alpha)}(\cdot)$ representing the joint spatial and label density, and   $r^{(\alpha)}$ the probability of existence of  track $\alpha$.
\end{Def}

\subsection{Multi-object Bayes Filter}
Multi-object Bayes filter is at the core of multi-object filtering in RFS framework. This subsection provides  a   review of  the multi-object Bayes filter in the formulation of labeled multi-object state, which is firstly presented in \cite{refr:label_1}.  To incorporate object tracks, objects are identified by an ordered pair of integers $\ell=(k,i)$, where $k$ is the time of birth, and $i\in\mathbb{N}$ is
a unique index to distinguish objects born at the same time.
The label space for objects born at time $k$, denoted as $\mathbb{L}_k$, is
then $\{k\}\times\mathbb{N}$. An object born at time $k$ has  a state $\bx\in\mathbb{X}\times\mathbb{L}_k$.
The label space for objects at time $k$ (including those born
prior to $k$), denoted as $\mathbb{L}_{0:k}$, is constructed recursively by
$\mathbb{L}_{0:k} =\mathbb{L}_{0:k-1}\cup\mathbb{L}_{k}$. A multi-object state $\bX$ at time $k$, is a
finite subset of $\mathbb{X}\times\mathbb{L}_{0:k}$. Note that $\mathbb{L}_{0:k-1}$ and $\mathbb{L}_k$ are disjoint.

The multi-object posterior density $\bpi_k$ is propagated forward recursively by the multi-object Bayes filter,
\begin{align}
\label{predict} \bpi_{k|k-1}(\bX_{k})&=\int \bff_{k|k-1}(\bX_{k}|\bX)\bpi_{k-1}(\bX)\delta\bX\\
\label{update}\bpi_k(\bX_k)&=\frac{g_k(\Upsilon_k|\bX_k)\bpi_{k|k-1}(\bX_k)}{\int g_k(\Upsilon_k|\bX)\bpi_{k|k-1}(\bX)\delta\bX}
\end{align}
where $\bpi_{k|k-1}$ is the  multi-object predicted density from time $k-1$ to time $k$; $\bff_{k|k-1}(\cdot|\cdot)$ is the multi-object transition density; $g_k(\Upsilon_k|\cdot)$ is the multi-object likelihood function and $\Upsilon_k$ denotes the observations of multi-object state  at time $k$. Note that $\Upsilon_k$ is a general notation which can represent a vector observation $\mathrm{z}_k$, or  a set  observation $Z_k$, depending on the observation model  adopted.

For convenience, in what follows we omit explicit references
to the time index $k$, and denote $\mathbb{L}\triangleq\mathbb{L}_{0:k}$,   $\mathbb{B}\triangleq\mathbb{L}_{k+1}$, $\mathbb{L}_+\triangleq\mathbb{L}\cup\mathbb{B}$, $\bpi\triangleq\bpi_k$, $\bpi_+\triangleq\bpi_{k+1|k}$, $g\triangleq g_k$, and $\bff\triangleq\bff_{k|k+1}$.
\subsection{Multi-object Transition Kernel}
This paper considers the standard multi-object transition model  \cite{refr:Mahler_book,refr:label_1}. Given a labeled multi-object state $\bX$, each state $(x,\ell)\in\bX$ either continues to exist at the next time step with probability $p_S(x,\ell)$ and evolves to a new state $(x_+,\ell_+) $ with probability density $f_+(x_+,\ell_+|x,\ell)=f_{+}(x_+|x,\ell)\delta_{\ell_+}(\ell)$, or dies with probability $1-p_S(x,\ell)$. According to Definition~\ref{definition:1}, The  set of new objects born at  the next time step defined on $\mathbb{X}\times\mathbb{B}$ is distributed according to
\begin{equation}\label{LMO born}
{\small
\bff_{B}(\bY)=\omega_B(\mathcal{L}(\bY))P_B(\bY).}
\end{equation}
Note that  the birth density $\bff_B$ also can be specified as an LMB density of form (\ref{LMB}) or a  GLMB density of form (\ref{GLMB}).

A multi-object state $\bX_+$ is  the superposition of surviving objects and  newly born objects. Assuming that the surviving and the newly born object states evolve independently, the multi-object transition function is given by~\cite{refr:label_2}
\begin{equation}\label{multi-object transition function}
{\small
\bff(\bX_+|\bX)=\bff_S(\bX_+\cap (\mathbb{X\times \mathbb{L}})|\bX)\bff_B(\bX_+\cap(\mathbb{X}\times\mathbb{B}))}
\end{equation} 
where
\begin{equation}
{\small
\begin{split}
\!\!\!\!\!\!\!\!\!\!\!\!\!\bff_S(\bW|\bX)=&\Delta(\bW)\Delta(\bX)1_{\mathcal{L}(\bX)}(\mathcal{L}(\bW)){[\Phi(\bW;\cdot)]}^\bX
\end{split}}
\end{equation} 
\begin{equation}
{\small
\begin{split}
\!\!\Phi(\!\bW;x,\ell)\!=\!&\left\{\begin{array}{l}\!\!\! p_S(x,\ell)\delta_{\ell_+}\!(\ell)f_+\!(x_+|x,\ell),\mbox{if}\, (x_+, \ell)\!\in\! \bW\\
\!\!\!1-p_S(x,\ell), \,\,\,\,\,\,\,\,\,\,\,\,\,\,\,\,\,\,\,\,\,\,\,\,\,\,\,\,\,\,\,\,\mbox{if}\,\,\ell\notin\mathcal{L}(\bW).\end{array}\right.
\end{split}}
\end{equation} 
\subsection{Generic Observation Model}
The standard formulation in the RFS based multi-object tracking is based on the standard observation model \cite{refr:tracking-2} where observation data is assumed to have been preprocessed into thresholded detections,  each object is assumed to cause at most one detection, and each detection is assumed to be either a false alarm (clutter) or generated from one object. The tracking filters under the standard observation model have  been well investigated. One remarkable development is    the GLMB family of densities are conjugate prior with respect to the standard multi-object likelihood. Utilizing this property, the GLMB filter is proposed as a closed form of the multi-object Bayes filter under the standard observation model, and its   performance has been well demonstrated in \cite{refr:label_1,refr:label_2}.

This paper considers the \textit{generic observation model }which  has the indication that  no simplifying assumptions on the multi-object likelihood are made. The terminology ``generic measurement (observation) model''  first arised in \cite{refr:label_6} for  the $\delta$-GLMB filter with the  generic multi-object likelihood.

The  considered GOM  covers both the standard and non-standard observation models.  As the first  category of sensor models has been well  investigated, this paper mainly focuses on the non-standard  sensor models.  Below, we present two typical examples for the non-standard observation models, namely, the pixled TBD model
 and the acoustic amplitude sensor model.
 
{\noindent \textit{\textbf{Example 1} - Pixeled TBD Model: }
 The surveillance
region is divided into $M$ cells. The observations at the current time step  are collected in
the vector $\textmd{z}=(z_1,\cdots,z_M)\in \mathbb{R}^M$, with $z_j$ being the intensity
observation obtained in the $j$th cell. An object $\bx$ can  illuminate several cells of its surroundings.
 Within the effective template of $\bx$,
   the intensity contribution from $\bx$ to the $j$th cell follows a point spread function \cite{refr:beyond_kalm_filer}  
\begin{equation}\label{The point spread function}
c_j(\bx)=\frac{\delta_x\delta_y\sigma_T}{2\pi\sigma^2_b}\!\exp\!\left(\!-\frac{(\delta_x a-p_x)^2+(\delta_y b-p_y)^2}{2\sigma^2_b}\right)
\end{equation}
  where $\sigma_T$ is the source intensity, $\sigma^2_b$ is the blurring factor, $\delta_x$ and $\delta_y$ are the cell side lengths, and $j=(a,b)$ denotes the position of the $j$th cell in a two-dimensional image of the surveillance region.  For the cells beyond the  effective template of $\bx$, the intensity contribution $c_j(\bx)=0$.
  
 The  observations obtained from different cells
are assumed to be independently distributed conditioned on the multi-object state $\bX$, and thus the multi-object likelihood is
\begin{equation}
{\small\label{likelihood-multiobject}
g(\textmd{z}|\bX)={\prod}_{j=1}^M g(z_j|\bX)} \end{equation}
where $g(z_j|\bX)$ denotes the likelihood of the $j$th cell.
The distribution of $g(z_{j}|\bX)$ varies from different applications. For instance, in the infrared/image application  
\cite{refr:MeMber_filter,refr:beyond_kalm_filer}, 
the likelihood $g(z_j|\bX)$ is assumed to be a Gaussian distribution, 
\begin{equation}\label{eq:TBD}
  g(z_j|\bX)=\mathcal{N}\left(z_j;{\sum}_{\bx\in \bX}c_j(\bx),\sigma^2_N\right)
\end{equation}
where $\mathcal{N}(z;\mu,\Gamma)$ denotes the Gaussian probability density
evaluated at $z$ with mean $\mu$ and covariance matrix $\Gamma$.
  

\noindent \textit{\textbf{Example 2} - Acoustic Amplitude Sensor:}
We consider a wireless sensor network consisting of $\mathcal{M}$ sensors \cite{refr:amplitude}. At a certain time step,   sensor $m$ ($m\in \{1,\cdots,\mathcal{M}\}$) acquires an observation $z_{m}\in\mathbb{R}$. 
Each object  $\bx$ emits a sound with amplitude $A$ that is assumed to be constant. For  sensor $m$ located at  position $\xi_{m}$, the received sound amplitude due to 
target $\bx$ is modelled as $A/\|\rho(\bx)-\xi_{m}\|^{\kappa}$, where   $\rho(\bx)$ is the position of object $\bx$,  and ${\kappa}$ is the path loss exponent. The scalar observation $z_{m}$ obtained by sensor $m$  is then given by
\begin{equation}\label{acoustic amplitude model}
z_{m}=h_{m}(\bX)+v_{m}
\end{equation}
with
\begin{equation}\label{acoustic amplitude model-parameter}
h_{m}(\bX)={\sum}_{\bx\in\bX}\frac{A}{\|\rho(\bx)-\xi_{m}\|^{\kappa}}
\end{equation}
where $v_{m}\sim\mathcal{N}(0,\sigma_{v}^{2})$ for $m=1,\cdots,\mathcal{M}$ are zero-mean Gaussian  noise variables of equal variance $\sigma_{v}^{2}$. Assume that $v_{1},\cdots,v_{\mathcal{M}}$ are mutually independent and independent of $\bX$. The likelihood function  between  observation vector $\textmd{z}=(z_{1}\cdots,z_{\mathcal{M}})\in\mathbb{R}^{\mathcal{M}}$ and  multi-object state $\bX$ is given by
\begin{equation}\label{likelihood-acoustic}
g(\textmd{z}|\bX)={\prod}_{m=1}^{\mathcal{M}} \mathcal{N}(z_{m};h_{m}(\bX),\sigma^{2}_{v}).
\end{equation}

Note that both (\ref{eq:TBD}) and  (\ref{likelihood-acoustic}) are highly non-linear likelihoods which are not closed under the Bayes  update equation (\ref{update}).


\section{The LMO-GOM Filter}\label{chp:4}
In this section, we derive the multi-object prediction and update equations of Bayes filter by specifying the multi-object prior and posterior as the product styled LMO densities provided in (\ref{P-LMO}). The result is an exact solution for labeled multi-object  Bayes filter with GOM under the standard multi-object transition model, and thus is called as  the  LMO-GOM filter. Furthermore, an SMC implementation of the LMO-GOM filter is presented. 
\subsection{Recursive Equations of the LMO-GOM Filter}
\begin{Pro}\label{pro:1}
Suppose that  the current multi-object prior density is a general LMO density of the form  (\ref{P-LMO})
and the birth density is also a general LMO density of the form (\ref{LMO born}),
then the multi-object predicted density under the multi-object transition function of the form (\ref{multi-object transition function}) is given by
\begin{align}\label{multi-object predicted density}
\bpi_{+}(\bX_+)=\omega_{+}(\mathcal{L}(\bX_+))P_{+}(\bX_+)
\end{align}
on state space $\mathbb{X}$ and label space $\mathbb{L}_+=\mathbb{L}\cup\mathbb{B}$,
where
\begin{align}
\label{w-add}\omega_{+}(L)&=\omega_{B}(L\cap\mathbb{B})\ \omega_S(L\cap\mathbb{L})\\
\label{p-add}P_{+}(\bX_+)&=P_B(\bX_+\cap\mathbb{X}\times\mathbb{B})\  P_S(\bX_+\cap\mathbb{X}\times\mathbb{L})\\
\label{ws}\omega_S(J)&={\sum}_{I\subseteq\mathbb{L}}1_{I}(J)\omega(I)\eta_{S,I}(J)\\
\label{Ps}P_S(\bW)&=\frac{\sum_{I\subseteq\mathbb{L}}1_{I}(\mathcal{L}(\bW))\omega(I)P_{S,I}(\bW)}{\omega_S(\mathcal{L}(\bW))}
\end{align}
with
\begin{equation}
{\small
\begin{split}
\label{Ps-I} P_{S,\{\ell_{1},\!\cdots\!,\ell_{n}\}}(\bW)&\!=\!
\!\!\int \!1_{\{\ell_{1},\cdots,\ell_{n}\}}(\mathcal{L}(\bW)) {\prod}^{n}_{i=1}\!\!\Phi(\bW; x_{i},\ell_i)\\
& \cdot P(\{(x_1,\ell_1)\cdots,(x_n,\ell_n)\})d (x_{1}\cdots  x_{n})
\end{split}}
\end{equation} 
\begin{equation}
{\small
\begin{split}
\label{where_0} &\!\!\!\!\!\eta_{S,\{\ell_1,\!\cdots\!,\ell_n\}}(\{\ell_{+,1},\cdots,\ell_{+,{n_+}}\})\!=
 \int \! \!P_{S,\{\ell_{1},\cdots,\ell_{n}\}}\cdot\\&(\{(x_{+,1},\!\ell_{+,1}),\cdots,(x_{+,n_+},\!\ell_{+,n_+})\})d (x_{+,1},\!\cdots\!\!,x_{+,n_{+}}).
\end{split}}
\end{equation}
\end{Pro}

\proproof{A}

Proposition \ref{pro:1} explicitly describes how to calculate  $\omega_+(\cdot)$ and $P_+(\cdot)$  of the  multi-object predicted density from  $\omega(\cdot)$ and $P(\cdot)$ of the multi-object  prior  density. We note that $\bpi_S(\bW)=\omega_S(\mathcal{L}(\bW))P_S(\bW)$ is  the density of the surviving objects with $\omega_S(\cdot)$ and $P_S(\cdot)$  shown in (\ref{ws}) and (\ref{Ps}).  For a  given label set $J$, $\omega_S(J)$ is the weighted sum  of the prior weights $\omega(I)$ over all subsets of $\mathbb{L}$ that contain the surviving set~$J$.  
The function $P_S(\bW)$ given a certain  label set $\mathcal{L}(\bW)$  is also a weighted sum of $P_{S,I}(\bW)$ terms over all subsets of $\mathbb{L}$  that contain the surviving set $\mathcal{L}(\bW)$. 
$P_{S,\{\ell_1,\cdots,\ell_n\}}(\bW)$  in (\ref{Ps-I})  for a certain $\mathcal{L}(\bW)$ is a non-normalized joint density evolved from the prior joint probability density $P(\{(x_1,\ell_1),\cdots,(x_n,\ell_n)\})$  with the ``pseudo'' transition density $\prod_{i=1}^n \Phi(\bW;x_i,\ell_i)$.  $P_{S,\{\ell_1,\cdots,\ell_n\}}(\bW)$ is conditional on that the previous label set is $I=\{\ell_1,\cdots,\ell_n\}$ and  only the objects with the label set $\mathcal{L}(\bW)\subseteq I$ exist after evolving. $\eta_{S,\{\ell_1,\cdots,\ell_n\}}(I_+)$ in (\ref{where_0}) is essentially the normalizing constant of $P_{S,\{\ell_1,\cdots,\ell_n\}}(\bW)$ with $\mathcal{L}(\bW)=I_+$.

Utilizing the independence of surviving objects and newly born objects, the multi-object predicted density can be obtained by multiplying   the weights  and the corresponding joint probability densities of newly born objects and surviving objects.
\begin{Pro}\label{pro:2}
Suppose that the current multi-object predicted density is a general LMO density of the form (\ref{multi-object predicted density}), then the multi-object posterior density under a generic multi-object likelihood $g(Z|\bX)$ is given by
\begin{equation}\label{posterior}
{\small
\bpi(\bX|\Upsilon)=\omega(\mathcal{L}(\bX);\Upsilon)P(\bX;\Upsilon)}
\end{equation}
on state space $\mathbb{X}$ and the label space ${\mathbb{L}}_+$,
where
\begin{align}\label{where_p}
   P(\bX;\Upsilon)=&\frac{g(\Upsilon|\bX)P_+(\bX)}{\eta_{\Upsilon}(\mathcal{L}(\bX))}\\
  \label{where_w} \omega(I_{+};\Upsilon)=&\frac{\eta_{\Upsilon}(I_{+})\omega_+(I_{+})}{\sum_{I_{+}\in\mathbb{L}_+}\eta_\Upsilon(I)\omega_+(I)}
 \end{align} 
with
  \begin{equation}
  {\small
    \begin{split}
    \label{where_eta} \eta_{\Upsilon}(\{\ell_1,\ell_2,\!\cdots\!,&\ell_n\})= \int g(\Upsilon|\{(x_1,\ell_1),\!\cdots\!,(x_n,\ell_n)\})\\&
\cdot P_+(\{(x_1,\!\ell_1),\!\cdots\!,(x_n,\!\ell_n)\})d(x_1,\!\cdots\!,x_n).
\end{split}}
\end{equation} 
\end{Pro}

\proproof{B}

Proposition \ref{pro:2} explicitly describes how to calculate the parameters $\omega(\cdot;\Upsilon)$ and $P(\cdot|\Upsilon)$  of the  multi-object posterior density from the parameters $\omega_+(\cdot)$ and $P_+(\cdot)$ of the multi-object  predicted   density. For a given label set $\mathcal{L}(\bX)$, the posterior joint probability density $P(\bX|\Upsilon)$ in (\ref{where_p}) is computed from the prior joint probability density $P_+(\bX)$    via ``Bayes' rule'' with likelihood $g(\Upsilon|\bX)$. For a given label set $I$, the posterior weight $\omega(I;\Upsilon)$ is  proportional to the predicted weight $\omega_+(I)$  scaled by the normalizing constant $\eta_\Upsilon(I)$.

 \subsection{The SMC Implementation of the LMO-GOM Filter}
In the above subsection, the combination of Propositions \ref{pro:1} and \ref{pro:2} provides an exact Bayesian solution  by adopting the decomposition of  the LMO  density  in the form of (\ref{P-LMO}). Hence, an intuitive  implementation of the LMO-GOM filter is to recursively compute the functions $\omega(\cdot)$ and $P(\cdot)$ at each time step. However, when implementing the LMO-GOM filter,   the  approximation of   $P(\mathbf{X})$  is not straightforward since  $P(\mathbf{X})$ (defined on $\mathcal{F}(\mathbb{X}\times\mathbb{L})$) is not a probability density.  To this end, we  represent the product styled of the LMO density of the form (\ref{P-LMO}) in another equivalent form  as  Remark \ref{mixture-LMO}.
\begin{Rem}\label{mixture-LMO}
An LMO density $\bpi$  on $\mathcal{F}(\mathbb{X}\times\mathbb{L})$ can be expressed as a mixture of multi-object densities, 
\begin{equation}\label{delta-LMO}
\bpi(\bX)={\sum}_{I\in\mathcal{F}(\mathbb{L})} \omega^{(I)}P^{(I)}(\bX)
\end{equation}
where
\begin{align}
\omega^{(I)}&\triangleq\omega(I),\\
P^{(I)}(\bX)&\triangleq\delta_{I}(\mathcal{L}(\bX))P(\bX)
\end{align}
in which the definitions of $P(\bX)$ and $\omega(I)$ are given in Definition \ref{definition:1}, 
$\omega^{(I)}$ denotes the  existence probability of the label set $I$ satisfying $\sum_{I\in\mathcal{F}(\mathbb{L})}\omega^{(I)}=1$, and  $P^{(I)}(\cdot)$ is the multi-object probability density (defined on $\mathcal{F}(\mathbb{X}\times\mathbb{L})$) conditional on the existence of the label set  $I$. Hence, $\bpi(\cdot)$ is completely characterized by a set of parameters $\{(\omega^{(I)}, P^{(I)}(\cdot))\}_{I\in\mathcal{F}(\mathbb{L})}$.
\end{Rem}


The integrals of $P^{(I)}(\bX)$ and $P(\bX)$ have the following relationship. For any $I=\{\alpha_1,\cdots,\alpha_n\}\in\mathcal{F}(\mathbb{L})$, and given an arbitrary function $\upsilon(\bX)$ on $\mathcal{F}(\mathbb{X}\times\mathbb{L})$, we have
\begin{equation}
{\small
\begin{split}
\label{integal-1} &\int P^{(I)}(\bX) \upsilon(\bX) \delta \bX\!=\int  \delta_{I}(\mathcal{L}(\bX))P(\bX) \upsilon(\bX)\delta\bX\\
 =&\!\int\! P(\{(x_1,\alpha_1),\cdots,(x_n,\alpha_n)\})\\
&\,\,\,\,\,\,\,\,\,\,\,\,\,\,\,\,\,\,\,\,\,\,\,\,\,\,\,\,\upsilon(\{(x_1,\alpha_1),\!\cdots\!,(x_n,\alpha_n)\}) d(x_1,\!\cdots\!,x_n).
\end{split}}
\end{equation}
Eq. (\ref{integal-1}) indicates that the set integral of $P^{(I)}(\cdot)$ is equivalent to  the Euclidean notion of integral of the joint probability density $P(\{(\cdot,\alpha_1),\cdots,(\cdot,\alpha_n)\})$ on $\mathbb{X}^{|I|}.$ 

Utilizing the formulas provided in Remark \ref{mixture-LMO},  implementing the LMO-GOM filter, based on Propositions \ref{pro:1} and \ref{pro:2}, amounts to computing the predicted parameter set   $\{(\omega_{+}^{(I_{+})},P_{+}^{(I_{+})}(\cdot))\}_{I_{+}\in\mathcal{F}(\mathbb{L}_{+})}$
with $\mathbb{L}_+\!=\!\mathbb{L}\cup\mathbb{B}$ 
and $$\omega_{+}^{(I_{+})}=\omega_{+}(I_{+}); \,\,P_{+}^{(I_{+})}(\bX_{+})\!=\!\delta_{I_{+}}(\mathcal{L}(\bX_{+}))P_{+}(\bX_{+}),$$
and the posterior parameter set $\{(\omega^{(I_+\!)}(\Upsilon),\!P^{(I_+\!)}(\cdot;\Upsilon))\}_{I_+\!\in\mathcal{F}(\mathbb{L}_+\!)}$ 
with $$\omega^{(I_+)}(\Upsilon)=\omega(I_+;\Upsilon); \,\,P^{(I_+)}(\bX;\Upsilon)=\delta_{I_+}(\mathcal{L}(\bX))P(\bX;\Upsilon)$$
forwards in time.

As  it was mentioned earlier, our algorithms are mainly designed for the non-standard observation model which usually involves the non-Gaussian/non-linear model and has no closed-from solution. Hence, in this subsection, we provide an SMC implementation of the LMO-GOM filter.  Each $P^{(I)}(\bX)$ is represented by a set of weighted particles. Associated weights, and normalizing constants can be computed from 
particles and their weights.

Suppose that the current prior parameter set is $\bpi=\{(\omega^{(I)},P^{(I)}(\cdot))\}_{I\in\mathcal{F}(\mathbb{L})}$ where each $P^{(I)}(\bX)$ is approximated with a set of particles $\{(w^{(I)}_j,\bX^{(I)}_j)\}_{j=1}^{N_p^{(I)}}$, i.e.,
\begin{align}
P^{(I)}(\bX)={\sum}_{j=1}^{N_p^{(I)}}
w^{(I)}_j \delta_{\bX_j^{(I)}}(\bX).\end{align}
Utilizing (\ref{integal-1}), the  quantities in the prediction step are computed as 
\begin{equation}
\begin{split}
\label{SMC-P-I-J} \!\!\!\!\!\!\!\!\!\!\!\!\!\!\!\!\!\!\!\!\!\!\!\!\!\!\!\!\!\!\!\!\!\!\!\!\!\!\!\!\!\!\!\!\!\!\!\!\!\!\!\!P_{S,I}(\bW)
=&\!\int \!1_{I}(\mathcal{L}(\bW))[\Phi(\bW;\cdot)]^{\bX} P^{(I)}(\bX) \delta\bX\\
=&{\sum}_{j=1}^{N_p^{(I)}} 1_{I}(\mathcal{L}(\bW))w^{(I)}_j [\Phi(\bW;\cdot)]^{\bX^{(I)}_j},
\end{split}
\end{equation}
\begin{equation}
\begin{split}
\label{SMC-eta-I-J}
\!\!\!\eta_{S,I}(J)
=&\!\!\int\! \delta_J(\mathcal{L}(\bW))P_{S,I}(\bW)\delta\bW\\
=&{\sum}_{j=1}^{N_p^{(I)}}\! w^{(I)}_{j}[p_S]^{\bX_j^{(I)}\cap\mathbb{X}\times J}[1\!-\!p_S]^{\bX_j^{(I)}\cap\mathbb{X}\times (I\!-\!J)}.
\end{split}
\end{equation}
Then  $\omega_S(J)$ is computed by substitution of (\ref{SMC-eta-I-J})  into (\ref{ws}), and  $\omega_+(L)$ is computed by substitution of the computed $\omega_S(J)$ into (\ref{w-add}).

For each label set $I_{+}\in\mathcal{F}(\mathbb{L}_{+})$,  firstly, choose a subset $\mathbb{U}$ of  $\mathcal{F}(\mathbb{L})$ according to  $$\mathbb{U}=\{I\in\mathcal{F}(\mathbb{L}): 1_{I}(I_{+}\cap\mathbb{L})=1\}.$$
Utilizing (\ref{p-add}), (\ref{Ps}) and (\ref{where_p}), the posterior parameter $P^{(I_{+})}(\cdot;\Upsilon)$ is computed as
\begin{equation}\label{update-P}
\begin{split}
\!\!\!&P_{+}^{(I_{+})}(\bX;\Upsilon)
\propto g(\Upsilon|\bX)\delta_{I_{+}\cap\mathbb{B}}(\mathcal{L}(\bX_{B})
)P_B(\bX_{B})\times\\
&\delta_{I_{+}\cap\mathbb{L}}(
\mathcal{L}(\bX_{S})){\sum}_{I\in\mathbb{U}}\overline\omega^{(I)}{\sum}_{j=1}^{N_p^{(I)}} w_{j}^{(I)}[\Phi(\bX_S;\cdot)]^{\bX^{(I)}_j}
\end{split}
\end{equation}
where
\begin{align}
\begin{split}
 \bX_S\triangleq&\bX_+\cap\mathbb{X}\times\mathbb{L},\,\,
\bX_B\triangleq\bX_+\cap\mathbb{X}\times\mathbb{B}
\end{split}\\
\begin{split}
\overline \omega^{(I)}=\frac{\omega^{(I)}}{\omega_{S}(I_{+}\cap\mathbb{L})}.
\end{split}
\end{align}
By employing the idea of the auxiliary particle filter \cite{refr:tbd-2,refr:tbd-3,refr:tbd-4}, sampling from (\ref{update-P}) can be achieved by sampling from the higher dimensional joint density
\begin{equation}
\begin{split}
P_{+}^{(I_{+})}(\bX, \mathcal{I},u;&\Upsilon)\!\propto  g(\Upsilon|\bX)\delta_{I_{+}\cap\mathbb{B}}(\mathcal{L}(\bX_{B})
)P_B(\bX_{B})\\
&\delta_{I_{+}\cap\mathbb{L}}(
\mathcal{L}(\bX_{S})) \overline\omega^{(\mathcal{I})} w_{u}^{(\mathcal{I})}[\Phi(\bX_S;\cdot)]^{\bX^{(\mathcal{I})}_u}.
\end{split}
\end{equation}
where the auxiliary variable $\mathcal{I}\in\mathbb{U}$ is the previous label set from which the current label set $I_{+}$ is evolved, and  the auxiliary variable $u\in\{1,\cdots,N_p^{(\mathcal{I})}\}$ is the index on the sample at the previous time step conditional on the previous label set $\mathcal{I}$.
The auxiliary variables aid in the sampling of suitable values of the multi-target state $\bX$. They are discarded after the sampling procedure is completed. States $\bX_{j_{+}}^{(I_{+})}$, the previous label set $\mathcal{I}_{j_{+}}$,
and particle indices ￼$u_{j_{+}}$ ￼ are drawn from an importance density $q^{(I_{+})}(\bX,\mathcal{I},u|\Upsilon)$ for $j_{+}=1,\cdots,N_p^{(I_{+})}$, and the un-normalized weight is computed as 
\begin{equation}
\begin{split}
&\widetilde w_{j_{+}}^{(I_{+})}=
g(\Upsilon|\bX)\delta_{I_{+}\cap\mathbb{B}}(\mathcal{L}(\bX_{B})
)P_B(\bX_{B}) \overline\omega^{(\mathcal{I})}\\&\delta_{I_{+}\cap\mathbb{L}}(
\mathcal{L}(\bX_{S})) w_{u}^{(\mathcal{I})}[\Phi(\bX_S;\cdot)]^{\bX^{(\mathcal{I})}_u}
/q(\bX_{j_{+}}^{(I_{+})},\mathcal{I}_{j_{+}},u_{j_{+}})
\end{split}
\end{equation}
A  feasible choice of the  proposal function $q^{(I_{+})}(\bX,\mathcal{I},u|\Upsilon)$ is as follows:
\begin{equation}\label{proposal-function}
\begin{split}
q^{(I_{+})}(\bX, \mathcal{I},u|&\Upsilon)\!=\delta_{I_{+}\cap\mathbb{B}}(\mathcal{L}(\bX_{B})
)P_{B}(\bX_{B})\\& \cdot\delta_{I_{+}\cap\mathbb{L}}(
\mathcal{L}(\bX_{S})) \overline\omega^{(\mathcal{I})} w_{u}^{(\mathcal{I})}[\Phi(\bX_S;\cdot)]^{\bX^{(\mathcal{I})}_u}.
\end{split}
\end{equation}
In this case, the un-normalized weight is computed as 
$$\widetilde w_{j_{+}}^{(I_{+})}=g(\Upsilon|\bX^{(I_{+})}_{j_{+}}).$$
Note that it is possible to design a more sophisticated proposal density than (\ref{proposal-function}), but it is beyond the scope of this paper.

Utilizing (\ref{integal-1}), the quantity $\eta_\Upsilon(I_+)$ is computed as 
\begin{equation}\label{SMC-eta}
{\small
\begin{split}
\!\!\!\!\!\eta_\Upsilon(I_+)\!=\!&\!\int\! \delta_{I_+}(\mathcal{L}(\bX))g(\bX|\Upsilon)P_+(\bX)\delta\bX
\!=\!{\sum}_{j_+=1}^{N_p^{(I_+)}}\!\widetilde{w}^{(I_+)}_{j_+}
\end{split}}
\end{equation}
and the posterior parameter $\omega^{(I_+)}(\Upsilon)=\omega(I_+;\Upsilon)$ is computed by substitution of (\ref{SMC-eta}) into (\ref{where_w}).

Resampling and Implementation Issues: After the update step, for each $P^{(I_+)}(\cdot;\Upsilon)$, perform resampling \cite{refr:SMC_Gordon} to obtain an evenly weighted particle
set. To reduce the growing number of parameters, the pair of posterior parameters $(w^{(I_+)}(\Upsilon),P^{(I_+)}(\cdot;\Upsilon))$ with existence probabilities $w^{(I_+)}(\Upsilon)$ below a threshold are discarded \cite{refr:label_1,refr:label_3,refr:label_6}.


\subsection{Discussions and Analysis}
The LMO-GOM filter provides an exact Bayesian solution for the labeled multi-object tracking problem under the GOM and the standard transition kernel. Nevertheless, in general, the LMO-GOM filter can be computationally prohibitive, especially for a large number of objects. Computing integrals on high-dimensional spaces (the integral of $P^{(I)}(\cdot)$) and exponential growth of the number of parameters with the number of objects are the two main reasons
  in many applications. 

Observing  Proposition \ref{pro:1}, the   prediction step of the LMO-GOM filter can be further simplified for particular multi-object priors such as the $\delta$-GLMB and LMB densities. 
This can be achieved due to the independence assumption between object motions when formulating the multi-object transition kernel. The $\delta$-GLMB-GOM  filter~\cite{refr:label_6} is essentially derived  by approximating the multi-object posterior as a principled $\delta$-GLMB density and assuming the $\delta$-GLMB prior. 
\begin{Rem}\label{remark:3}
The implementation of the LMO-GOM  filter only involves  one source of inaccuracy which is the numerical error caused by the Monte Carlo (MC) approximation of the high-dimensional integral. 
Based on the convergence properties of the MC approximation~\cite{refr:convergence_partical}, when the number of samples approaches infinity, the numerical errors of the integral computations approach zero, and the LMO-GOM filter is implemented with perfect accuracy.  Hence, with sufficient computing resources, the LMO-GOM filter is expected to exhibit the optimal performance, and 
possibly served as a performance benchmark in labeled multi-object tracking with the standard transition kernel. 
In comparison,  the implementation of the $\delta$-GLMB filter involves two sources of inaccuracy. One is related to the approximation of posteriors and the other to the MC approximation.   Due to the first source of inaccuracy, even if the number of particles approaches infinity, the $\delta$-GLMB filter is not exact.
\end{Rem}

\section{The LMB-GOM Filter}\label{chp:5}
As an efficient approximation of the LMO-GOM filter,  the $\delta$-GLMB-GOM filter  alleviates the computation burden by simplifying the prediction equation. However, the $\delta$-GLMB prediction can be still intractable (both memory- and computational-wise) due to the exponential growth of the number of  terms of multi-object exponentials in the predicted $\delta$-GLMB density  with the number of objects.

 In this section, we further explore  more tractable approximations of the LMO-GOM filter.  \cite{refr:label_5} proposed an LMB  filter for the standard multi-object likelihood which approximates the $\delta$-GLMB multi-object posterior as a principled LMB density preserving the first-order moment, and further proposed a dynamic grouping procedure based implementation  which drastically reduces execution time with slight accuracy promise by exploiting the mathematical formulation of the LMB prior.
Motivated by the LMB filter  and its fast implementation proposed in \cite{refr:label_5}, in this section, we seek the ``best'' LMB   approximation to   replace the full  multi-object posterior under the GOM, and consequently,  develop an extension of the LMB filter for  the GOM,  referred to as the LMB-GOM filter.  Furthermore, we also present an efficient implementation   for the  LMB-GOM filter based on a dynamic grouping procedure.


\subsection{The ``Best'' LMB Approximation}
In this subsection,  we  derive  the ``best'' LMB approximation of the general LMO density.   Herein, the ``best'' approximation means the  best information-theoretic fit in terms of the minimal KLD. Propositions \ref{pro:3} and \ref{pro:4}, respectively, derive the explicit formulas of the labeled PHDs for the general LMO density and the LMB density, which are the basis of  the derivation of the ``best" LMB approximation.  Proposition~\ref{pro:5}  provides the explicit formula for the  ``best'' LMB  approximation of the general LMO density.
\begin{Pro}\label{pro:3}
Given an arbitrary LMO density $\bpi(\bX)=\omega(\mathcal{L}(\bX))P(\bX)$ on state space $\mathbb{X}$ and label space $\mathbb{L}$, the labeled PHD of $\bpi$ is
\begin{equation}\label{labeled-PHD-LMO}
v(x,\ell)={\sum}_{I\in\mathcal{F}(\mathbb{L})}1_{I}(\ell)\omega(I)p_{I-\{\ell\}}(x,\ell)
\end{equation} 
where ``$-$'' denotes set difference, and 
\begin{align}
\label{marginal_label}
\begin{split}
p_{\{\ell_1,\cdots,\ell_n\}}(x,\ell) =\int  P(&\{(x,\ell),(x_1,\ell_1),\cdots,\\ &(x_n,\ell_n)\})  d(x_1,\cdots,x_n).
\end{split}
\end{align} 
\end{Pro}
\proproof{C}

\begin{Pro}\label{pro:4}
Given an LMB RFS  with the LMB parameters $\bpi=\{(r^{(\alpha)},p^{(\alpha)}(\cdot))\}_{\alpha\in\mathbb{L}}$, the labeled PHD of $\bpi$ is
\begin{equation}\label{labeled-PHD-LMB}
\begin{split}
v(x,\ell)={\sum}_{\alpha\in\mathbb{L}}r^{(\alpha)}p^{(\alpha)}(x,\ell)=r^{(\ell)}p(x,\ell)
\end{split}
\end{equation}
with 
$p^{(\alpha)}(x,\ell)=\delta_{\alpha}(\ell)p(x,\ell)$.
\end{Pro}

\proproof{D}


\begin{Pro}\label{pro:5}
Given an arbitrary LMO density with the parameter set $\bpi=\{(\omega^{(I)},P^{(I)}(\cdot))\}_{I\in\mathcal{F}(\mathbb{L})}$,
the LMB  density in the class defined in (\ref{LMB}) which minimizes the Kullback-Leibler divergence from $\bpi$, and  preserves the first-order moment of $\bpi$ is given by $$\hat{\bpi}_{\emph{LMB}}=\{(\hat r^{(\alpha)},\hat p^{(\alpha)}(\cdot))\}_{\alpha\in\mathbb{L}}$$ where 
\begin{align}
\label{LMB_where_r} \hat r^{(\alpha)}&={\sum}_{I\in\mathcal{F}(\mathbb{L})}1_I(\alpha)\omega(I)\\
\label{LMB_where_p} \hat p^{(\alpha)}(x,\!\ell)&\!=\!\frac{1}{\hat r^{(\alpha)}}{\sum}_{I\in\mathcal{F}(\mathbb{L})}1_I(\ell)\omega(I)\delta_{\alpha}(\ell) p_{I-\{\ell\}}(x,\!\ell).
\end{align}
The density $\hat{\bpi}_{\emph{LMB}}$ is referred to as the ``best'' LMB approximation of $\bpi$.
\end{Pro}

\proproof{E}
\subsection{Recursive Equations of the LMB-GOM Filter}
In this subsection, we apply  the derived ``best'' LMB approximation  to the labeled multi-object filtering problem, and develop the LMB-GOM filter.  The following proposition provides the update equations of the LMB-GOM filter.
\begin{Pro}\label{pro:6}
 Suppose that the current multi-object predicted density is an LMB density with the LMB parameters $\bpi_+=\{(r_+^{(\alpha_{+})},p_+^{(\alpha_{+})}(\cdot))\}_{\alpha_{+}\in\mathbb{L}_+}$. Under a generic multi-object likelihood $g(\Upsilon|\bX)$, the best ``LMB'' approximation of the multi-object posterior is $\hat\bpi(\cdot|\Upsilon)=\{(\hat r^{(\alpha_{+})}(\Upsilon),\hat p^{(\alpha_{+})}(\cdot;\Upsilon))\}_{\alpha_{+}\in\mathbb{L}_+}$, where
\begin{equation}
{\small
\begin{split}
\label{update_r}\!\!\!\!\!\!\!\!\!\!\!\!\!\!\!\!\!\!\!\!\!\!\!\!\!\!\!\!\!\!\hat r^{(\alpha_{+})}(\Upsilon)&={\sum}_{I_{+}\in\mathcal{F}(\mathbb{L}_+)}1_{I_{+}}(\alpha_{+})\omega(I_{+};\Upsilon)
\end{split}}
\end{equation}
\begin{equation}
{\small
\begin{split}
\label{update_p}\hat p^{(\alpha_{+})}(x,\ell;\Upsilon)&\!=\!\frac{\delta_{\alpha_{+}}(\ell)}{ r^{(\alpha_{+})}(\Upsilon)}\!\!\sum_{I_{+}\in\mathcal{F}(\mathbb{L}_+)}\!\!1_{I_{+}}(\ell)\omega(I_{+};\Upsilon)p_{I_{+}\!-\{\ell\}}(x,\ell;\Upsilon)
\end{split}}
\end{equation}
with
\begin{align}
\notag p_{\{\ell_1,\cdots,\ell_n\}}(x,\ell;\Upsilon)=&\int P(\{(x,\ell),(x_1,\ell_1),\cdots, \\ 
\label{update_where}&\,\,\,\,(x_n,\ell_n)\};\Upsilon)d(x_1,\cdots,x_n)\\
\label{update_w-lmb} \omega(I_{+};\Upsilon)= &\frac{\eta_{\Upsilon}(I_{+})\omega_+(I_{+})}{\sum_{I_{+}\in\mathcal{F}(\mathbb{L}_+)}\eta_\Upsilon(I_{+})\omega_+(I_{+})}\\
\label{update_P-lmb} P(\bX_{+};\Upsilon)=&\frac{{[p_+]}^\bX g(\Upsilon|\bX_{+})}{\eta_{\Upsilon}(\mathcal{L}(\bX_{+}))}\\
\label{predict_w-lmb}\omega_+(I_{+})=&\!{\prod}_{i\in\mathbb{L}_+}\left(1-r_+^{(i)}\right)\!{\prod}_{\ell\in I_{+}}\frac{1_{\mathbb{L}_+}(\ell)r_+^{(\ell)}}{1-r_+^{(\ell)}}\\
\notag  \eta_{\Upsilon}(\{\ell_1,\ell_2,\!\cdots\!,\ell_n\})=\!\!&\int g(\Upsilon|\{(x_1,\ell_1),\cdots,(x_n,\ell_n)\}) \\ 
\label{where_2}&\!\!\!\! \left({\prod}_{i=1}^n p_{+}(x_i,\ell_{i})\right)d(x_1,\cdots,x_n).
\end{align}
\end{Pro}

\proproof{F}

Proposition \ref{pro:6} explicitly describes how to calculate the posterior 
LMB parameters  $\{(\hat r^{(\alpha_{+}))}(\Upsilon),\hat p^{(\alpha_{+})}(\cdot;\Upsilon)\}_{\ell\in\mathbb{L}_+}$  from the predicted multi-Bernoulli parameters $\{(r_+^{(\alpha_{+})},p_+^{(\alpha_{+})}(\cdot))\}_{\ell\in\mathbb{L}_+}$. The update stage of the LMB-GOM filter  has three steps:

\noindent-- Write the predicted LMB density in the general LMO density form, i.e., $\bpi_+(\bX)=\Delta(\bX)\omega_+(\mathcal{L}(\bX))[p_+]^\bX$;

\noindent-- Compute the full multi-object posterior density $\bpi(\bX_{+}|\Upsilon)$ from the general LMO density form according to Proposition \ref{pro:2}, resulting in $\bpi(\bX_{+}|\Upsilon)=\omega(\mathcal{L}(\bX_{+});\Upsilon)P(\bX_{+};\Upsilon)$; 

\noindent-- Approximate $\bpi(\bX_{+}|\Upsilon)$  with its ``best'' LMB approximation $\hat\bpi(\cdot|\Upsilon)$ according to Proposition \ref{pro:5}.
\begin{Rem}\label{remark:4}
From Propositions \ref{pro:5} and \ref{pro:6}, we can deduce that in the LMB filter proposed in \cite{refr:label_5}, the LMB RFS which matches the first-order moment of the $\delta$-GLMB posterior, also minimizes the KLD from the $\delta$-GLMB posterior, among all the LMB densities.
\end{Rem}

Utilizing Proposition \ref{pro:6} and the prediction equations of the LMB filter in~\cite{refr:label_5}, we   can obtain the recursive equations of the LMB-GOM filter. 
Under the standard object motion model, the  multi-object predicted density 
$\bpi_+$ is  an LMB density if the multi-object prior is an LMB density~\cite{refr:label_5}. Moreover, based on  Proposition \ref{pro:6}, the  multi-object posterior
density $\bpi(\cdot|\Upsilon)$ can  be approximated as a principled LMB density under the GOM, if the  multi-object predicted density $\bpi_+$ is an LMB density. The specified  prediction  and update steps of the LMB-GOM filter are given via the following:
\\
\\
\textit{\textbf{LMB prediction}}: Given the current prior LMB density with the LMB parameters $\bpi=\{(r^{(\alpha)},p^{(\alpha)}(\cdot))\}_{\alpha\in\mathbb{L}}$ and the LMB multi-object birth with the LMB parameters  $\bpi_B=\{(r_B^{(\alpha')},p_B^{(\alpha')}(\cdot))\}_{\alpha'\in\mathbb{B}}$,  the multi-object prediction is  another LMB density on state space $\mathbb{X}$ and finite label space $\mathbb{L}_+=\mathbb{B}\cup\mathbb{L}$ given by 
 \begin{equation}\label{prediction-LMB} \bpi_{+}=\{(r^{(\alpha)}_{+,S},p^{(\alpha)}_{+,S}(\cdot))\}_{\alpha\in\mathbb{L}}\cup\{(r^{(\alpha')}_{B},p^{(\alpha')}_{B}(\cdot))\}_{\alpha'\in\mathbb{B}}
 \end{equation}
 where
 \begin{align}
 \label{r-s} r^{(\alpha)}_{+,S}&=\eta_S^{(\alpha)}r^{(\alpha)} \\
  \label{p-s}p_{+,S}^{(\alpha)}(x_+,\ell_+)&=\delta_{\alpha}(\ell_+)\big<p_S(\cdot)f_+(x_+|\cdot),p^{(\alpha)}(\cdot)\big>\big/\eta_S^{(\alpha)}\\
\label{where_1}\eta_S^{(\alpha)}&=\big<p_S(\cdot),p^{(\alpha)}(\cdot)\big>.
 \end{align}
\textit{\textbf{LMB update}}: Given the current predicted LMB density
  $\bpi_+=\{(r_+^{(\alpha_+)},p_+^{(\alpha_+)}(\cdot))\}_{\alpha_+\in\mathbb{L}_+}$ and 
  the generic multi-object likelihood function $g(\Upsilon|\cdot)$, the  approximate multi-object  posterior   is another LMB density   $\hat\bpi(\cdot;\Upsilon)=\{(\hat r^{(\alpha_+)}(\Upsilon),\hat p^{(\alpha_+)}(\cdot;\Upsilon))\}_{\alpha_+\in\mathbb{L}_+}$  computed by  (\ref{update_r}) and (\ref{update_p}).
\begin{Rem}\label{remark:5}
Compared with the $\delta$-GLMB-GOM filter, the LMB-GOM filter involves  less computation in its prediction step because it not only reduces the integration space to single-object space, but also involves a number of integrals that increases linearly with the object number.  Actually, the computational efficiency of the LMB-GOM filter can be achieved because  the ``best'' LMB approximation completely loses correlation between object states,  while  the ``best''  $\delta$-GLMB approximation still preserves part of the correlation between object states.  The $\delta$-GLMB density has the ability to depict the statistical dependence between points \cite{refr:label_6}. However, unlike the general LMO density,  the points in a $\delta$-GLMB RFS are assumed statistically independent conditional on their existences with a set of distinct labels. This  assumption can lead to a scarification of some part of information on correlation between object states when approximating the full multi-object posterior as the ``best'' $\delta$-GLMB  approximation. As for the LMB density, the points (including object states and their labels) are  assumed to be statistically independent.  Hence, information on correlation between object states is completely discarded when approximating the full multi-object posterior as the ``best'' LMB approximation. 
\end{Rem}
\subsection{The SMC Implementation of the LMB-GOM Filter}

Suppose that the current  LMB prior is parameterised by  $\bpi=\{(r^{(\alpha)},p^{(\alpha)}(\cdot))\}_{\alpha\in\mathbb{L}}$, where each single object density 
$p^{(\alpha)}(x,\ell)$  is approximated by a set of weighted particles. 

At the prediction stage, for each label $\alpha\in\mathbb{L}$ of the surviving objects, the predicted existence probability $r^{(\alpha)}_{+,S}$ and the probability density  $p^{(\alpha)}_{+,S}(\bx)$ are evaluated using the particles and the corresponding weights of $p^{(\alpha)}(x,\ell)$. For explicit calculation formulas, refer to the SMC implementation of multi-Bernoulli filter\cite{refr:MeMber_filter1}.

At the update stage, in the first place, we evaluate the parameter set $\{\omega^{(I_{+})}, P^{(I_{+})}(\bX;\Upsilon) \}_{I_{+}\subseteq \mathbb{L}_{+}}$ of the full multi-object posterior $\bpi_{+}(\bX|\Upsilon)$.  Similar to the SMC implementation of the LMO-GOM filter presented in Subsection III-B, for each label set $I_{+}\subseteq \mathbb{L}_{+}$, the multi-object density  $P^{(I_{+})}(\bX;\Upsilon)$ is approximated  by a set of weighted particles  $\{(w_{j_{+}}^{(I_{+})},\bX_{j_{+}}^{(I_{+})})\}^{N_{p}^{(I_{+})}}_{j_{+}=1}$,
where each particle $\bX_{j_{+}}^{(I_{+})}$ for $j=1,\cdots, N_{p}^{(I_{+})}$ is drawn from a   properly designed  importance density.

Then for each label $\alpha_{+}\in\mathbb{L}_{+}$, the updated LMB parameters $\hat r^{(\alpha_+)}(\Upsilon)$ and $\hat p^{(\alpha_+)}(x_+,\ell_+;\Upsilon)$ can be calculated from the parameter set $\{\omega^{(I_{+})}, P^{(I_{+})}(\bX;\Upsilon) \}_{I_{+}\subseteq \mathbb{L}_{+}}$  utilizing the particles and the corresponding weights of  each  $P^{(I_{+})}(\bX;\Upsilon)$. A key term when calculating the single object density $p^{(\alpha)}(x,\ell;\Upsilon)$ is   $\delta_{\alpha_{+}}(\ell)p_{I_+-\{\ell\}}(x,\!\ell)$ in (\ref{update_p}).  By utilizing (\ref{update_where}) and (\ref{integal-1}), this term  is evaluated  as,
\begin{equation}\label{key-term}
\begin{split}
&\delta_{\alpha_{+}}(\ell)p_{I_+\!-\{\ell\}}(x,\ell)\!\\
= &\delta_{\alpha_{+}}(\ell)\!\int\!\! P^{(I_+)}(\bX_{+}\cup\mathbb{X}\!\times\!\{\ell\};\Upsilon)\delta\bX_+\\
 \propto&\delta_{\alpha_{+}}(\ell){\sum}_{j_{+}=1}^{N_{p}^{(I_+)}}\widetilde\omega^{(I_{+})}_{j_{+}} \int \delta_{\bX^{(I_{+})}_{j_+}}(\bX_{+}\cup\mathbb{X}\times\{\ell\})\delta\bX_+\\
 \propto&{\sum}_{j_{+}=1}^{N_{p}^{(I_+)}}\widetilde\omega^{(I_{+})}_{j_{+}} \delta_{\bX^{(I_+)}_j\cap\mathbb{X}\times \{\alpha_+\}}(x_+,\ell_+).\\
 \end{split}
\end{equation}
After the update step, the resampling and truncation processes are also applied similar to the SMC implementation of the LMO-GOM filter.

\subsection{Grouping based LMB-GOM Filter}
The proposed LMB-GOM filter can be seen as  an extension of the LMB filter proposed in~\cite{refr:label_5} that accommodates generic multi-object likelihood. To enhance the implementation efficiency of the LMB filter for the  standard observation model, the parallel group update via the construction of the so called ``groups'' was proposed in~\cite{refr:label_5}. Each group contains only closely spaced objects and their associated measurements. This method can achieve significant reductions in computation because updating independent groups in parallel is usually much faster than updating the entire multi-target state. In this subsection, we also extend the parallel group update to the LMB-GOM filter.
Combining the prediction step of the LMB-GOM filter with the parallel group update leads to a variant of the LMB-GOM filter,  called the grouping based LMB-GOM (G-LMB-GOM) filter.

In this subsection, the observation set is considered as an RFS $Z$ defined on the observation space $\mathbb{Z}$.  
By  exploiting the  mathematical formulation of LMB RFSs, the LMB predicted density $\bpi_+=\{(r^{(\alpha)},p^{(\alpha)}(\cdot))\}_{\alpha\in\mathbb{L}_+}$ admits an exact decomposition based  an arbitrary partition of tracks in the  label space $\mathbb{L}_{+}$, denoted by $\{\mathbb{L}_+^{1},\cdots, 
\mathbb{L}_+^{N}\}$, i.e.,
\begin{equation}\label{De-LMB-Prediction}
{\small
\begin{split}
\bpi_+(\bX)=
{\prod}_{i=1}^N\bpi^{i}_+(\bX\cap\mathbb{X}\times\mathbb{L}_+^{i})
\end{split}}
\end{equation}
where  $\bpi^{i}_+=\{(r^{(\alpha)},p^{(\alpha)}(\cdot))\}_{\alpha\in\mathbb{L}^{i}_+}$. The decomposition in (\ref{De-LMB-Prediction}) is achieved by utilizing the independence between  Bernoulli components and the convolution formula given in [1, p.385].

Having the flexible decomposition of the LMB prediction, as long as there exist one  partition of the tracks such that the multi-object likelihood can be decomposed as
\begin{equation}\label{De-likelihood}
{\small
\begin{split}
g(Z|\bX)= g(Z^{0}|\emptyset){\prod}_{i=1}^N g(Z^{i}|\bX\cap\mathbb{X}\times\mathbb{L}_+^{i}),
\end{split}}
\end{equation}
where $Z^{i}\subseteq Z$ for $i=1,\cdots, N$ denotes the observation subset associated with the tracks in $\mathbb{L}^{i}_{+}$ and $Z^0=Z-\cup_{i=1}^N Z^i$, then the parallel group update can be achieved, i.e., 
\begin{equation}\label{De-Posterior}
{\small
\begin{split}
&\bpi(\bX|Z)\propto\bpi_+(\bX)g(Z|\bX)
\\ =&{\prod}_{i=1}^N\bpi^{i}_+(\bX\cap\mathbb{X}\times\mathbb{L}_+^{i}) {\prod}_{i=1}^N g(Z^{i}|\bX\cap\mathbb{X}\times\mathbb{L}_+^{i})\\
\propto &{\prod}_{i=1}^N \bpi^{(i)}(\bX\cap\mathbb{X}\times\mathbb{L}_+^{i}|Z^{i}).
\end{split}}
\end{equation}

The decomposition of the multi-object likelihood  in (\ref{De-likelihood}) essentially demands that the effects of different multi-object subsets $\bX\times\mathbb{X}\cap\mathbb{L}_{+}^{i}$ on the observations can be separated. Specifically, each observation subset $Z^{i}$ is only correlated with the multi-object subset $\bX\cap\mathbb{L}\times\mathbb{L}_{+}^{i}$.
Nevertheless, this demand is not necessarily valid  for the GOM. Hence, in the following, we firstly discuss the constrains on the observation model. Then we provide a principled method to partition  tracks and observations for which the decomposition in (\ref{De-likelihood}) holds approximately. Finally, the parallel group update is formulated. 
\subsubsection{Decomposition of the Likelihood
} 
The following assumptions on the observation model are made.

\textit{A.1}: The observations $z\in Z$ are  conditionally independent under   the multi-object state $\bX$;

\textit{A.2}: The object with state $\bx$  only contributes to the observations within a region $T(\bx)\subset\mathbb{Z}$.

The first assumption is common in multi-object tracking (see, for example, \cite{refr:MeMber_filter,refr:tbd-5,refr:superpositional-2}).  
The second assumption indicates that each observation $z\in Z$ is generated by a set of objects $\bX_z=\{\bx\in\bX: z\in T(\bx)\}$ ($\bX_z$  can also be an empty  set).  $T(\bx)$ is referred to as the valid observation region (VOR) of  object $\bx$, and $T(\bX)\triangleq \cup_{\bx\in\bX} T(\bx)$ is referred to as the VOR of the state set $\bX$.  The VOR is related to the sensing characteristic of a sensor.

\begin{Pro}\label{pro:7}
Given an observation model characterized by multi-object likelihood $g(Z|\bX)$,  and satisfing Assumptions A.1 and A.2, if a subset of object states,  $\bX'\subseteq\bX$ satisfies 
\begin{equation}\label{isolated-object-cluster}
T(\bX')\cap T(\bX-\bX')=\emptyset,
\end{equation}
 then the observation subset $Z\cap T(\bX)$ is statistically independent of  object states $\bX-\bX'$, and the observation subset $Z-T(\bX)$ is statistically independent of  object states $\bX$, i.e., the multi-object likelihood can be represented as
\begin{equation}\label{Decomposition-likelihood}
g(Z|\bX)=g(Z\cap T(\bX')|\bX')g(Z- T(\bX')|\bX-\bX')
\end{equation} 
Herein, $\bX'$ is called as an  isolated object cluster. 
\end{Pro}
\proproof{G}

According to Proposition \ref{pro:7},  as long as  one isolated object cluster arises,  the multi-object likelihood can be further  decomposed as (\ref{Decomposition-likelihood}). Observing (\ref{isolated-object-cluster}), one can easily obtain that the smaller the size of the VOR $T(\bx)$ is, the more likely it is  for an isolated object cluster to arise. Generally, the sensor models can be divided into three categories in terms of different types of VORs.

$\bullet$ \textit{Type I: Completely confined VOR. The size of  $T(\bx)$ is relatively small compared with the observation space $\mathbb{Z}$, namely,  an object $\bx$ can only affect  the observations in a  very limited region,  and then the  contribution of $\bx$ on the observations beyond this region is zero. } For example, in video tracking \cite{refr:MeMber_filter,refr:computer-vision,refr:vedio-tracking}, a  rigid body   can only occupy several  pixels of its surroundings. For this category of sensors, it is easy to produce  isolated object clusters  and then exactly decompose the likelihood according to Proposition \ref{pro:7}. The pixeled TBD observation model employed in subsection III-E belongs to this type.

Except for Type I sensors, there also exist sensors  whose VOR is  the whole observation space (or a region having a comparable  size with the whole observation space). Hence,  all the objects   contribute to almost all  the observations, making the  observations correlated with all the objects. These sensors    can be further classified into two types as follow.

$\bullet$ \textit{Type II: Approximately confined VOR. Correlation between  observation $z$ and  object $\bx$ decays as the ``distance'' between $z$ and $\bx$ increases. } The acoustic sensor network  observation model \cite{refr:amplitude} shown in Section III-E is a typical example. When the distance $\|\rho(\bx)-\xi_{m}\|$ between the sensor  and the object   is sufficiently large, the received sound amplitude  at sensor $\xi_{m}$ due to the object $\bx$ decays rapidly according to $\frac{A}{\|\rho(\bx)-\xi_{m}\|}$. Hence, the contribution of  the object $\bx$ to the observation  at sensor $m$ can be negligible. Consequently,   by suitably truncating the complete VOR,  the decomposition of the multi-object likelihood according to (\ref{Decomposition-likelihood}) can be achieved with an affordble approximation error. 

$\bullet$  \textit{Type III: Full VOR.  In this case, observations are strongly correlated to all the objects, and the decomposition of likelihood  is  not possible.}
For instance, when estimating the slowly diffusing sources using a sensor network, the received observations  at a certain sensor are strongly affected by  all the remote sources \cite{diffusion_source}. 

\subsubsection{Grouping and Parallel Group Update}
For  the standard observation model, track grouping is based on a standard gating procedure which also partitions the observation set~\cite{refr:JPDA,refr:label_5}.  
Inspired by this, this subsection provides  a principled method to construct independent groups of tracks and observations for a wide variety of observation models. 
%
%
%
%
The following two definitions  will be used in formulating our method. 

\begin{Def}\label{definition:3}
%
%
%
  Let $f_{\Theta}(x)$ be a density function of a random variable $\Theta$. A measurable subset of  the sample space $\mathbb{O}$ of $\Theta$, denoted by $R$ is called the highest  density  region (H.D.R.) of confidence $\lambda$ if
 
a) $\Pr\{\Theta\in R\} = \int_{R} f_\Theta (x) d x=\lambda$;
 
b) for $x_1\in R$ and $x_2\notin R$, $f_{\Theta}(x_1)\geqslant f_{\Theta}(x_2)$.
\end{Def}

\begin{Rem}\label{remark:6} The concept of H.D.R. is provided in \cite{refr:HPD,refr:HDR}. The posterior density for every point inside the H.D.R.  is greater than that for every point outside of region. Thus, the region includes the more probable values of $\Theta$. Usually, the confidence $\lambda$ is set to be very close to one, e.g. $\lambda= 0.99$. Thus $f_\Theta(\!x\!)$ is negligible for $x\!\notin\!R$ and can be approximated with 0.
%
\end{Rem}

\begin{Def}\label{definition:4}
Consider an LMB density with the LMB parameters $\bpi=\{(r^{(\alpha)},p^{(\alpha)}(\cdot))\}_{\alpha\in\mathbb{L}}$.  Denote the H.D.R. of
confidence $\lambda$ for $p^{(\alpha)}(\cdot)$  by $\overline{\mathbb{X}}^{(\alpha)}$, with $\overline{\mathbb{X}}^{(\alpha)}\subset\mathbb{X}\times\mathbb{L}$.   $T^{(\alpha)}=\bigcup_{\bx\in \overline{\mathbb{X}}^{(\alpha)}}T(\bx)$ is called as the VOR of track 
$\alpha$. Tracks $\alpha$ and $\alpha'$ are referred to as the coupling tracks   if their  VORs  have intersection, i.e., $T^{(\alpha)}\cap T^{(\alpha')}\neq\emptyset$.
\end{Def}

 Given the LMB prediction with the LMB parameters $\bpi_+=\{r^{(\alpha)},p^{(\alpha)}(\cdot)\}_{\alpha\in\mathbb{L}_+}$,  the predicted label set is partitioned as $\mathbb{L}_+ = \biguplus_{i=1:N}\ \mathbb{L}_+^{i}$ such that no track in $\mathbb{L}_+^{i}$ is coupled with any track in $\mathbb{L}_+^{j}$ for any $i\neq j$, where $\biguplus$ denotes the disjoint union.  In other words, $\forall (i,j)\in[1,N]^2$,
\begin{equation}\label{grouping-c}
{\small
\begin{split}
 i\neq j\ \Rightarrow \ \ \left(\cup_{\alpha\in\mathbb{L}_+^{i}}T^{(\alpha)}\right)\bigcap \left(\cup_{\alpha'\in\mathbb{L}_+^{j}}T^{(\alpha')}\right)=\emptyset.
 \end{split}}
\end{equation}
Accordingly,  the multi-object observation set $Z$ is partitioned as $\{Z^{0},Z^{1},\cdots, Z^{N}\}$ where 
\begin{equation}\label{grouping-Z-1}
{\small
\begin{split}
Z^{i}=Z\bigcap\left(\cup_{\alpha\in\mathbb{L}^{i}_+}T^{(\alpha)}\right)
\end{split}}
\end{equation}
  denotes the observation subset related to the group of tracks with label subset $\mathbb{L}_+^{(i)}$,  $i=1,\cdots,N$, and 
\begin{equation}\label{grouping-Z-2}
{\small
\begin{split}
Z^{0}=Z-{\bigcup}_{i=1}^N Z^{i}
\end{split}}
\end{equation}
 denotes the observation subset  having no associated tracks.

The above partitions of the predicted label set and the observation set naturally produce a set of  pairs $$\{(\mathbb{L}^{1}_+,Z^{1}),\cdots,(\mathbb{L}_+^{N},Z^{N})\}$$ with each $(\mathbb{L}^{i}_+,Z^{i})$, $i=1,\cdots,N$,  referred to as a group. 

Consider the multi-object state $\bX\subseteq\bigcup_{\alpha\in\mathbb{L}_{+}}\overline{\mathbb{X}}^{(\alpha)}$ with  confidence $\lambda$ sufficiently large. According to Definition \ref{definition:4}, for each group $(\mathbb{L}_{+}^{i},Z^{i})$, we have 
$$\cup_{\alpha\in\mathbb{L}_{+}^{i}}T^{(\alpha)}\supseteq T(\bX\cap\mathbb{X}\times\mathbb{L}_{+}^{i}),$$ then by the combination of (\ref{grouping-c}), the observation subset $Z^{j}$ of  any other group with $j\neq i$ has the following relationship,
$$Z^{j}\subseteq Z-\cup_{\alpha\in\mathbb{L}_{+}^{i}}T^{(\alpha)}\subseteq Z-T(\bX\cap\mathbb{X}\times\mathbb{L}_{+}^{i}).$$ Under Assumption \textit{A.1}, by utilizing the independence between any $Z^{j}$ and $\bX\cap\mathbb{X}\times\mathbb{L}_{+}^{i}$ ($i\neq j$), the multi-object  likelihood can be  decomposed 
 as
\begin{equation} 
{\small
\begin{split}
g(Z|\bX)\cong g(Z^{0}|\emptyset){\prod}_{i=1}^N g(Z^{i}|\bX\cap\mathbb{X}\times\mathbb{L}_+^{i}),
\end{split}}
\end{equation}
and consequently the posterior density is decomposed as
\begin{equation}\label{De-Posterior}
{\small
\begin{split}
&\bpi(\bX|Z)
\propto {\prod}_{i=1}^N \bpi^{i}(\bX\cap\mathbb{X}\times\mathbb{L}_+^{i}|Z^{i})
\end{split}}
\end{equation}
where $\bpi^{i}(\cdot|Z^{i})$ denotes the posterior density of the $i$th group.

For the multi-object state $\bX\nsubseteq\bigcup_{\alpha\in\mathbb{L}_{+}}\overline{\mathbb{X}}^{(\alpha)}$ with  confidence $\lambda$ sufficiently large, the predicted density $ \bpi_{+}(\bX)=\Delta(\bX)\omega_{+}(\mathcal{L}(\bX))p_{+}^{\bX}$ is negligible, and consequently the corresponding posterior density $\bpi(\bX|Z)$ is negligible.

As a result,  the full Bayes update  can be approximated as  a group  of parallel  updates.  Specifically, the LMB prediction $\bpi^{i}_+$ for the $i$th group, 
is updated by the  likelihood $g(Z^{i}|\cdot)$  resulting in the posterior density $\bpi^{i}(\cdot|Z^{i})$ of the $i$th group.
\subsubsection{Partition Criterion}
An important issue of the partition procedure is  the choice of the criterion used to  judge whether two tracks are coupling or not.  A straightforward criterion according to the previous subsections  is 
the  predicted tracks $\alpha$ and $\alpha'$ exhibit significant coupling  if their VORs have the intersection, i.e., $T^{(\alpha)}\cap T^{(\alpha')}\neq\emptyset.$


In practice, the criterion can be simplified by the combination of the specific observation model.  Taking the two observation models provided in Section III-E as examples, we provide principled criterions as follow.
%

 -- For the pixeled TBD model, 
 as suggested by \cite{refr:tbd-2,refr:tbd-3, refr:tbd-4},  the  predicted tracks $\alpha$ and $\alpha'$ exhibit significant coupling 
if their distance is small, i.e.,
\begin{equation}\label{grouping}
d(\alpha,\alpha')\leqslant \Lambda
\end{equation}
where  $\Lambda$ is a grouping threshold and $d(\cdot,\cdot)$ is a distance function which depends on the way in which observations are acquired and the statistics of the predicted tracks.  A feasible  distance function is  
\begin{equation}\label{distance-function}
d(\alpha,\alpha')=\| \hat z_+^{(\alpha)}-\hat z_+^{(\alpha')}\|,
\end{equation}
where $\hat z_+^{(\alpha)}$ is the  predicted position of the track $\alpha$, and $\|\cdot\|$ denotes 2-norm distance~\cite{refr:tbd-2,refr:tbd-3, refr:tbd-4}. 
In this case, the threshold $\Lambda$ is mainly decided by both the covariance of   $p^{(\alpha)}(\cdot)$ and the VOR $T(\bx)$. Analytical details of the selection of the threshold can be found in~\cite{refr:tbd-3}. 
 Another suitable distance function can be the Mahalanobis distance (MHD) which depicts the impacts of both state and covariance estimate, and then the threshold is mainly  decided by the VOR $T(\bx)$.
 
-- For the acoustic amplitude sensor model, a feasible    criterion is   the predicted tracks $\alpha$ and $\alpha'$ exhibit coupling if
\begin{equation}\label{grouping-threshold}
\{(z_{m},\xi_{m}): \|\xi_{m}-\hat z^{(\alpha)}\|\leqslant \beta\}\cap \{(z'_{m},\xi'_{m}): \|\xi'_{m}-\hat z^{(\alpha')}\|\leqslant \beta\}
\end{equation}
 where  $\beta$ is a given threshold for which the value $\frac{A}{\beta^{\kappa}}$ is sufficiently small. 

After the criterion is established, we can obtain the partition of tracks by adopting suitable clustering algorithms~\cite{refr:clustering_method}. 
Then according to (\ref{grouping-Z-1}) and (\ref{grouping-Z-2}), the associated observation subset of each group can be obtained.


\begin{algorithm}[t]\label{algorithm: generate lgmb}
\caption{The G-LMB-GOM filter.}
\begin{minipage}{0.93\columnwidth}
{\small
\textbf{Input: } the prior LMB density with the LMB parameters $\bpi\!=\!\{(r^{(\alpha)},p^{(\alpha)}(\cdot))\}_{\alpha\in\mathbb{L}}$.  
\begin{enumerate} [$1.$]
\item  
Perform LMB prediction  and
 compute the predicted LMB density 
with the LMB parameters 
 $\bpi_+\!=\!\{r^{(\alpha)}_+,p_+^{(\alpha)}(\cdot)\}_{\alpha\in\mathbb{L}_+}$ using (\ref{prediction-LMB})-(\ref{where_1}).
\item 
Partition the predicted tracks in $\mathbb{L}_+$ and the observation set $Z$ into  groups  $\{(\mathbb{L}^{1}_+,Z^{1}),\cdots,(\mathbb{L}^{N}_+,Z^{N})\}$ according to (\ref{grouping-c})$-$(\ref{grouping-Z-2}).
\item \textbf{for }\textit{each group} $(\mathbb{L}_+^{i},Z^{i})$, $i=1,\cdots,N$ \textbf{do}

\item Perform LMB update under the multi-object likelihood $g(Z^{i}|\cdot)$ using (\ref{update_r})-(\ref{where_2}) and get the ``best'' LMB  approximation of posterior multi-object density $ \hat \bpi^{i}(\cdot|Z^{i})=\{(\hat r^{(\alpha^i)}(Z^{i}),\hat p^{(\alpha^i)}(\cdot;Z^{i}))\}_{\alpha^i\in\mathbb{L}_+^{i}}$.
\item \textbf{end}
\end{enumerate} 
\textbf{Output:} the posterior LMB density with the LMB parameters $\hat \bpi(\cdot|Z)=\cup_{i=1}^N \hat\bpi^{i}(\cdot|Z^{i})$.}
\end{minipage}
\end{algorithm}

  \begin{Rem}\label{remark:8}
The G-LMB-GOM filter can be extended to the case of vector observations easily, because a random vector can be equivalently transformed to a labeled RFS having a constant cardinality~\cite{set-jpda}. 
\end{Rem} 

\begin{Rem}\label{remark:8}
If a Type III sensor is used or all tracks are too close to be isolated, then the partition of tracks and observations is not possible. In this case, the G-LMB-GOM filter degenerates to the LMB-GOM filter automatically. 
\end{Rem} 
\subsubsection{Summary}
Algorithm 1 summarizes the steps through which the G-LMB-GOM filter can be implemented.  The advantages of  the G-LMB-GOM filter are two-fold:

\noindent-- Firstly, it  improves the computational efficiency dramatically by exploiting the  parallel  implementation. The detailed computational complexity is analyzed later in Section V. 

\noindent-- Secondly, it has the potential to improve the tracking performance especially when computing and memory resources (e.g., the  number of particles that can be handled in real-time applications) are limited. On one hand, the performance compromise incurred by the grouping procedure  is slight  when the grouping threshold is sufficiently large. On the other hand,   the densities required to be approximated (by particles) after grouping at the update stage have much lower dimensions than those in the original LMB-GOM and $\delta$-GLMB-GOM filters.  Since the number of particles required to keep a certain tracking performance increases exponentially with the dimension of the state space to be sampled \cite{refr:curse_dimension,refr:tbd-3}, the performance improvement stemmed from the better numerical approximation of the lower dimensional densities (given a fixed number of particles)  can sometimes go   beyond the inaccuracy due to the grouping procedure.


\section{Computational Complexity Analyses and Schematics}\label{chp:6}
In this section, we compare the LMO-GOM, $\delta$-GLMB-GOM, LMB-GOM and G-LMB-GOM filters
 in terms of 
  computational complexities of their respective prediction and update equations, as shown in Table I. Fig.~\ref{fig:flow} shows how these filters operate at the conceptual level.  
All these algorithms can accommodate the GOM because they all embed the  LMO-GOM update (or the parallel group LMO-GOM update)  which is  an exact solution with the generic multi-object likelihood. 

\begin{table}[h]
\caption{Computational complexity analysis}
\hrule\vspace{0.7mm}\hrule
\vspace{2mm}
$\bullet\,\,$
\textit{\textbf{LMO-GOM filter}}:  In the prediction equations, a dominant portion of computation is for calculating the quantities $\eta_{S,I}(J) , J\subseteq I, I\subseteq\mathbb{L}$ in (\ref{where_0}) which involve $ \sum_{a=n}^{|\mathbb{L}|}C_a^n$ integrals each to be computed on $\mathbb{X}^n$ with $n$ varying from $1$ to $|\mathbb{L}|$, where  $C_a^n$ denotes the number of possible combinations of $n$ objects from a set of $a$ objects. In the update equation, computation is dominated by calculation of the quantities $\eta_{\Upsilon}(I), I\subseteq\mathbb{L_+}$ in (\ref{where_eta}) which involve computing $C_{|\mathbb{L}_+|}^n$ integrals on $\mathbb{X}^n$ with $n$ varying from $1$ to $|\mathbb{L}_+|$. 
\vspace{2mm}

$\bullet\,\,$
\textit{\textbf{$\delta$-GLMB-GOM filter}}:  In the prediction equations, the main part of computation is for the quantities $\eta_{S}^{(I)}(\ell), \ell\in I, I\subseteq\mathbb{L}$ which involve computing $\sum_{n=1}^{|\mathbb{L}|}n\cdot C_{|\mathbb{L}|}^n$ integrals on  $\mathbb{X}$. In the update equation, computation is mainly for the quantities $\eta_{\Upsilon}(I), I\subseteq\mathbb{L_+}$ with its computational complexity being the same as that of the LMO-GOM filter. 
\vspace{2mm}

$\bullet\,\,$
\textit{\textbf{LMB-GOM filter}}:  In the prediction equations, a major part of computation is for the quantities $\eta_{S}(\ell), \ell\in\mathbb{L}$ in (\ref{where_1}) which involve computing $|\mathbb{L}|$ integrals on  $\mathbb{X}$. In the update equation, the main part of computation is for the quantities $\eta_{\Upsilon}(I), I\subseteq\mathbb{L}_+$ in (\ref{where_2}) whose computational complexity are also the same as that of the LMO-GOM filter. 
\vspace{2mm}

$\bullet\,\,$
\textit{\textbf{G-LMB-GOM filter}}:  The computational complexity of prediction equations is same as that of the LMB-GOM filter. If the label space $\mathbb{L}_+$ is  partitioned into $\{\mathbb{L}^{1}_+,\cdots,\mathbb{L}_+^{N}\}$, then  the main part of computation in the update step is for calculating $N$ groups of  quantities $\{\eta_{\Upsilon^{i}}(I^i), I^i\subseteq\mathbb{L}^{i}_+\}_{i=1}^N$. For a certain group $i$,  it involves computing $C_{{\small{|\mathbb{L}^{i}_+}}|}^n$ integrals on $\mathbb{X}^n$ with $n$ varying from $1$ to $|\mathbb{L}^{i}_+|$.  Another computation lies in the clustering algorithm  for the partition procedure. 
Taking the  hierarchical clustering    algorithm \cite{refr:clustering_method} as an example,
the computational expense  is $\mathcal{O}(|\mathbb{L}_{+}|^{2}\log(|\mathbb{L}_{+}|))$, which is much cheaper than the computational expense for filtering.
\vspace{2mm}\hrule
\end{table}
\begin{figure}[t]
  \centering
  \includegraphics[width=9cm]{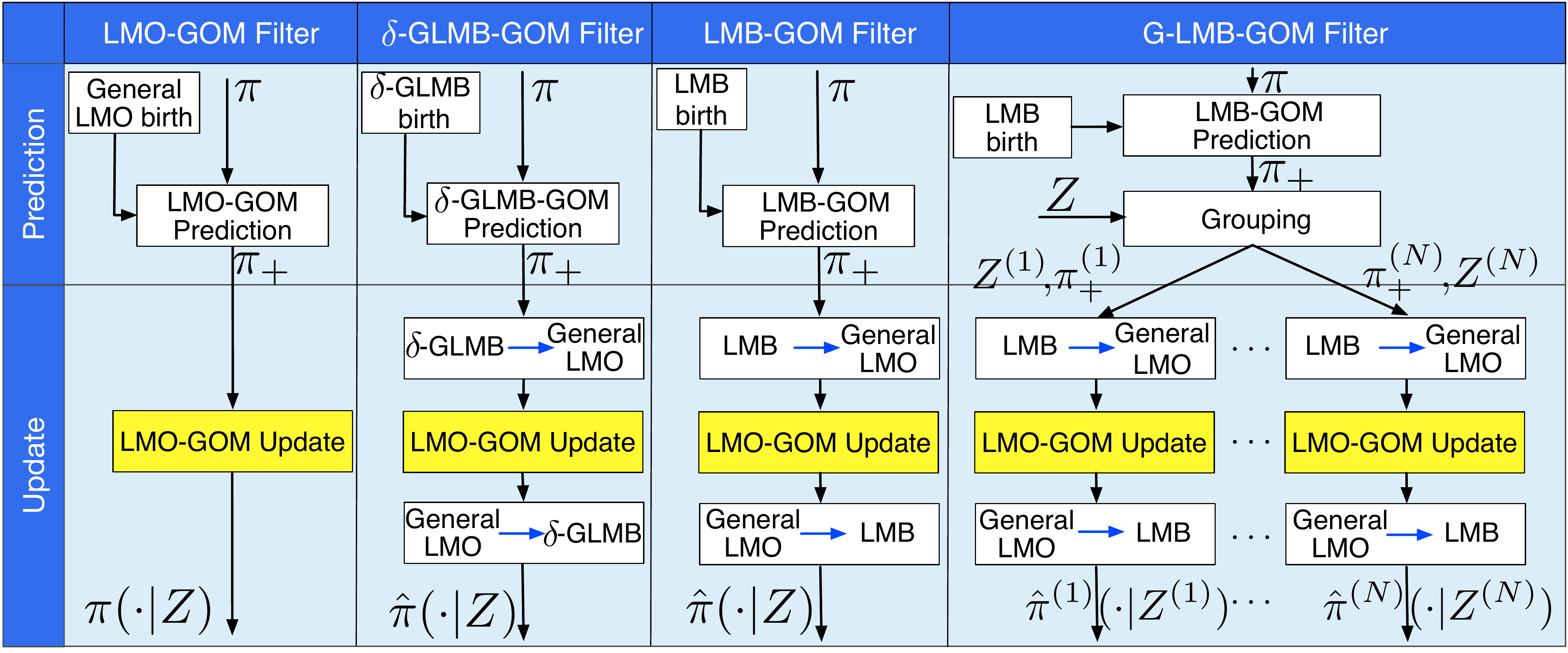}\\
  \caption{Schematic presentation of how the LMO-GOM, $\delta$-GLMB-GOM, LMB-GOM and G-LMB-GOM filters operate in  prediction and update steps.}
  \label{fig:flow}
\end{figure}

 \section{Performance Assessment}\label{chp:7}
 In this section,  the performance of the  proposed algorithms including the LMO-GOM,  LMB-GOM, and G-LMB-GOM filters is examined and compared with the state-of-the-art in comprehensive numerical experiments. The two observation models listed in Subsection II-E, i.e., the pixeled TBD model and the acoustic amplitude model are considered in our experiments.  As we  analysed in Subsection IV-D, these two observation models are two typical examples of Type I and Type II sensors, respectively.  All the algorithms are implemented using the SMC approximation method.

\begin{table}
\caption{ Particle Number of  Parameters for  Different Algorithms}
\begin{tabular}{p{0.85cm}|p{1.375cm}|p{1.375cm}|p{1.375cm}|p{1.375cm}}
\hline\hline
$\,$&
LMO-GOM Filter & $\delta$-GLMB-GOM Filter&LMB-GOM Filter& G-LMB-GOM Filter\\
\hline
\multirow{3}{*}{Prior} 
&  $P^{(I)}(\cdot)$ & $p^{(I,\alpha)}(\cdot)$ & $p^{(\alpha)}(\cdot)$ & $p^{(\alpha)}(\cdot)$ \\
\cline{2-5}
& $2\times10^{6}$ & $N_{p}$ & $N_{p}$ &$N_{p}$ \\
\hline
\multirow{3}{*}{Posterior} 
&  $P^{(I_{+})}(\cdot;\Upsilon)$ & $P^{(I_{+})}(\cdot;\Upsilon)$ & $P^{(I_{+})}(\cdot;\Upsilon)$ & $P^{(I^{i}_{+})}(\cdot;\Upsilon^{i})$ \\
\cline{2-5}
& $2\times10^{6}$ & $N_{p}$ & $N_{p}$ & $N_{p}$ \\
\hline
\end{tabular}
\end{table}

The standard  multi-object transition kernel provided in Section II-D is adopted. The kinematic object state variable is a vector of the plannar position and velocity $x=
\begin{bmatrix}
p_{x} & p_{y} & \dot{p}_{x} & \dot{p}_{y}
\end{bmatrix}
^{\top}$, where ``$^{\top}$'' denotes matrix transpose. The single-object transition model is  linear Gaussian with
\begin{equation} \label{Gaussian-model}
{\small
\mathbf{F}=
 \begin{bmatrix}
\mathbf{I}_{2}& \Delta \mathbf{I}_{2} \\ \mathbf{0}_{2} & \mathbf{I}_{2}
 \end{bmatrix}
,\,\,\,\,\,\,\, \mathbf{Q}=\sigma_v^{2}
\begin{bmatrix}
\frac{\Delta^4}{3}\mathbf{I}_{2}& \frac{\Delta^3}{2}\mathbf{I}_{2} 
\\ 
\frac{\Delta^3}{2}\mathbf{I}_{2} & \Delta^2\mathbf{I}_{2}
\end{bmatrix}}
\end{equation}
where $\mathbf{I}_{2}$ and $\mathbf{0}_{2}$ denote the $2\times 2$ identity and zero matrices respectively,  $\Delta=1$s is the sampling period, and $\sigma_{v}$ is the standard deviation of the process noise.  The probability of object survival  $P_{S}$ is set to be 0.98.

The optimal sub-pattern assignment (OSPA) error \cite{refr:OSPA} serves as the main performance metric with the cut-off value $c = 30$\,m and the order parameter $p = 1$. All performance metrics are averaged over 100 MC runs.
\subsection{Pixeled TBD Model}
 The  efficacy of the proposed algorithms   is  first evaluated  in a typical TBD scenario which presents object crossing, objects in a close proximity for a long time,  and  well-separated objects.  Observations are collected on a  $50\times50$ array of cells with cell lengths $\delta_x\!=\!\delta_y\!=\!1\,\text{m}$. The blurring factor for the Gaussian point spread function is set to be $\delta^2_b=1$. The effective template  is the $7\!\times\!7$ pixel  square region whose center is closest to $(p_x,p_y)$.  The SNR value of each object is set to be 15\,dB.  Figs. \ref{fig: scenario and observation map}(a) and (b) show the trajectories of five objects and an  observation map at a certain time step, respectively. 
 The duration of this scenario is  $T_s=28$\,s.
 
We compare our methods with  the $\delta$-GLMB-GOM filter and the  MB-TBD filter~\cite{refr:MeMber_filter}. The SMC implementation for the MB-TBD filter adopts $5\times 10^4$ particles for each Bernoulli component.   The particles employed by 
the other algorithms are assigned according to Table II, with  $N_p=5\times 10^{4}$.  With the  G-LMB-GOM filter, we choose the partition criterion given in  (\ref{grouping}) with the distance function  (\ref{distance-function}), and the grouping threshold is set to be $\Lambda=10$\,m.

 One of the main purposes of this experiment is  to verify that the LMO-GOM filter is possibly served as the theoretical performance upper bound under the standard observation model as we analysed in Remark \ref{remark:3}. Hence, in order to  guarantee a negligible numerical error with a sufficiently large but tractable number of particles  (i.e., $2\times 10^{6}$), the uncertainties of parameters are set to be relatively low. Specifically, all filters assume no object births and are initialized from the regions around  the correct object positions. Also the five trajectories are considered with only slight maneuverability, i.e., $\sigma_\nu=0.01\,\text{m}/\text{s}^{2}$.  
The aim of this setting is to ensure a controlled experiment in which the objects can approach each other in a small distance for a relatively long period.  The duration of this scenario is  $T_s=60$\,s.

   \begin{figure}[ht]
\setlength{\abovecaptionskip}{-7pt}
\setlength{\belowcaptionskip}{-7pt}
\begin{minipage}[b]{0.48\linewidth}
  \centering
\centerline{\includegraphics[width=4.9cm]{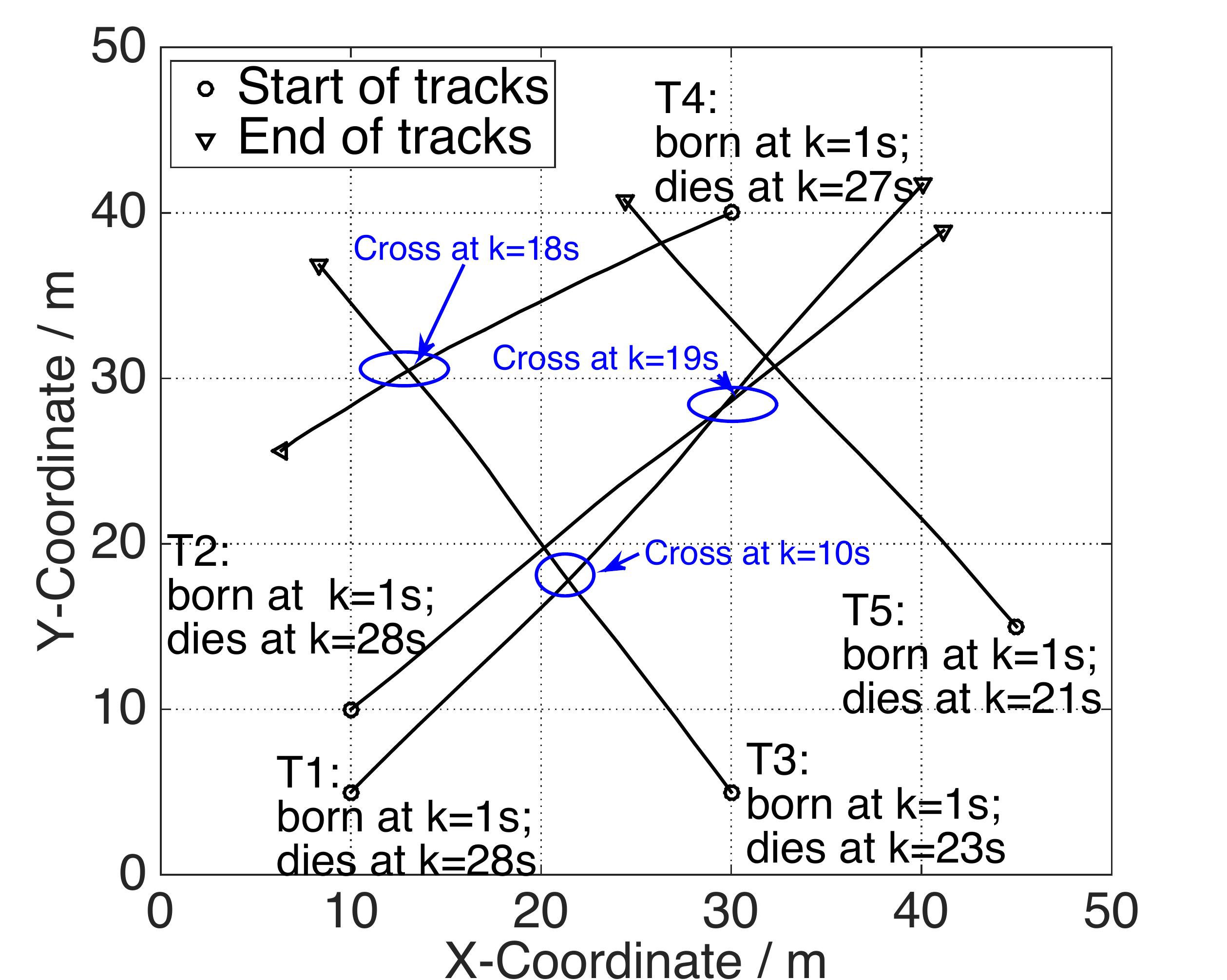}}
\centerline{\small{\small{(a)}} }\medskip
\end{minipage}
\hfill
\begin{minipage}[b]{0.48\linewidth}
  \centering
  \centerline{\includegraphics[width=4.95cm]{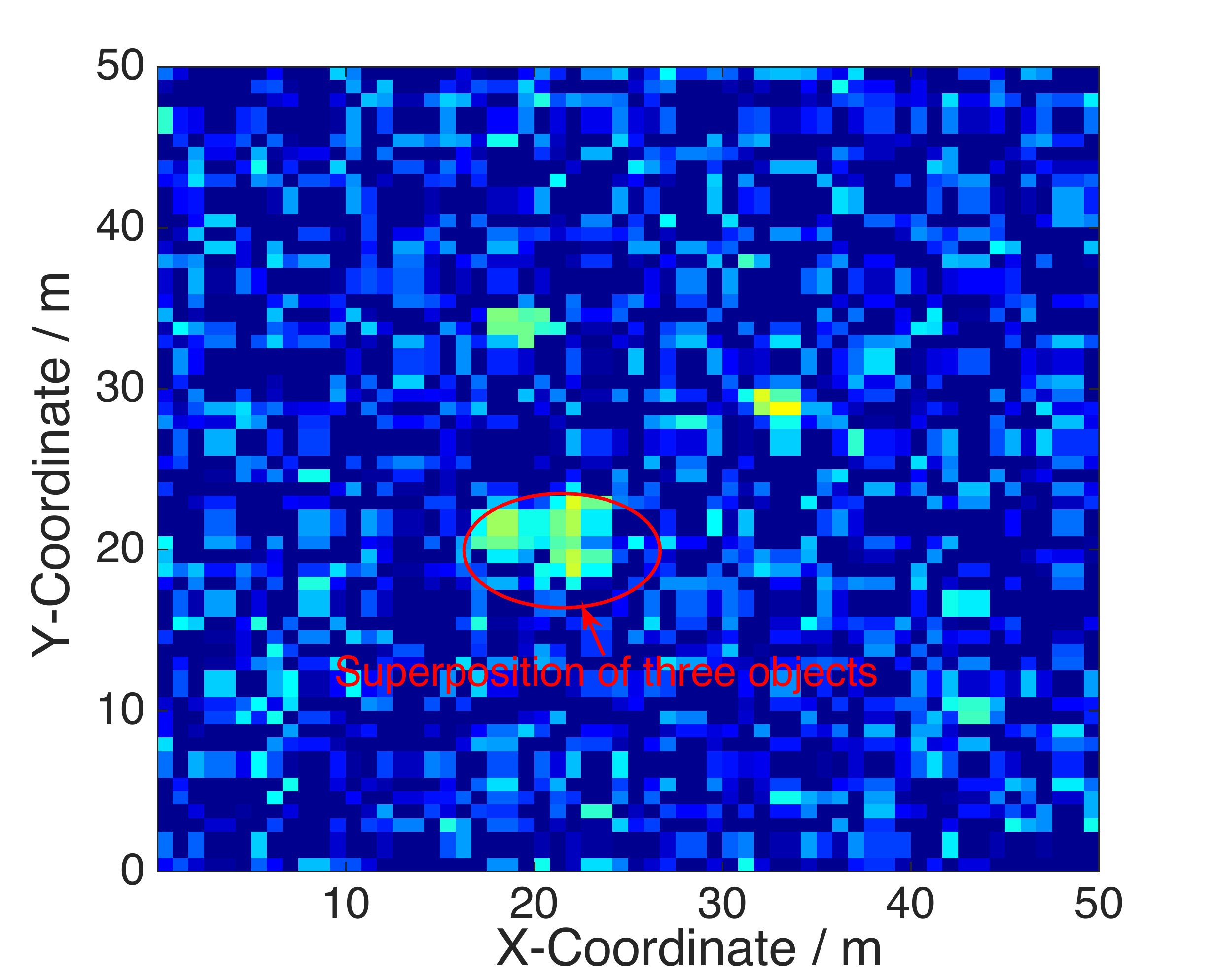}}
\centerline{\small{\small{(b)}}}\medskip
\end{minipage}
\caption{(a) The trajectories of five objects in $x-y$ plane with the initial positions of objects indicated by several crosses; (b) An observation map  at time $k=12$\,s.\label{fig: scenario and observation map}}
\end{figure}
\begin{figure}[ht]
\begin{minipage}[b]{0.49\linewidth}
  \centering
\centerline{\includegraphics[width=4.85cm]{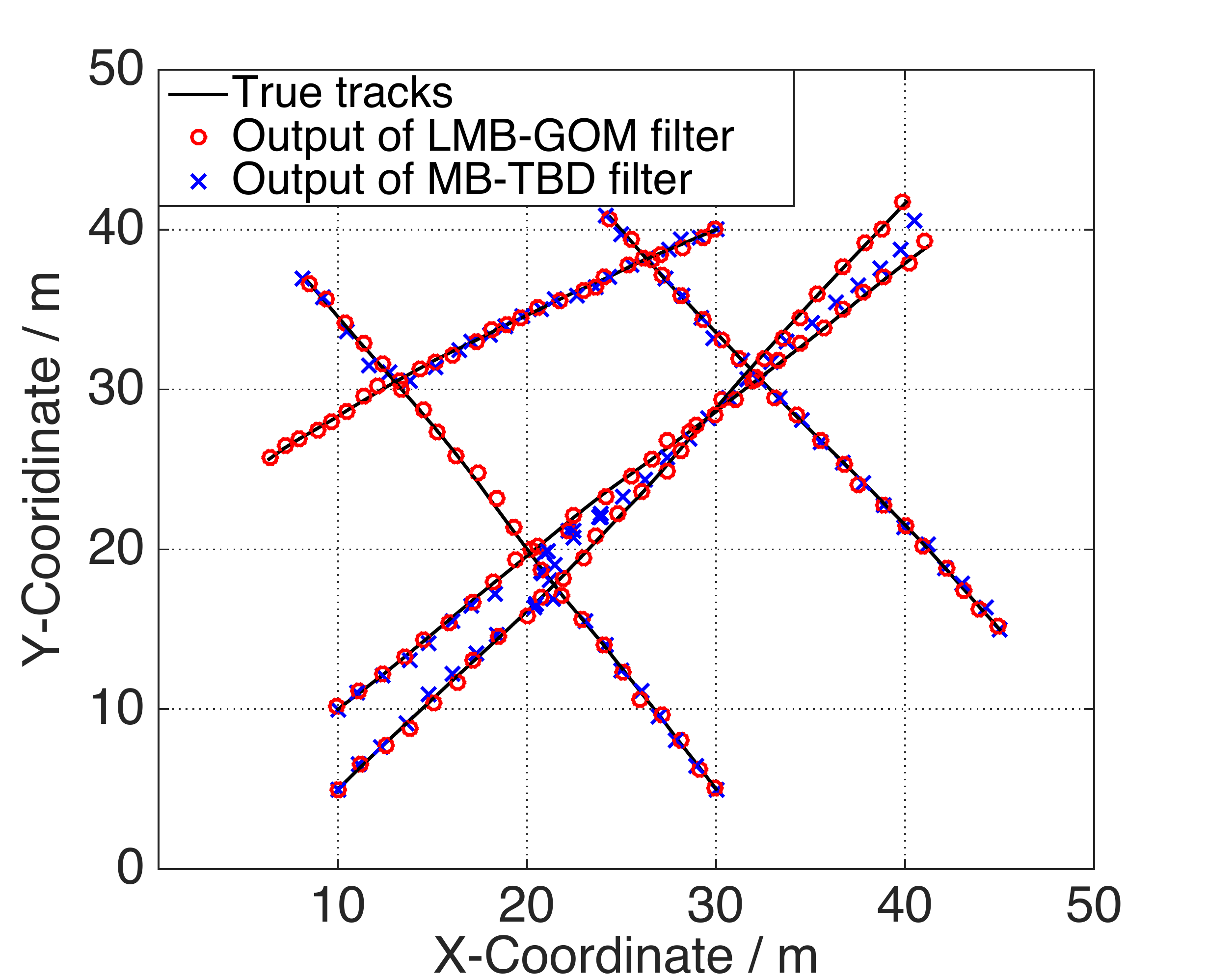}}
\centerline{\small{\small{(a)}} }\medskip
\end{minipage}
\hfill
\begin{minipage}[b]{0.49\linewidth}
  \centering
\centerline{\includegraphics[width=4.85cm]{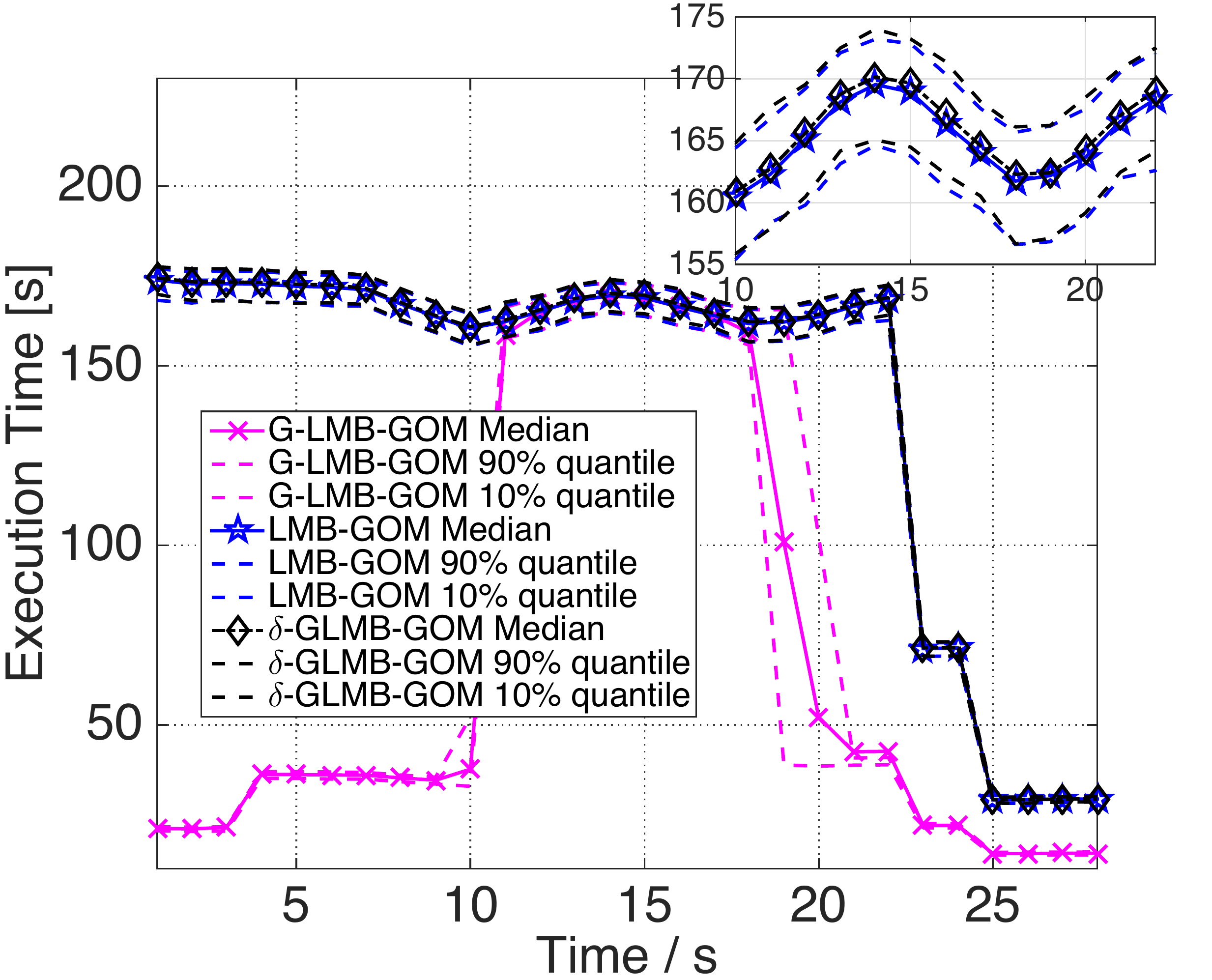}}
\centerline{\small{\small{(b)}}}\medskip
\end{minipage}
\caption{ (a) The respective outputs   of the LMB-GOM and MB-TBD filters for a single MC run; (b)  Execution time per frame for the $\delta$-GLMB-GOM, LMO-GOM and G-LMB-GOM filters.\label{fig: computing time and single-run output}}
\end{figure}
   \begin{figure}[ht]
\setlength{\abovecaptionskip}{-7pt}
\setlength{\belowcaptionskip}{-7pt}
\begin{minipage}[b]{0.48\linewidth}
  \centering
\centerline{\includegraphics[width=5.5cm]{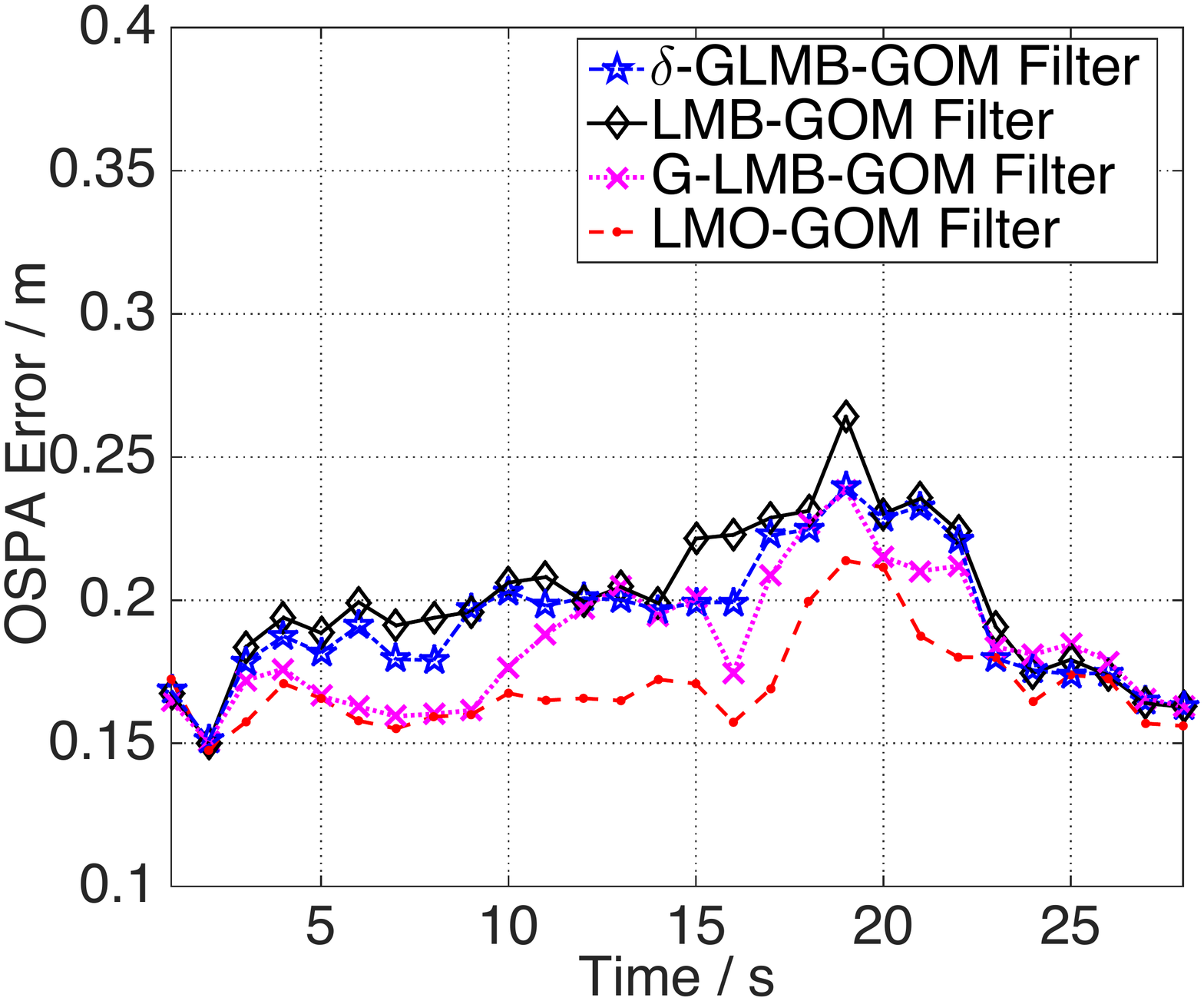}}
\centerline{\small{\small{(a)}} }\medskip
\end{minipage}
\hfill
\begin{minipage}[b]{0.48\linewidth}
  \centering
\centerline{\includegraphics[width=5.5cm]{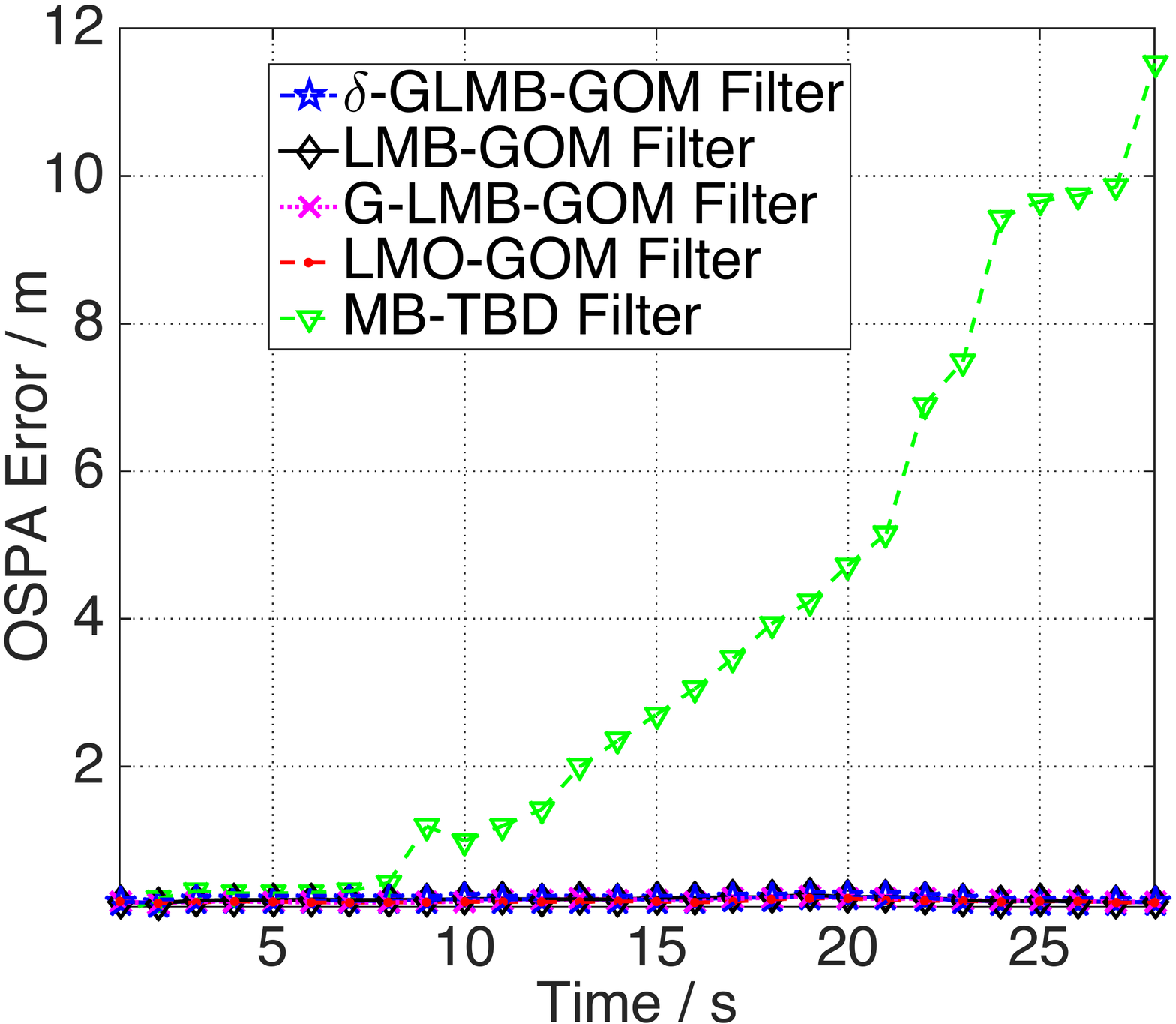}}
  \centerline{\small{\small{(b)}}}\medskip
\end{minipage}
\caption{(a) OSPA errors: (a) the $\delta$-GLMB-GOM, LMB-GOM, G-LMB-GOM and LMO-GOM filters; (b) all five algorithms.}
  \label{fig: performance of five targets}
\end{figure}

Fig.~\ref{fig: computing time and single-run output}(a) shows the respective outputs of the LMB-GOM  and MB-TBD filters for a single MC run.   It can be seen that the LMB-GOM filter performs accurately and consistently for the entire scenario
in the sense that it maintains locking on all tracks and correctly
estimates object positions.
  On the other hand, the MB-TBD filter performs considerably worse.  Specifically,  it loses object tracks very quickly after object crossing since  object superpositions are 
not formulated in the MB-TBD filter.

Fig. \ref{fig: computing time and single-run output}(b) shows the execution times per frame for the $\delta$-GLMB-GOM, LMB-GOM and G-LMB-GOM filters. It can be seen that the execution  time of the LMB-GOM filter is only slightly less than the $\delta$-GLMB-GOM filter since the scenario only considers a relative small and fixed number of objects without object birth, i.e., $|\mathbb{L}|=5$. However, due to the utilization of  parallel group updates, the execution time of the G-LMB-GOM filter is dramatically less than its other competitors especially when more separated objects exist during periods 1\,--\,10\,s and 19\,--\,28\,s.

Fig.~\ref{fig: performance of five targets}(a) shows  the estimation errors over time in terms of average OSPA errors   for the LMO-GOM,  LMB-GOM,  G-LMB-GOM and $\delta$-GLMB-GOM  filters.  We observe comparable performance from the LMB-GOM and $\delta$-GLMB-GOM filters at all times except for the periods 7\,--\,12\,s and 16\,--\,21\,s during which objects the very close to each other. As the performance upper bound, the LMO-GOM filter still performs the best. Moreover, the G-LMB-GOM filter  has even better performance than both the LMB-GOM and $\delta$-GLMB-GOM filter, because the grouping of objects alleviates the combinational and  high-dimension problem at the update stage.

Fig.~\ref{fig: performance of five targets}(b)  shows average OSPA errors for the MB-TBD filter and others. The results observed are consistent  with that of the single run of the MB-TBD filter. When objects are far away from each other before the time of 8\,s, the MB-TBD filter has decent accuracy, then its error begins to increase as objects get close to each other, and finally it diverges.

\subsection{Acoustic Amplitude Model}
To further assess  the capabilities of the LMB-GOM and G-LMB-GOM filter,  the scenario considers the problem of tracking an unknown and time varying number of objects using acoustic amplitude sensors.   
 A number of  961 acoustic sensors  are dispersed evenly over a two-dimensional surveillance region $[0\,\, 300]\,\text{m}\times[0\,\,300]\,\text{m}$  as shown in Fig. \ref{fig: scenario and time}(a).   At most four objects appear and travel with the standard deviation of the process noise $\sigma_{v}=0.7\,\text{m}/\text{s}^{2}$.  The path loss exponent is set to be $\kappa=1$.  The duration of this scenario is  $T_s=60$\,s.

This case is quite different from the pixeled TBD observation model in the sense that the VOR $T(\bx)$ is able to cover the whole observation space, which can be reflected from Fig. \ref{fig: scenario and time}(b) drawing the received sound amplitude at each acoustic sensor. For 
the G-LMB-GOM filter, the grouping criterion  (\ref{grouping-threshold}) is utilized  with the  threshold $\beta=45$\,m  by approximately  truncating $T(\bx)$. 
The particles employed by each algorithm are set according to Table II  with $N_{p}= 10^{4}$.  The birth procedure for each algorithm  is as follows.  At each time step, the birth process   is an LMB RFS with the parameter set $\bpi_{B}=\{(r_{B}^{(i)}, p_{B}^{(i)})\}_{i=1}^{2}$ where $r_{B}^{(i)}=0.02$ and $p_{B}^{(i)}=\mathcal{N}(x; m_{B}^{(i)}, P_{B})$ with $m_{B}^{(1)}=[50\,\, 180\,\,0\,\,0]^{\top}$, $m_{B}^{(2)}=[200\,105\,\,\,0\,\,0]^{\top}$ and $P_{B}=\mbox{diag}([2\,\,2\,\,2\,\,2])$.  

\begin{table}[b]
\renewcommand{\arraystretch}{1.2}
	\caption{The average OSPA errors (m) for   different sound amplitudes.\label{tab_scenario_2}}

\begin{center}
	\footnotesize
\begin{tabular*}{0.48\textwidth}{@{\extracolsep{\fill}}c| c c c }
\hline\hline
Sound Amplitude $A$	&	10         &  7.9        & 5.6      \\
		\hline
$\delta$-GLMB-GOM      &     2.0296 &2.4284     &3.7133                \\
\hline
LMB-GOM                      & 2.1012        &2.4510   &3.7404\\
\hline
G-LMB-GOM                  &1.9852        & 2.4092   &3.8196\\
\hline
\end{tabular*}
\label{tabone}
	\normalsize
\end{center}
\end{table}

\begin{figure}[h]
\setlength{\abovecaptionskip}{-7pt}
\setlength{\belowcaptionskip}{-7pt}
\begin{minipage}[b]{0.48\linewidth}
  \centering
\centerline{\includegraphics[width=4.9cm]{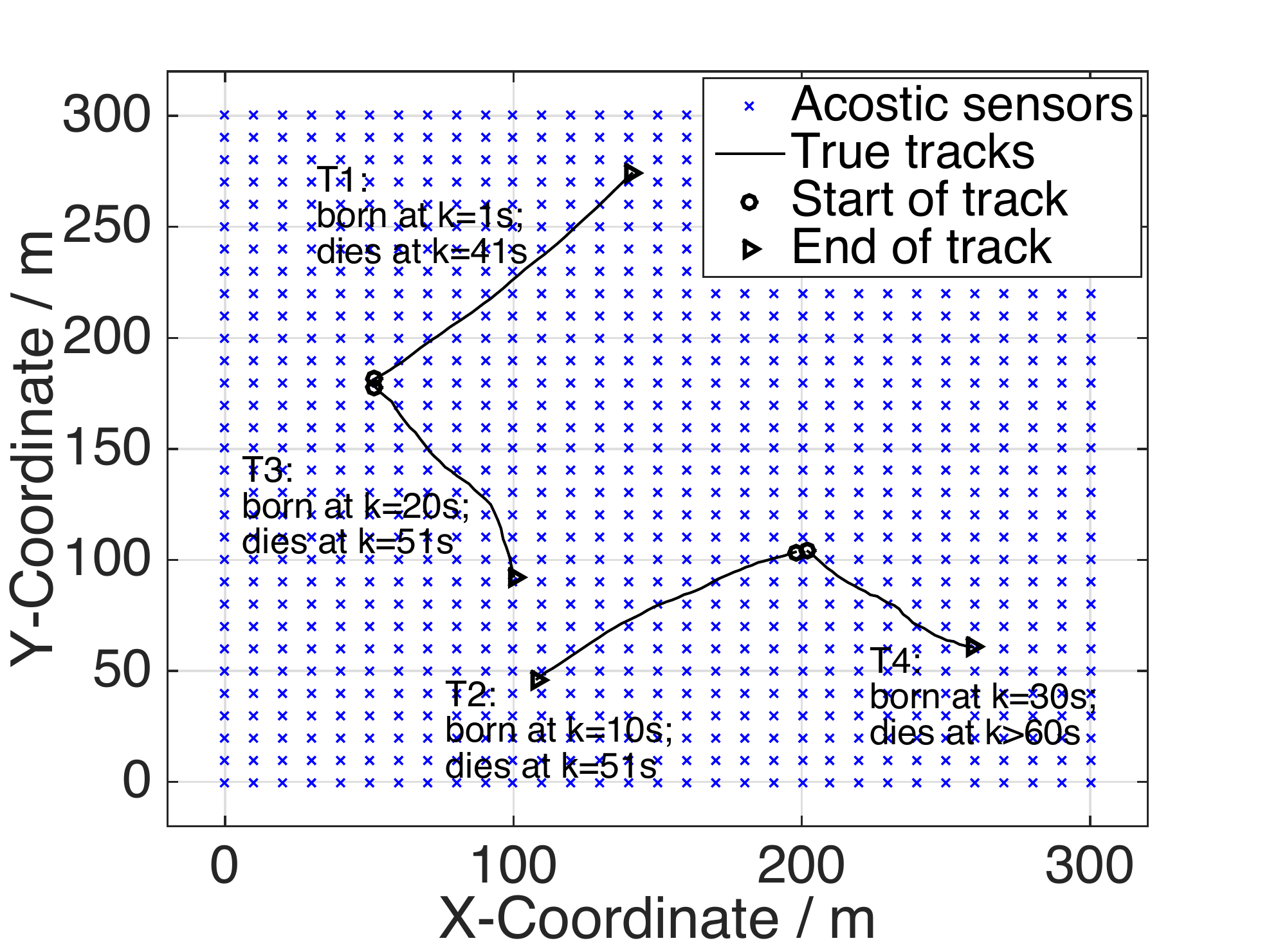}}
\centerline{\small{\small{(a)}} }\medskip
\end{minipage}
\hfill
\begin{minipage}[b]{0.48\linewidth}
  \centering
\centerline{\includegraphics[width=4.9cm]{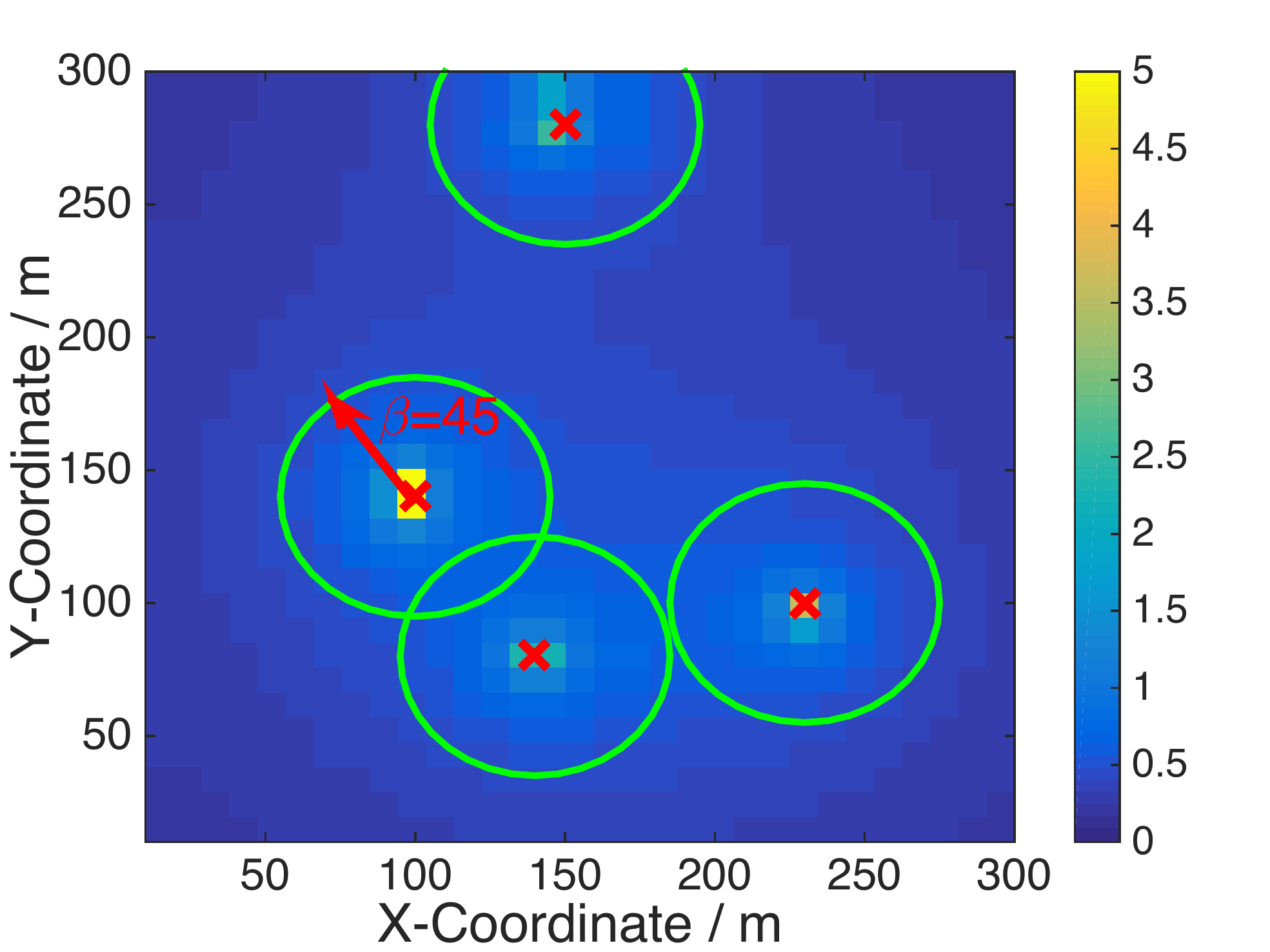}}
  \centerline{\small{\small{(b)}}}\medskip
\end{minipage}
\caption{(a)  the trajectories of four objects in $x-y$ plane; (b) the received sound amplitude at each acoustic sensor under $A=7.9$.}
  \label{fig: scenario and time}
\end{figure}
\begin{figure}[h]
\setlength{\abovecaptionskip}{-7pt}
\setlength{\belowcaptionskip}{-7pt}
\begin{minipage}[b]{0.48\linewidth}
  \centering
\centerline{\includegraphics[width=5.0cm]{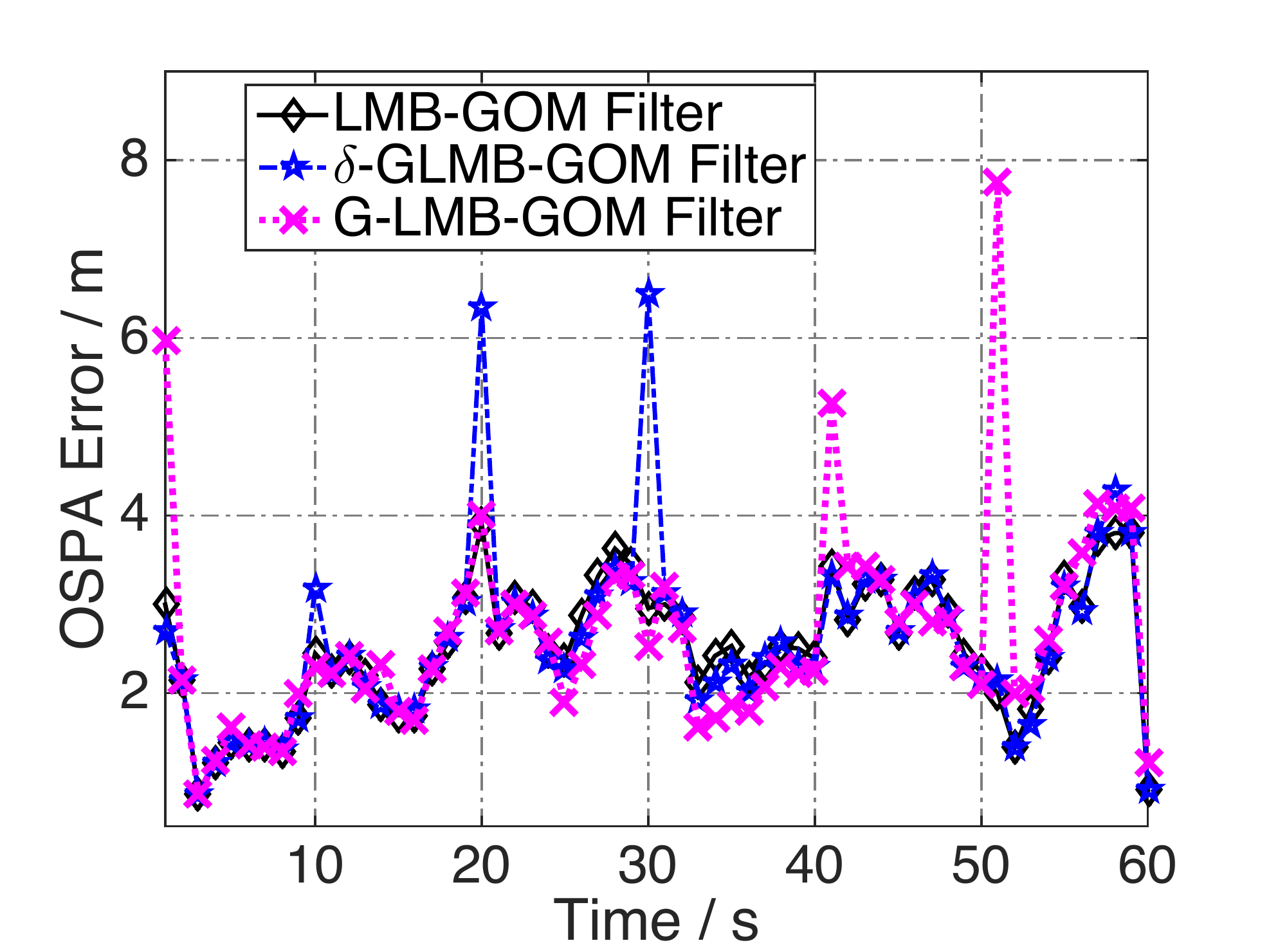}}
\centerline{\small{\small{(a)}} }\medskip
\end{minipage}
\hfill
\begin{minipage}[b]{0.47\linewidth}
  \centering
\centerline{\includegraphics[width=5.cm]{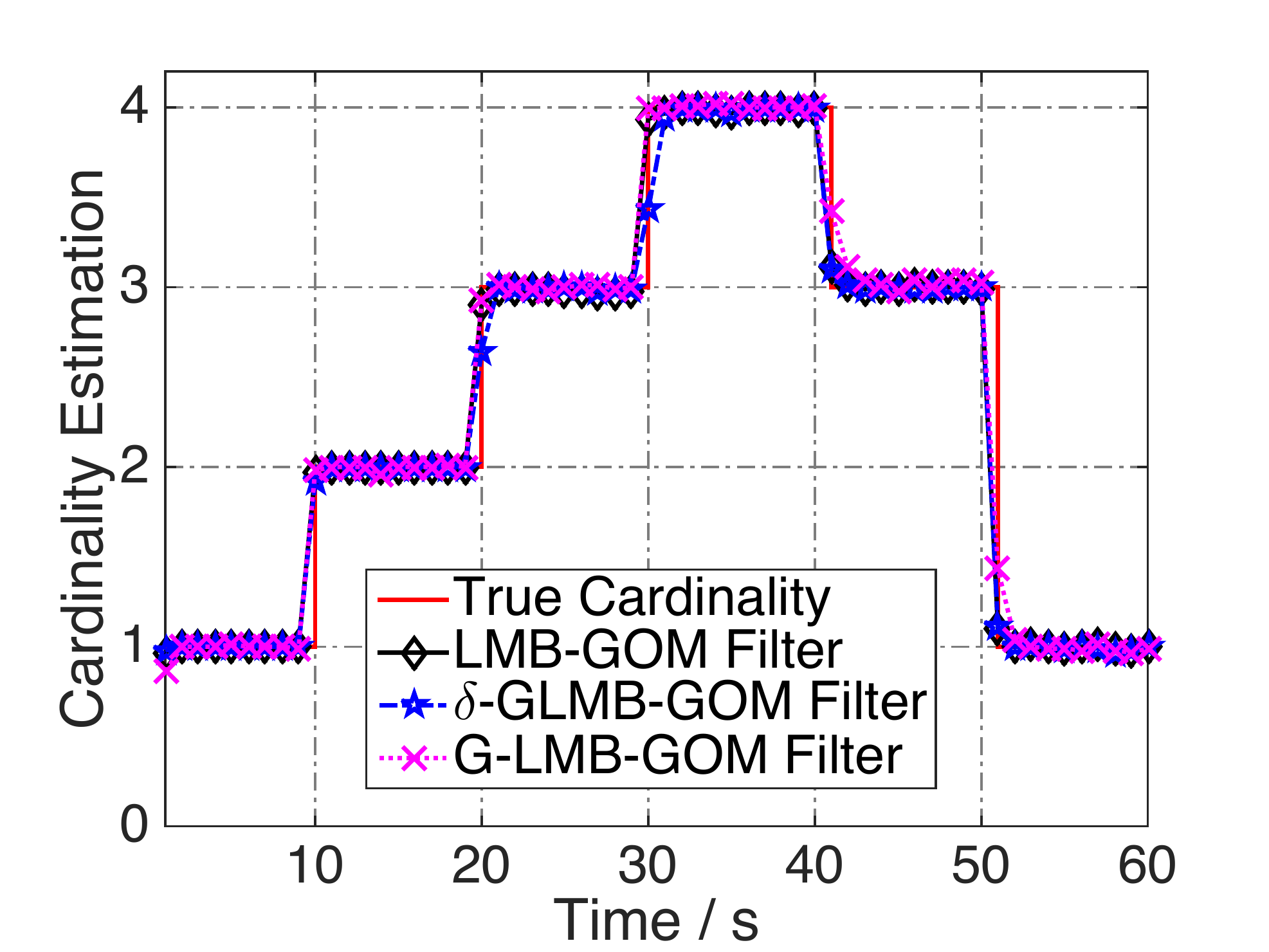}}
  \centerline{\small{\small{(b)}}}\medskip
\end{minipage}
\caption{Performance metrics for the $\delta$-GLMB-GOM, LMB-GOM, G-LMB-GOM filters under $A=7.9$: (a) average OSPA errors; (b) cardinality estimates and true cardinalities.}
  \label{fig: performance of four targets}
\end{figure}

\begin{figure}[t]
  \centering
  \includegraphics[width=8cm,height=4cm]{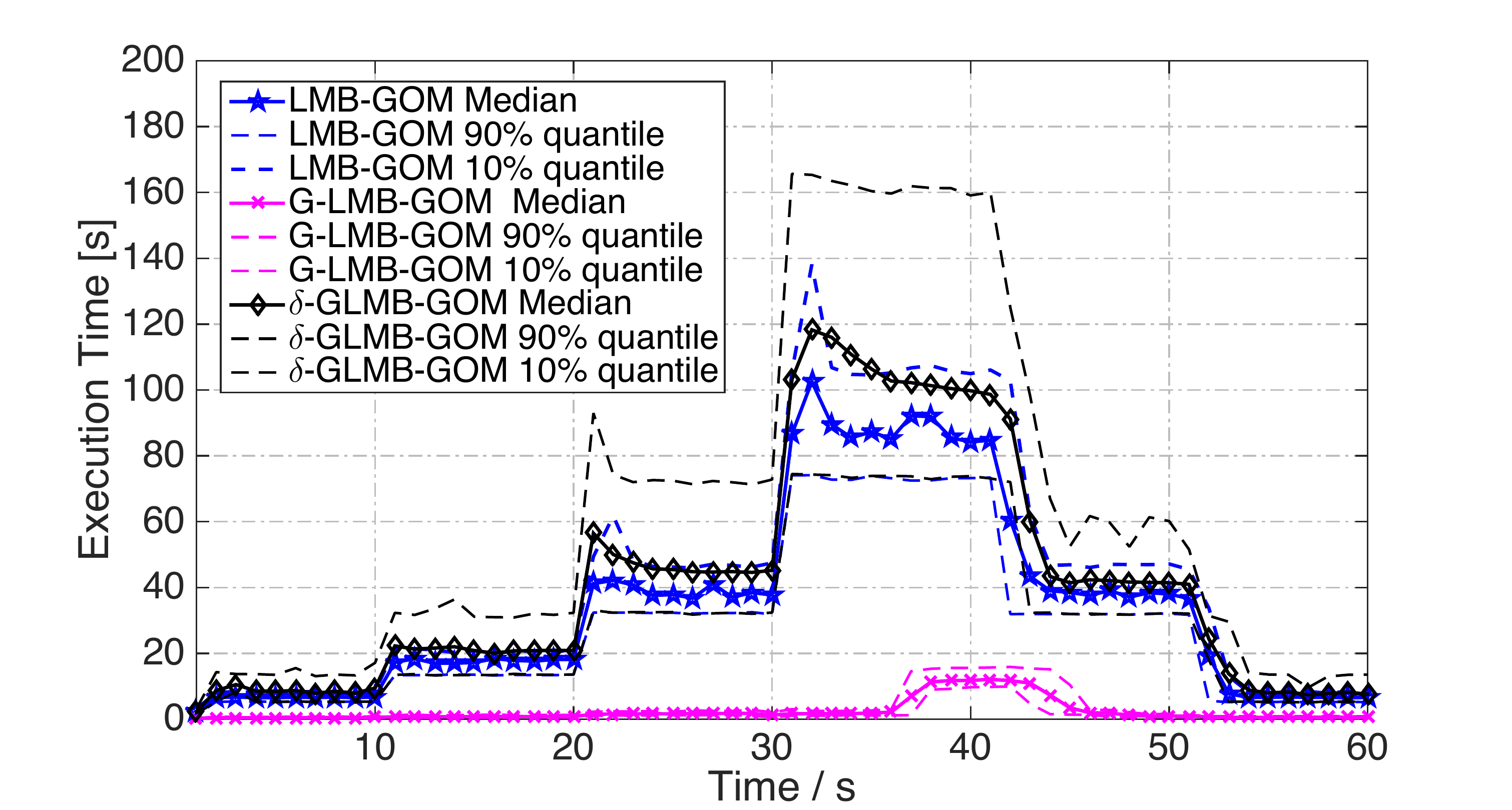}\\
  \caption{Average execution times  for the $\delta$-GLMB-GOM, LMB-GOM, G-LMB-GOM filters under  $A=7.9$.}
  \label{fig:time comparison}
\end{figure}

Figs. \ref{fig: performance of four targets} and \ref{fig:time comparison} show the execution times, average OSPA errors and the cardinality estimates over time for the LMB-GOM, G-LMB-GOM and $\delta$-GLMB filters under the sound amplitude $A=7.9$, respectively. We observe a comparable performance from the LMB-GOM and $\delta$-GLMB-GOM filters  in terms of both the cardinality estimates and the OSPA errors, while the LMB-GOM filter achieves a more evident reduction in the execution time compared to  the $\delta$-GLMB-GOM filter  due to the incorporation of the object birth process in this scenario. On one hand, whenever an object is born, the tracking error of the $\delta$-GLMB-GOM filter  sharply increases but retracts to the normal level quickly, while the LMB-GOM filter can handle the births of objects well.  On the other hand, the tracking errors of the LMB-GOM filter are slightly higher than the $\delta$-GLMB-GOM filter at the stable stage.   The tracking performance of the G-LMB-GOM filter is also comparable with the other two filters. More importantly, the OSPA errors  of the G-LMB-GOM filter is even lower than the other two algorithms during 20\,s\,--\,40\,s when more isolated  tracks have appeared. Also, the execution time for the G-LMB-GOM filter is dramatically reduced compared with the other two algorithms. However, one can also observe that when objects die (at times 40\,s and 50\,s), the OSPA error of the G-LMB-GOM sharply increases but retracts to the normal level quickly, while the other two algorithms can handle the deaths of objects better. 
The reason is that 
the performance loss arising from the grouping error can be larger than the improvement in the numerical accuracy due to the parallel group update, when an object dies.
  The results of this experiment also verify that the G-LMB-GOM filter can also be effective for a Type II sensor.

Further,  we investigate how the performances of different algorithms are affected by  different values of the sound amplitude $A$. The  post-transient values of the OSPA errors under $A=10,\,7.9, 5.6$ averaged over 100 MC runs and 60 time steps are presented in Table III.

\section{Conclusion}\label{chp:8}
An exact  Bayesian filtering solution using labeled random finite sets, for the multi-object tracking problem under the generic observation model (GOM) and the  standard transition kernel, was presented. 
The proposed exact solution can be served as the theoretical performance benchmark in multi-object tracking under the standard transition kernel.  We also proposed a generalization of the  LMB filter, named LMB filter for GOM (LMB-GOM filter) which is derived by approximating the full multi-object density with the closest  LMB density  in terms of Kullback-Leibler divergence (and it is proven to preserve the first moment  as well). A variant of the LMB-GOM filter, called grouping based LMB-GOM (G-LMB-GOM) filter was devised    and presented through a step-by-step algorithm.  The G-LMB-GOM filter can be viewed as a computationally tractable way to implement the LMB-GOM filter. The efficacy of the proposed algorithms is demonstrated using the sequential Monte Carlo implementation  under  two types of   non-standard observation models. 

Possible future works  incorporate the study on the numerical implementation methods of the proposed algorithms, e.g., the unscented  Kalman filter, the cubature Kalman filter.
\appendices
\section{Proof of Proposition \ref{pro:1}}
The density of the surviving multi-object state at the next time is given by the Chapman-Kolmogorov equation
\begin{equation}\label{surving-density}
{\small
\begin{split}
\!\!&\bpi_S(\bW)=\int \bff_S(\bW|\bX)\bpi(\bX)\delta\bX\\
=&\sum_{n=0}^{\infty}\frac{1}{n!}\sum_{(\ell_1,\cdots,\ell_n)\in\mathbb{L}^n}\omega(\{\ell_1,\cdots,\ell_n\})1_{\{\ell_1,\cdots,\ell_n\}}(\mathcal{L}(\bW))\\&
\!\!\!\!\!\!\! \int \prod_{i=1}^n \Phi(\bW;x_i,\ell_i) P(\{(x_1,\ell_1),\!\cdots\!,(x_n,\ell_n)\})d(x_1,\!\cdots\!,x_n).
\end{split}}
\end{equation}
Substituting   $P_{S,\{\ell_,\cdots,\ell_n\}}$ of form (\ref{Ps-I}) into (\ref{surving-density}), we have
\begin{equation}
{\small
\begin{split}
\label{surving-density-2}
\bpi_S(\bW)
=&
{\sum}_{I\subseteq\mathbb{L}}1_{I}(\mathcal{L}(\bW))\omega(I)P_{S,I}(\bW).
\end{split}}
\end{equation}
According to Definition \ref{definition:1}, we can compute the joint probability of the label set $\{\ell_{+,1},\cdots,\ell_{+,n_+}\}$ for $\bpi_S(\bW)$ as,
\begin{equation}
{\small
\begin{split}
&\omega_S(\{\ell_{+,1},\cdots,\ell_{+,n_+}\})\\
=&{\sum}_{I\subseteq\mathbb{L}}1_{I}(\{\ell_{+,1},\cdots,\ell_{+,n_+}\})\omega(I)\int P_{S,I}(\{(x_{+,1},\\&\ell_{+,1}),\cdots,(x_{+,n_+},\ell_{+,n_+})\})d(x_{+,1},\cdots,x_{+,n_+}).\\
\end{split}
}
\end{equation}
Substitution of $\eta_{S,I}(\{\ell_{+,1},\cdots,\ell_{+,n_+}\})$ in (\ref{where_0}) results in
\begin{equation}
{\small
\begin{split}
&\omega_S(\{\ell_{+,1},\cdots,\ell_{+,n_+}\})\\=&\sum_{I\subseteq\mathbb{L}}1_{I}(\{\ell_{+,1},\cdots,\ell_{+,n_+}\})\omega(I)\eta_{S,I}(\{\ell_{+,1},\cdots,\ell_{+,n_+}\}).
\end{split}}
\end{equation}
Also, we can compute the joint probability density of the states $x_{+,1},\!\cdots\!,x_{+,n_+}$ conditional on $\ell_{+,1},\!\cdots\!,\ell_{+,n_+}$ by Definition \ref{definition:1}, 
\begin{equation}\notag
{\small
\!\!
\begin{split}
&P_S(\{(x_{+,1},\ell_{+,1}),\cdots,(x_{+,n_+},\ell_{+,n_+})\})=\\
&\frac{\sum_{I\subseteq\mathbb{L}}\!1_{I}(\{\ell_{+\!,1},\!\cdots\!,\ell_{+\!,n_+}\})\omega(I)P_{S,I}(\{(x_{+\!,1},\!\ell_{+,1}),\!\cdots\!,(x_{+,n_+},\!\ell_{+\!,n_+})\})}{\omega_{S}(\{(\ell_{+\!,1}),\!\cdots\!,(\ell_{+,n_+})\})}.
\end{split}
}
\end{equation}
Hence, $\bpi_S(\bW)$  can be presented as
\begin{equation}\label{surving-density-3}
{\small
\bpi_S(\bW)=\omega_S(\mathcal{L}(\bW))P_S(\bW).
}
\end{equation}

For the predicted multi-object density, recall the birth density (\ref{LMO born}), then we have
\begin{equation}\label{predict-density}
{\small
\begin{split}
\bpi_+(\bX_+)
=&\bff_B(\bX_+\cap\mathbb{X}\times\mathbb{B})\bpi_S(\bX_+\cap\mathbb{X}\times\mathbb{L})\\
=&\omega_B(\mathcal{L}(\bX_+)\cap\mathbb{B})\omega_S(\mathcal{L}(\bX_+)\cap\mathbb{L}) \\&\cdot P_B(\bX_+\cap\mathbb{X}\times\mathbb{B})P_S(\bX_+\cap\mathbb{X}\times\mathbb{L}).
\end{split}}
\end{equation}
Using (\ref{w-add}) and (\ref{p-add}), (\ref{predict-density}) can be computed by
\begin{equation}\label{predict-density-2}
{\small
\bpi_+(\bX_+)=\omega_+(\mathcal{L}(\bX_+))P_+(\bX_+).
}
\end{equation}
\section{Proof of Proposition \ref{pro:2}}
Based on the Bayes' rule, the numerator of the multi-object posterior  density $\bpi(\bX|\Upsilon)$ can be computed as 
\begin{equation}\label{fenzi}
{\small
\begin{split}
g(\Upsilon|\bX)\bpi_+(\bX)=&g(\Upsilon|\bX)\omega_+(\mathcal{L}(\bX))P_+(\bX).\\
\end{split}}
\end{equation}
Substitution of   $\eta_\Upsilon(\cdot)$ in (\ref{where_eta}), and   $P(\bX|\Upsilon)$ in (\ref{where_p}), (\ref{fenzi}) can be further computed by
\begin{equation}\label{fenzi-2}
{\small
\begin{split}
g(\Upsilon|\bX)\bpi_+(\bX)
=&\eta_{\Upsilon}(\mathcal{L}(\bX))\omega_+(\mathcal{L}(\bX))\frac{g(\Upsilon|\bX)P_+(\bX)}{\eta_\Upsilon(\mathcal{L}(\bX))}\\
=&\eta_{\Upsilon}(\mathcal{L}(\bX))\omega_+(\mathcal{L}(\bX))P(\bX;\Upsilon).
\end{split}}
\end{equation}
Then, the denominator of    (\ref{update}) can be computed by
\begin{equation}\label{fenmu}
{\small
\begin{split}
%
\int g(\Upsilon;\bX)\bpi_+(\bX)\delta \bX=&{\sum}_{I_{+}\subseteq\mathbb{L}_+}\eta_{\Upsilon}(I_{+})\omega_+(I_{+}).\end{split}
}
\end{equation}
Hence, the multi-object posterior density  is
\begin{equation}
\notag
{\small
\begin{split}
\bpi(\bX|\Upsilon)=&\frac{\eta_{\Upsilon}(\mathcal{L}(\bX))\omega_+(\mathcal{L}(\bX))P(\bX;\Upsilon)}{\sum_{I_{+}\in\mathcal{F}(\mathbb{L}_+)}\eta_{\Upsilon}(I_{+})\omega_+(I_{+})}\!=\!
\omega(\mathcal{L}(\bX);\Upsilon)P(\bX;\Upsilon)
\end{split}}
\end{equation}
where $\omega(I_{+};\Upsilon)$ is given in (\ref{where_w}).
\section{Proof of Proposition \ref{pro:3}}
According to the definition of the PHD~\cite{refr:Mahler_book}, the labeled PHD of $\bpi(\bX)=\omega(\mathcal{L}(\bX))P(\bX)$ can be computed as
\begin{equation}\label{labeled-PHD-LMO-proof_1}
{\small
\begin{split}
& v(x,\ell)=\int\omega(\mathcal{L}(\{(x,\ell)\cup\mathbf{X}\}))P(\{(x,\ell)\}\cup\mathbf{X})\delta\bX\\
   =&\sum_{n=0}^\infty\sum_{(\!\ell_1,\!\cdots\!,\ell_n\!)\in\mathbb{L}^n}\!\!\!(1\!-\!1_{\{\ell_1,\cdots,\ell_n\}}(\ell))\delta_n(\!\{\ell_1,\!\cdots\!,\ell_n\}\!)\omega(\!\{\ell,\ell_1,\!\cdots\!,\ell_n\}\!)\!\!\\&\cdot  \frac{1}{n!} \int\! P(\{(x,\ell),(x_1,\ell_1),\!\cdots\!,(x_n,\ell_n)\})d(x_1,\!\cdots\!,x_n).\\
  \end{split}}
  \end{equation}
 Substituting $p_{\{\ell_1,\cdots,\ell_n\}}(\cdot)$ of form (\ref{marginal_label})  into (\ref{labeled-PHD-LMO-proof_1}), (\ref{labeled-PHD-LMO-proof_1}) leads to 
  \begin{equation}\label{labeled-PHD-LMO-proof_2}
\begin{split}
\!\!&v(x,\ell)
   \\
   \!\!=&\sum_{n=0}^\infty \frac{1}{n!}\!\sum_{(\ell_1,\cdots,\ell_n)\in\mathbb{L}^n}\delta_n(\{\ell_1,\cdots,\ell_n\})(1-1_{\{\ell_1,\cdots,\ell_n\}}(\ell)) \\
   \!\!&\,\,\,\,\,\,\,\,\,\,\,\,\,\,\,\,\,\,\,\,\,\,\,\,\,\,\,\,\,\,\,\,\,\,\,\,\,\,\,\,\,\,\,\,\cdot\omega(\{\ell,\ell_1,\cdots,\ell_n\}) p_{\{\ell_1,\cdots,\ell_n\}}(x,\ell)\\
   \!\!=& {\sum}_{I\in\mathcal{F}(\mathbb{L})}1_{I}(\ell)\omega(I)p_{I-\{\ell\}}(x,\ell).\\
\end{split}
\end{equation}
Hence, the Proposition holds.
\section{Proof of Proposition \ref{pro:4}}
The LMB density can  be presented as the form of (\ref{P-LMO}) with
\begin{small}
\begin{align}
\label{LMB-w}\omega(\mathcal{L}(\bX))=&{\prod}_{i\in\mathbb{L}}(1-r^{(i)}){\prod}_{j\in \mathcal{L}(\bX) }\frac{1_{\mathbb{L}}(j)r^{(j)}}{1-r^{(j)}}\\
\label{LMB-p}P(\bX)=&\Delta(\bX)[p]^\bX.
\end{align}
\end{small}
According to Proposition \ref{pro:3}, substituting (\ref{LMB-w}) and (\ref{LMB-p})  in (\ref{labeled-PHD-LMO}), we can obtain the labeled PHD of LMB density $\bpi$ as,
\begin{equation}\label{labeled-PHD-LMB-proof}
{\small
\begin{split}
\!\!v(x,\ell)=&
{\sum}_{I\in\mathcal{F}(\mathbb{L})}1_{I}(\ell){\prod}_{j\in I }r^{(j)}{\prod}_{i\in\mathbb{L}-I}(1\!-\!r^{(i)})p(x,\ell)\\
=&r^{(\ell)}p(x,\ell)\sum_{I'\in\mathcal{F}(\mathbb{L}-\{\ell\})}\prod_{j\in I' }r^{(j)}\!\!\!\prod_{i\in \mathbb{L}-\{\ell\}-I'}\!\!\!(1-r^{(i)})
\\
=&r^{(\ell)}p(x,\ell)={\sum}_{\alpha\in\mathbb{L}}r^{(\alpha)}p^{(\alpha)}(x,\ell).
\end{split}}
\end{equation}
Hence, the Proposition holds.
%

\section{Proof of Proposition \ref{pro:5}}
Given an arbitrary LMO density $\bpi(\bX)=\omega(\mathcal{L}(\bX))P(\bX)$ of the form (\ref{P-LMO}) on state space $\mathbb{X}$ and label space $\mathbb{L}$, we can easily obtain  the  LMB density  $\hat \bpi_{\text{LMB}}(\bX)$ 
matching the   labeled PHD  of $\bpi(\bX)$ by comparing  the labeled PHDs  of the general labeled RFS and the LMB RFS shown in (\ref{labeled-PHD-LMO}) and (\ref{labeled-PHD-LMB}) respectively.
Specifically, the parameters of  $\hat\bpi_{\text{LMB}}(\bX)$ of the form (\ref{LMB})
can be computed by
\begin{align}
\label{existence_probability}
  \hat{r}^{(\ell)}&=\int v(x,\ell) dx={\sum}_{I\in\mathcal{F}(\mathbb{L})}1_{I}(\ell)\omega(I)\\
\label{probability_density}
 \hat{p}(x,\ell)&=\frac{v(x,\ell)}{ \hat{r}^{(\ell)}}=\frac{1}{ \hat{r}^{(\ell)}}{\sum}_{I\in\mathcal{F}(\mathbb{L})}1_{I}(\ell)\omega(I)p_{I-\{\ell\}}(x,\ell)
\end{align}
where $v(x,\ell)$ is the labeled PHD of $\bpi$.

In the following, we   prove that  $\hat \bpi_{\text{LMB}}$ which matches the labeled PHD of $\bpi$ also  minimizes the KLD from $\bpi$ over the class of LMB RFS family. 

The KLD from $\bpi$ and any LMB density  $\overline \bpi_{\text{LMB}}$ of the form (\ref{LMB}) with the parameters $\overline r^{(\ell)}$ and $p(x,\ell)$, is given by
\begin{equation}\label{D_KL_LMB}
{\small
\begin{split}
&D_{\text{KL}}(\bpi;\overline\bpi_{\text{LMB}})
\\=&\int \log\left(\frac{\omega(\mathcal{L}(\bX))P(\bX)}{\overline\omega(\mathcal{L}(\bX))\overline P(\bX)}\right)\omega(\mathcal{L}(\bX))p(\bX)\delta\bX\\
=&\int \log\left(\frac{P(\bX)}{\overline P(\bX)}\right)\omega(\mathcal{L}(\bX))P(\bX)\delta\bX+D_{\text{KL}}(\omega;\overline\omega)
\end{split}}
\end{equation}
where 
\begin{equation}
{\small
\begin{split}
\overline\omega(I)&={\prod}_{\ell\in I} \overline r^{(\ell)} {\prod}_{\ell'\in\mathbb{L}-I} \left(1-\overline r^{(\ell')}\right)\\
\overline P(\bX)&=\Delta(\bX)[{\overline p}]^{\bX}.
\end{split}}
\end{equation}
Observing (\ref{D_KL_LMB}), one can find that $D_{\text{KL}}(\bpi;\overline\bpi_{\text{LMB}})$ is the sum of two parts. We define the first part as
 \begin{equation}C(\overline P)
 {\small
\triangleq \int \log\left(\frac{P(\bX)}{\overline P(\bX)}\right)\omega(\mathcal{L}(\bX))P(\bX)\delta\bX}
\end{equation}
and the second part as
\begin{equation}
{\small
C(\overline\omega)\triangleq D_{\text{KL}}(\omega;\overline\omega).}
\end{equation}
First, we consider the part $C(\overline P)$, and it can be computed by
\begin{equation}\label{D_KL_p}
{\small
\begin{split}
C(\overline P)=&K_1-\int \omega(\mathcal{L}(\bX))P(\bX){\sum}_{\bx\in\bX}\log \overline p(\bx)\delta\bX
\end{split}}
\end{equation}
where $K_1$ is  a constant  having no functional dependence on $\overline\bpi_{\text{LMB}}(\bX)$.

According to  Proposition 2a in~\cite{refr:tracking-1}, i.e.,
\begin{equation}
{\small
\int {\sum}_{y\in Y}h(y)\pi(Y)\delta Y=\int h(y)v(y)dy}
\end{equation}
with $v(y)$ being the PHD of $\pi$, we have
\begin{equation}\label{log_p}
{\small
\begin{split}
&\int \omega(\mathcal{L}(\bX))p(\bX)\sum_{\bx\in\bX}\log\overline p(\bx)\delta\bX\!=\!\sum_{\ell\in\mathbb{L}}\int\!v(x,\ell)\log\overline p(x,\ell)d x.
\end{split}}
\end{equation}

According to Proposition \ref{pro:3}, $v(x,\ell)$  has the form of (\ref{labeled-PHD-LMO}). Substituting (\ref{labeled-PHD-LMO}) and (\ref{log_p}) into (\ref{D_KL_p}), we have
\begin{equation}\label{D_KL_p2}
{\small
\begin{split}
&C(\overline P)=K_1\!\!-\!\!\sum_{\ell\in\mathbb{L}}\int\!\!\sum_{I\in\mathcal{F}(\mathbb{L})}\!\!1_{I}(\ell)\omega(I)p_{I-\{\ell\}}(x,\ell)\log\overline p(x,\ell)dx.
\end{split}}
\end{equation}

The substituting (\ref{existence_probability}) and (\ref{probability_density}) into (\ref{D_KL_p2}), we have
\begin{equation}
{\small
\begin{split}
C(\overline P)=&K_1+K_2+{\sum}_{\ell\in\mathbb{L}}\hat r^{(\ell)}D_{\text{KL}}(\hat p(\cdot,\ell);\overline p(\cdot,\ell))
\end{split}}
\end{equation}
where
\begin{equation}
{\small
K_2=-{\sum}_{\ell\in\mathbb{L}}\hat r^{(\ell)} \int \hat p(x,\ell) \log \hat p(x,\ell)dx}
\end{equation}
which is  a constant that
has no functional dependence on  $\overline p(\cdot,\ell), \ell\in\mathbb{L}$.
Hence,  $C(\overline P)$ is minimized only if
$ \overline p(\cdot,\ell)=\hat p(\cdot,\ell)$ for each $\ell\in\mathbb{L}$.

Secondly, consider the part $C(\overline\omega)$.
According to the definition of KLD, we have
\begin{equation}\label{D_KL_w}
{\small
\begin{split}
C(\overline\omega)=&K_3-{\sum}_{\ell'\in\mathbb{L}}{\sum}_{I\in\mathcal{F}(\mathbb{L)}}1_{\mathbb{L}-I}(\ell')\omega(I)\log\left(1-\overline r^{(\ell')}\right)\\&\,\,\,\,\,\,\,\,\,\,\,\,\,\,\,\,\,\,\,\,\,\,\,\,\,\,\,\,\,\,\,\,\,-{\sum}_{\ell\in\mathbb{L}}{\sum}_{I\in\mathcal{F}(\mathbb{L)}}1_{I}(\ell)\omega(I)\log{\overline r^{(\ell)}}
\end{split}}
\end{equation}
where $K_3$ is a constant independent of $\overline\omega(\cdot)$.

It is obvious that
\begin{equation}
{\small
\sum_{I\in\mathcal{F}(\mathbb{L)}}1_{\mathbb{L}-I}(\ell')\omega(I)=1-\sum_{I\in\mathcal{F}(\mathbb{L)}}1_{I}(\ell')\omega(I)=1-\hat r^{(\ell')}}
\end{equation}
with $\hat r^{(\ell')}$ shown in (\ref{existence_probability}).
Thus (\ref{D_KL_w}) can be  presented as
\begin{equation}\label{D_KL_w2}
{\small
\begin{split}
&\!\!C(\overline\omega)\!=\!K_3\!-\!{\sum}_{\ell\in\mathbb{L}}\left(\left(1\!-\!\hat r^{(\ell)}\right)\log\left({1\!-\!\overline r^{(\ell)}}\right)\!+\!\hat r^{(\ell)}\log{\overline r^{(\ell)}}\right).
\end{split}}
\end{equation}

We define two Bernoulli distributions $\hat E_\ell$ and $\overline E_{\ell}$ for each $\ell\in\mathbb{L}$ as
\begin{small}
\begin{align}
&\Pr(\hat E_\ell=1)=\hat r^{(\ell)};\,\,\,\,\,\,\, \Pr(\hat E_\ell=0)=1-\hat r^{(\ell)}\\
&\Pr(\overline E_\ell=1)=\overline r^{(\ell)};\,\,\,\,\,\,\Pr(\overline E_\ell=0)=1-\overline r^{(\ell)}.
\end{align}\end{small}
Then, Eq. (\ref{D_KL_w2}) yields to
\begin{equation}\label{D_KL_w3}
{\small
\begin{split}
&C(\overline\omega)\!=\!
K_3\!+\!K_4\!+\!{\sum}_{\ell\in\mathbb{L}}\!D_{\text{KL}}\left(\Pr(\hat E_\ell=e);\Pr(\overline E_\ell=e)\right)\\
\end{split}}
\end{equation}
where 
\begin{equation}
{\small
K_4\!=\!-{\sum}_{\ell\in\mathbb{L}}\left(\!(1\!-\!\hat r^{(\ell)})\log({1\!-\!\hat r^{(\ell)}})\!+\!\hat r^{(\ell)}\log{\hat r^{(\ell)}}\right)}
\end{equation}
which is a constant having no functional dependence on any $\overline r^{(\ell)}, \ell\in\mathbb{L}$. Hence,  $C(\overline\omega)$ is minimized only if
$\overline r^{(\ell)}\!\!=\!\hat r^{(\ell)}$  for each $\ell\!\in\!\mathbb{L}$.

According to (\ref{D_KL_LMB})$, D_{\text{KL}}(\bpi;\overline\bpi)$ is minimized only if both $C(\overline P)$ and $C(\overline\omega)$  are minimized. Hence, $D_{\text{KL}}(\bpi;\overline\bpi)$ is minimized by $\overline\bpi_{\text{LMB}}=\hat \bpi_{\text{LMB}}$ over the class of LMB RFS family.
\section{Proof of Proposition 6}
Firstly, one can write the LMB prediction  in the general LMO density form,
\begin{equation}\label{LMP-LMO}
{\small
\bpi_+(\bX)=\Delta(\bX)\omega_+(\mathcal{L}(\bX)){[p_+]}^\bX)}
\end{equation}
with $\omega_+(\cdot)$  shown as (\ref{w-add}) and $p_+(x,\ell)=p_+^{(\ell)}(x)$.

Then, according to Proposition 2,   we can obtain the following multi-object posterior under the generic observation likelihood $g(\Upsilon|\bX)$,
\begin{equation}\label{update_equation}
{\small
\begin{split}
  \bpi(\bX|\Upsilon)
=&\Delta(\bX)\omega(\mathcal{L}(\bX);\Upsilon)P(\bX|\Upsilon)
\end{split}}
\end{equation}
where  $\omega(\cdot;\Upsilon)$ and $P(\cdot|\Upsilon)$ are computed  using (\ref{update_w-lmb}) and (\ref{update_P-lmb}), respectively.

According to Proposition 5, the LMB RFS that matches exactly the labeled first-order moment of $\bpi(\bX|\Upsilon)$ as well as minimizes the Kullback-Leibler divergence  from  $\bpi(\bX|\Upsilon)$ can be computed by
 \begin{equation}
 {\small
 \hat\bpi(\cdot|\Upsilon)=\{\hat r^{(\ell)}(\Upsilon),\hat p^{(\ell)}(\cdot;\Upsilon)\}_{\ell\in\mathbb{L}_+},}\end{equation}
where $\hat r^{(\ell)}(\Upsilon)$ and $\hat p^{(\ell)}(\cdot;\Upsilon)$ is computed by (\ref{LMB_where_r}) and (\ref{LMB_where_p}).
\section{Proof of Proposition 7}
According to Assumption \textit{A.1}, for a subset of $\bX$, denoted by $\bX'$, the belief mass function [1] of the observation $Z$ can be presented as
\begin{equation}\label{belief-mass-function}
\begin{split}
\beta(S|\bX)=&\Pr(Z\subseteq S|\bX)\\
=&\Pr(Z\cap T(\bX')\subset S|\bX)\Pr(Z- T(\bX')\subset S|\bX).
\end{split}
\end{equation}

Also, according to Assumption \textit{A.2}, $T(\bX')\cap T(\bX-\bX')=\emptyset$  which indicates that  observations $z\in T(\bX')$ are generated only by object states in $\bX'$, and hence are independent from $\bX-\bX'$, i.e.,
\begin{equation}\label{belief-mass-function-subset-1}
\begin{split}
\Pr(Z\cap T(\bX')\subseteq S|\bX)
=&\Pr(Z\cap T(\bX')\subseteq S|\bX')\\
=&\int_{S} 1_{T(\bX')}(Z)g(Z|\bX') \delta Z.
\end{split}
\end{equation}
Similarly, the observations $z\in Z-T(\bX)$ are independent from $\bX-\bX'$. 
As a result, the belief mass function given in (\ref{belief-mass-function}) can be calculated  as follows:
\begin{equation}
\begin{split}
\beta_{Z}(S|\bX)=&
\int_{S} 1_{T(\bX')}(Z)g(Z|\bX') \delta Z  \\&\cdot \int_{S} (1-1_{T(\bX')}(Z))g(Z|\bX-\bX') \delta Z.
\end{split}
\end{equation}

By computing the set derivative of the above mass believe function, the multi-object likelihood can be represented as 
\begin{equation}
\begin{split}
g(Z|\bX)
=&\!\!\sum_{Z'\subseteq Z}1_{T(\bX')}(Z')g(Z'|\bX')(1-1_{T(\bX')}(Z\!-\!Z'))\\
&\cdot g(Z\!-\!Z'|\bX\!-\!\bX')\\
=&g(Z\cap T(\bX')|\bX')g(Z-T(\bX')|\bX-\bX').
\end{split}
\end{equation}

\bibliographystyle{IEEEtran}
\bibliography{LMB_TBD}

\begin{thebibliography}{10}
\providecommand{\url}[1]{#1}
\csname url@samestyle\endcsname
\providecommand{\newblock}{\relax}
\providecommand{\bibinfo}[2]{#2}
\providecommand{\BIBentrySTDinterwordspacing}{\spaceskip=0pt\relax}
\providecommand{\BIBentryALTinterwordstretchfactor}{4}
\providecommand{\BIBentryALTinterwordspacing}{\spaceskip=\fontdimen2\font plus
\BIBentryALTinterwordstretchfactor\fontdimen3\font minus
  \fontdimen4\font\relax}
\providecommand{\BIBforeignlanguage}[2]{{%
\expandafter\ifx\csname l@#1\endcsname\relax
\typeout{** WARNING: IEEEtran.bst: No hyphenation pattern has been}%
\typeout{** loaded for the language `#1'. Using the pattern for}%
\typeout{** the default language instead.}%
\else
\language=\csname l@#1\endcsname
\fi
#2}}
\providecommand{\BIBdecl}{\relax}
\BIBdecl

\bibitem{refr:Mahler_book}
R.~Mahler, \emph{Statistical Multisource-Multitarget Information Fusion}.\hskip
  1em plus 0.5em minus 0.4em\relax Norwell, MA, USA: Artech House, 2007.

\bibitem{refr:biology}
V.~Marmarelis and T.~Berger, ``General methodology for nonlinear modeling of
  neural systems with {Poisson} point-process inputs,'' \emph{Mathematical
  biosciences}, vol. 196, no.~1, pp. 1--13, Jul. 2005.

\bibitem{refr:physics}
D.~L. Snyder, L.~J. Thomas, and M.~M. Ter-Pogossian, ``A mathematical model for
  {P}ositron-emission tomography systems having time-of-flight measurement,''
  \emph{IEEE Trans. Nuclear Science}, vol.~28, no.~3, pp. 3575--3583, Jun.
  1981.

\bibitem{refr:vedio-tracking}
R.~Hoseinnezhad, B.-N.Vo, and B.-T. Vo, ``Visual tracking in background
  subtracted image sequences via multi-{B}ernoulli filtering,'' \emph{IEEE
  Trans. on Signal Process.}, vol.~61, no.~2, pp. 392--397, Jan. 2013.

\bibitem{refr:tracking-1}
R.~Mahler, ``Multitarget {B}ayes filtering via first-order multitarget
  moments,'' \emph{IEEE Trans. on Aerosp. Electron. Syst}, vol.~39, no.~4, pp.
  1152--1178, Oct. 2003.

\bibitem{refr:tracking-2}
------, \emph{Advances in Statistical Multisource-Multitarget Information
  Fusion}.\hskip 1em plus 0.5em minus 0.4em\relax Norwell, MA, USA: Artech
  House, 2014.

\bibitem{refr:GCI-MB}
B.~L. Wang, W.~Yi, R.~Hoseinnezhad, S.~Q. Li, L.~J. Kong, and X.~B. Yang,
  ``Distributed fusion with multi-{B}ernoulli filter based on generalized
  {C}ovariance {I}ntersection,'' \emph{IEEE Trans. on Signal Process.},
  vol.~65, no.~1, pp. 242--255, Jan. 2017.

\bibitem{refr:robotics}
J.~Mullane, B.~N. Vo, M.~Adams, and B.~T. Vo, ``A random-finite-set approach to
  {B}ayesian {SLAM},'' \emph{IEEE Trans. on Robotics}, vol.~27, no.~2, pp.
  268--282, Apr. 2011.

\bibitem{refr:PHD}
B.~T. Vo and W.~K. Ma, ``The {G}aussian mixture probability hypothesis density
  filter,'' \emph{IEEE Trans. on Signal Process.}, vol.~54, no.~11, pp.
  4091--4104, Nov. 2006.

\bibitem{refr:CPHD}
R.~Mahler, ``{PHD} filters of higher order in target number,'' \emph{IEEE
  Trans. on Aerosp. Electron. Syst.}, vol.~43, no.~4, pp. 1523--1543, Oct.
  2007.

\bibitem{refr:CPHD-2}
B.~T. Vo, B.~N. Vo, and A.~Cantoni, ``Analytic implementations of the
  cardinalized probability hypothesis density filter,'' \emph{IEEE Trans. on
  Signal Process.}, vol.~55, no.~7, pp. 3553--3567, Jun. 2007.

\bibitem{refr:MeMber_filter1}
------, ``The cardinality balanced multi-target multi-{B}ernoulli filter and
  its implementations,'' \emph{IEEE Trans. on Signal Process.}, vol.~57, no.~2,
  pp. 409--423, Feb. 2009.

\bibitem{refr:MeMber_filter}
B.~T. Vo, B.~N. Vo, N.~T. Pham, and D.~Suter, ``Joint detection and estimation
  of multiple objects from image observations,'' \emph{IEEE Trans. on Signal
  Process.}, vol.~58, no.~10, pp. 5129--5141, Oct. 2010.

\bibitem{refr:label_1}
B.~T. Vo and B.~N. Vo, ``Labeled random finite sets and multi-object conjugate
  priors,'' \emph{IEEE Trans. on Signal Process.}, vol.~61, no.~13, pp.
  3460--3475, Jul. 2013.

\bibitem{refr:label_2}
B.~N. Vo, B.~T. Vo, and D.~Phung, ``Labeled random finite sets and the {B}ayes
  multi-target tracking filter,'' \emph{IEEE Trans. on Signal Process.},
  vol.~62, no.~24, pp. 6554--6567, Dec. 2014.

\bibitem{refr:label_3}
C.~Fantacci, B.~T. Vo, F.~Papi, and B.~N. Vo, ``The marginalized
  $\delta$-{GLMB} filter,'' \emph{arXiv preprint arXiv:1501.00926}, 2015.

\bibitem{refr:label_4}
M.~Beard, B.~T. Vo, and B.~N. Vo, ``Bayesian multi-target tracking with merged
  measurements using labelled random finite sets,'' \emph{IEEE Trans. Signal
  Process.}, vol.~63, no.~6, pp. 1433--1447, Aug. 2015.

\bibitem{refr:label_5}
S.~Reuter, B.~T. Vo, B.~N. Vo, and K.~Dietmayer, ``The labeled
  multi-{B}ernoulli filter,'' \emph{IEEE Trans. on Signal Process.}, vol.~62,
  no.~12, pp. 3246--3260, Jun. 2014.

\bibitem{refr:label_6}
F.~Papi, B.~N. Vo, B.~T. Vo, C.~Fantacci, and M.~Beard, ``Generalized labeled
  multi-{B}ernoulli approximation of multi-object densities,'' \emph{IEEE
  Trans. on Signal Process.}, vol.~63, no.~20, pp. 5487--5497, 2015.

\bibitem{refr:label_7}
F.~Papi and D.~Y. Kim, ``A particle multi-target tracker for superpositional
  measurements using labeled random finite sets,'' \emph{IEEE Trans. on Signal
  Process.}, vol.~63, no.~16, pp. 4348--4358, Jun. 2015.

\bibitem{Vo-Vo-JMS}
W.~Yi, M.~Jiang, and R.~Hoseinnezhad, ``The multiple model {V}o-{V}o filter,''
  \emph{IEEE Trans. Aerosp. Electron. Syst.}, vol.~53, no.~2, pp. 1045--1054,
  Apr. 2017.

\bibitem{robust-distributed-fusion}
S.~Q. Li, W.~Yi, R.~Hoseinnezhad, G.~Battistelli, B.~L. Wang, and L.~J. Kong,
  ``Robust distributed fusion with labeled random finite sets,'' \emph{IEEE
  Trans. on Signal Process.}, accepted, DOI: 10.1109/TSP.2017.2760286, Sep.
  2017.

\bibitem{refr:tbd-2}
A.~F. Garc\'{\i}a-Fern\'{a}ndez, J.~Grajal, and M.~R. Morelande, ``Two-layer
  particle filter for multiple target detection and tracking,'' \emph{IEEE
  Trans. on Aerosp. Electron. Syst.}, vol.~49, no.~3, pp. 1569--1588, Jul.
  2013.

\bibitem{refr:tbd-3}
M.~R. Morelande, C.~M. Kreucher, and K.~Kastella, ``A {B}ayesian approach to
  multiple target detection and tracking,'' \emph{IEEE Trans. on Signal
  Process.}, vol.~55, no.~5, pp. 1589--1604, May 2007.

\bibitem{refr:tbd-4}
W.~Yi, M.~R. Moreland, L.~Kong, and J.~Yang, ``A computationally efficient
  particle filter for multitarget tracking using an independence
  approximation,'' \emph{IEEE Trans. on Signal Process.}, vol.~61, no.~4, pp.
  843--856, Feb. 2013.

\bibitem{refr:tbd-5}
W.~Yi, M.~Morelande, L.~Kong, and J.~Yang, ``An efficient multi-frame
  track-before-detect algorithm for multi-target tracking,'' \emph{IEEE J. Sel.
  Top. Signal Process.}, vol.~7, no.~3, pp. 421--434, Jun. 2013.

\bibitem{haichao1}
H.~C. Jiang, W.~Yi, T.~Kirubarajan, L.~J. Kong, and X.~B. Yang, ``Multiframe
  radar detection of fluctuating targets using phase information,'' \emph{IEEE
  Trans. Aerosp. Electron. Syst.}, vol.~53, no.~2, pp. 736--749, Apri. 2017.

\bibitem{haichao2}
------, ``Track-before-detect strategies for radar detection in
  {G}0-distributed clutter,'' \emph{IEEE Trans. Aerosp. Electron. Syst.},
  vol.~PP, no.~99, May. 2017.

\bibitem{refr:superpositional-1}
R.~Mahler, ``{CPHD} filters for superpositional sensors,'' in \emph{Proc. SPIE
  Optical Engineering and Applications}, 2009, pp. 74\,450E--74\,450E.

\bibitem{refr:superpositional-2}
S.~Nannuru, M.~Coates, and R.~Mahler, ``Computationally-tractable approximate
  {PHD} and {CPHD} filters for superpositional sensors,'' \emph{IEEE J. Sel.
  Top. Signal Process.}, vol.~7, no.~3, pp. 410--420, Jun. 2013.

\bibitem{refr:superpositional-3}
S.~Nannuru and M.~Coates, ``Hybrid multi-{B}ernoulli and {CPHD} filters for
  superpositional sensors,'' \emph{IEEE Trans. on Aerosp. Electron. Syst.},
  vol.~51, no.~4, pp. 2847--2863, Oct. 2015.

\bibitem{refr:extended-1}
C.~Lundquist, K.~Granstrm, and U.~Orguner, ``An extended target {CPHD} filter
  and a {G}amma {G}aussian inverse {W}ishart implementation,'' \emph{IEEE J.
  Sel. Top. Signal Process.}, vol.~7, no.~3, pp. 472--483, Jun. 2013.

\bibitem{refr:amplitude}
O.~Hlinka, O.~Sluciak, F.~Hlawatsch, P.~M. Djuric, and M.~Rupp, ``Likelihood
  consensus and its application to distributed particle filtering,'' \emph{IEEE
  Trans. on Signal Process.}, vol.~60, no.~8, pp. 4334--4349, Aug. 2012.

\bibitem{refr:computer-vision}
R.~Hoseinnezhad, B.-N. Vo, B.-T. Vo, and D.~Suter, ``Visual tracking of
  numerous targets via multi-{B}ernoulli filtering of image data,''
  \emph{Pattern Recognition}, vol.~45, no.~10, pp. 3625--3635, Oct. 2012.

\bibitem{refr:conference-GOM}
S.~Q. Li, W.~Yi, B.~L. Wang, and L.~J. Kong, ``Labeled multi-object tracking
  algorithms for generic observation model,'' in \emph{Proc. IEEE Int. Fusion
  Conf.}, Jul. 2016, pp. 1125--1131.

\bibitem{refr:beyond_kalm_filer}
B.~Ristic, S.~Arulampalam, and N.~J. Gordon, \emph{Beyond the Kalman Filter:
  Particle Filters for Tracking Applications}.\hskip 1em plus 0.5em minus
  0.4em\relax Artech House, 2004.

\bibitem{refr:SMC_Gordon}
N.~J. Gordon, D.~J. Salmond, and A.~F.~M. Smith, ``Novel approach to
  nonlinear/non-{G}aussian {B}ayesian state estimation,'' in \emph{Proc. Inst.
  Elect. Eng. F}, vol. 140, no.~2, 1993, pp. 107--113.

\bibitem{refr:convergence_partical}
D.~Crisan and A.~Doucet, ``Convergence of sequential {M}onte {C}arlo methods,''
  University of Cambridge, CUED/F-INFENG/TR.381, Tech. Rep., 2000.

\bibitem{diffusion_source}
G.~Battistelli, L.~Chisci, N.~Forti, G.~Pelosi, and S.~Selleri, ``Point source
  estimation via finite element multiple-model {K}alman filtering,'' in
  \emph{2015 IEEE 54th Annual Conference on Decision and Control (CDC)}, Dec.,
  pp. 4984--4989.

\bibitem{refr:JPDA}
J.~Dezert and Y.~Bar-Shalom, ``Joint probabilistic data association for
  autonomous navigation,'' \emph{IEEE Trans. on Aerosp. Electron. Syst.},
  vol.~29, no.~4, pp. 1275--1286, Oct. 1993.

\bibitem{refr:HPD}
G.~E. Box and G.~C. Tiao, \emph{Bayesian Inference in Statistical
  Analysis}.\hskip 1em plus 0.5em minus 0.4em\relax Addison Wesley, 1973.

\bibitem{refr:HDR}
R.~J. Hyndman, ``Computing and graphing highest density regions,'' \emph{The
  American Statistician}, vol.~50, no.~2, pp. 120--126, 1996.

\bibitem{refr:clustering_method}
L.~Rokach and O.~Maimon, ``Clustering methods,'' in \emph{Data mining and
  knowledge discovery handbook}.\hskip 1em plus 0.5em minus 0.4em\relax
  Springer US, 2005, pp. 321--352.

\bibitem{set-jpda}
D.~Svensson, M.~Guerriero, and P.~Willett, ``Set {JPDA} filter for multitarget
  tracking,'' \emph{IEEE Trans. on Signal Process.}, vol.~59, no.~10, pp.
  4677--4691, Oct. 2011.

\bibitem{refr:curse_dimension}
F.~Daum and J.~Huang, ``Curse of dimensionality and particle filters,'' in
  \emph{Proc. IEEE Aerosp. Conf.}, vol.~4, Mar. 2003, pp. 1979--1993.

\bibitem{refr:OSPA}
D.~Schumacher, B.~T. Vo, and B.~N. Vo, ``A consistent metric for performance
  evaluation of multi-object filters,'' \emph{IEEE Trans. on Signal Process.},
  vol.~56, no.~8, pp. 3447--3457, Aug. 2008.

\end{thebibliography}
\begin{IEEEbiography}
[{\includegraphics[width=0.9\columnwidth,draft=false]{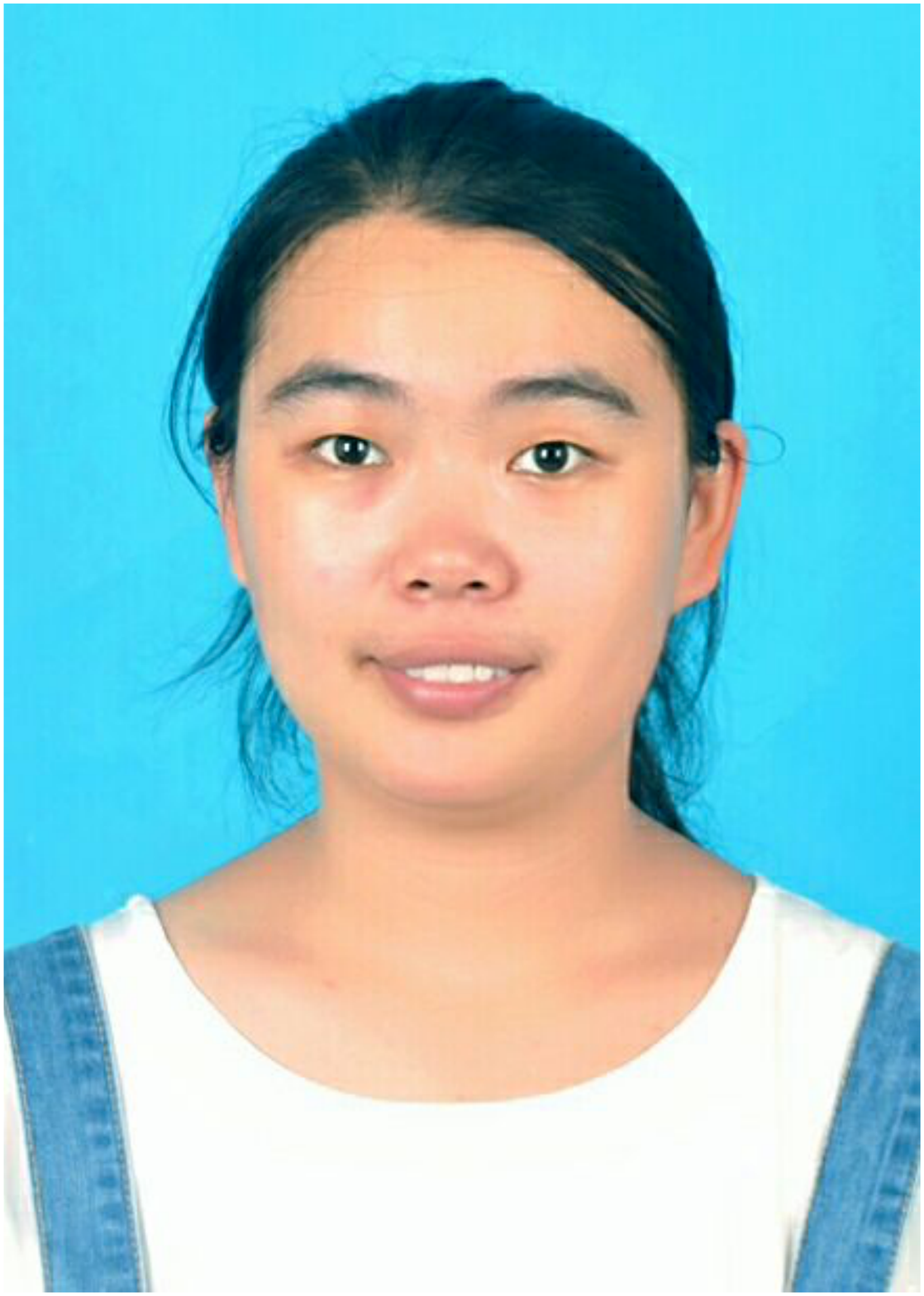}}]{Suqi Li}  is born in 1990.  She received the B.E. degree in electronic engineering from the University of Electronic Science and Technology of China, Chengdu, in 2011. 
Since September 2011, she has been pursuing the Ph.D. degree at the School of Electronic Engineering, University of Electronic Technology and Science of China.  
Currently, she is a visiting student with the Dipartimento di Ingegneria dell' Informazione (DINFO), Universit$\grave{\mbox{a}}$ degli Studi di Firenze, Italy. Her research interests include the random finite set, multi-target tracking, nonlinear filtering, sensor networks and data fusion. 
\end{IEEEbiography}
 \begin{IEEEbiography}
[{\includegraphics[width=0.9\columnwidth,draft=false]{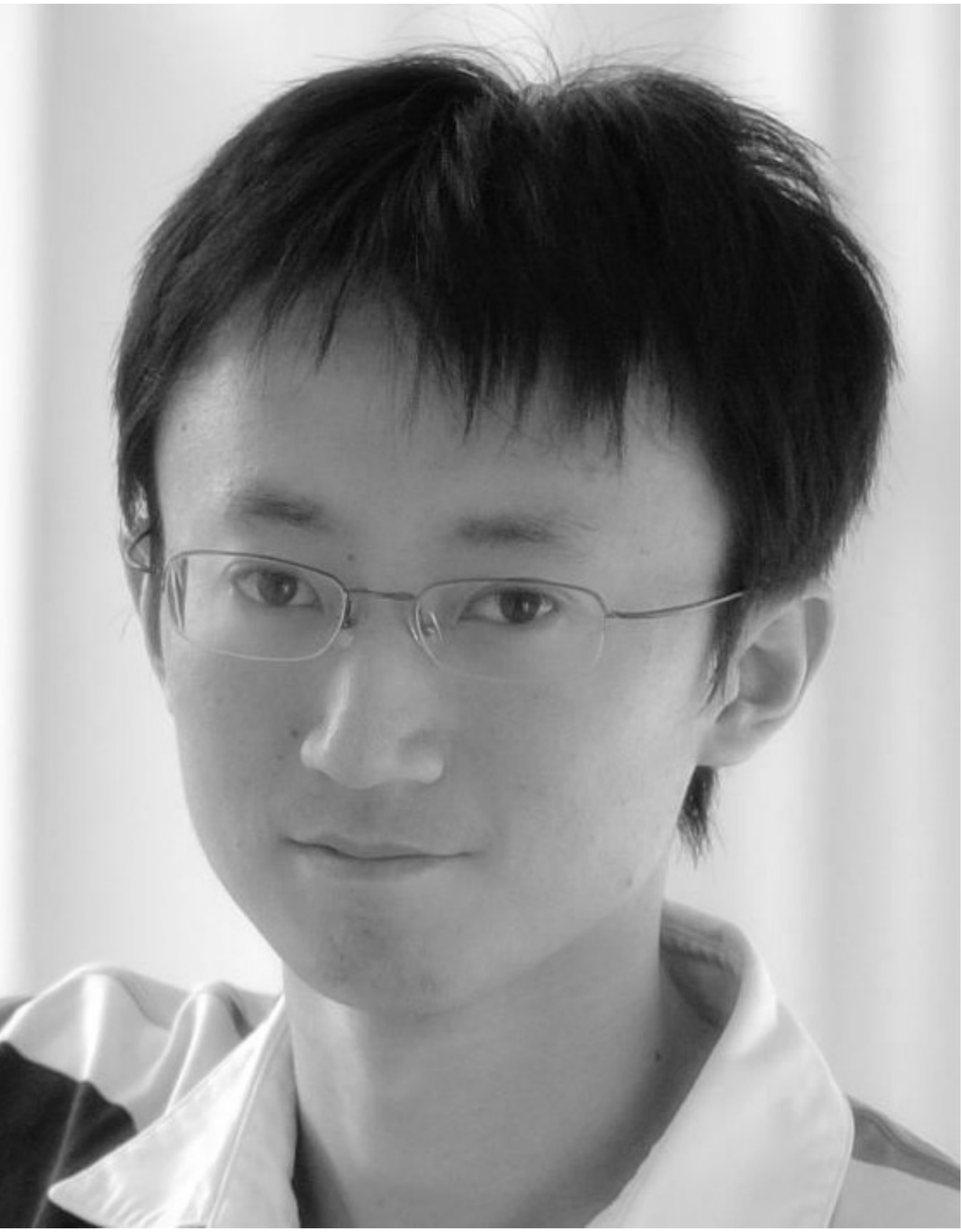}}]{Wei Yi}
 received the B.E. degree in electronic engineering from the University of Electronic Science and Technology of China, Chengdu, in 2006.

Since 2007, he has been pursuing the Ph.D. degree at the School of Electronic Engineering of the University of Electronic Technology and Science of China. 

From March 2010 to February 2012, he was a visiting student in the Melbourne Systems Laboratory, University of Melbourne, Australia. His research interests include particle filtering and target tracking (particular emphasis on multiple target tracking and track-before-detect techniques).

Mr. Yi received the ``Best Student Paper Award'' at the 2012 IEEE Radar Conference, Atlanta, United States and the ``Best Student Paper Award'' at the 15th International Conference on Information Fusion, Singapore, 2012.
\end{IEEEbiography}
 \begin{IEEEbiography}
[{\includegraphics[width=0.9\columnwidth,draft=false]{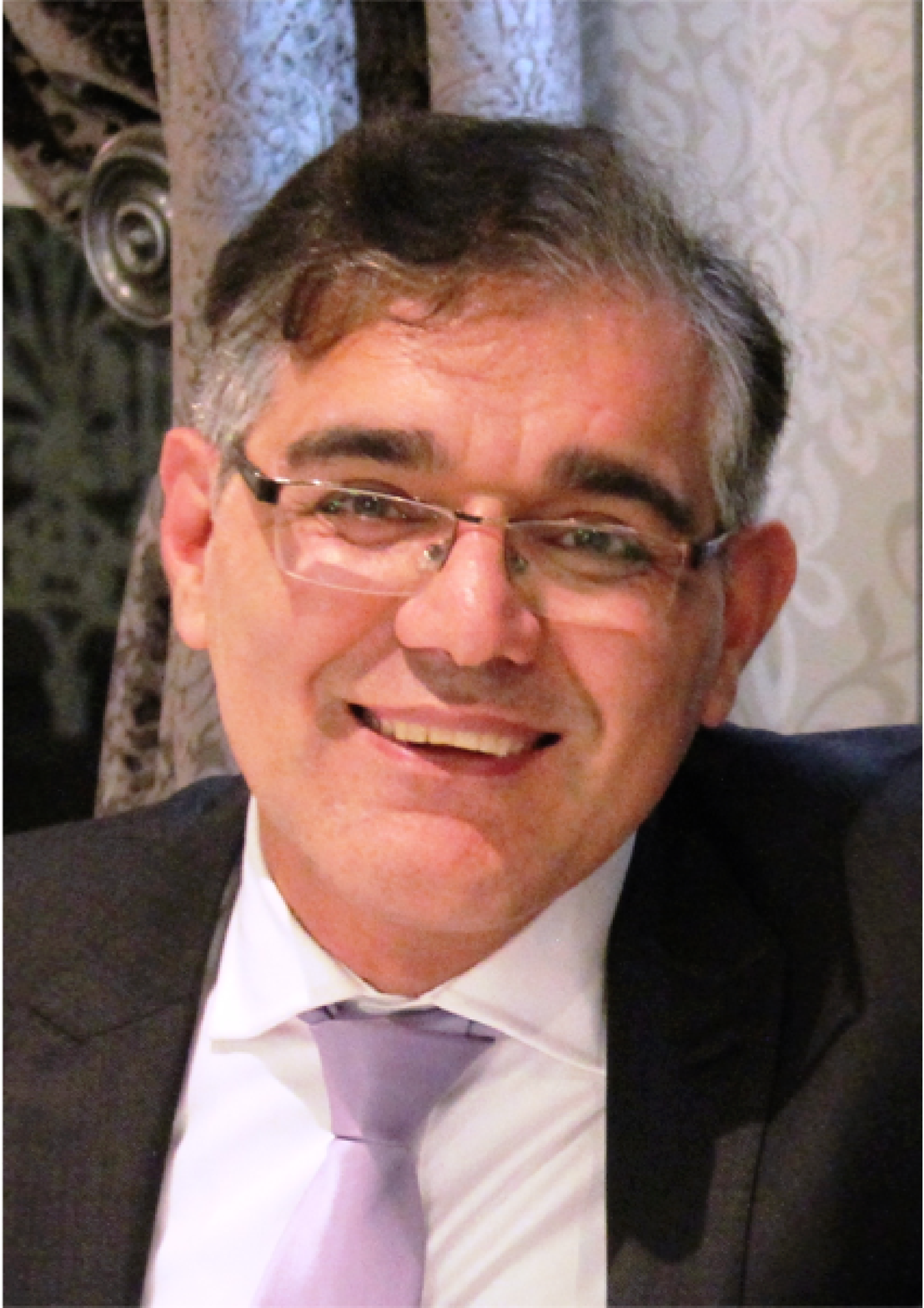}}]{Reza Hoseinnezhad}
received his B.Sc., M.Sc. and Ph.D. degrees in Electronic, Control and Electrical Engineering all from the University of Tehran, Iran, in 1994, 1996 and 2002, respectively. Since 2002, he has held various academic positions at the University of Tehran, Swinburne University of Technology, the University of Melbourne and RMIT University. He is currently an Associate Professor with the School of Aerospace, Mechanical and Manufacturing Engineering, RMIT University, Victoria, Australia. His research is currently focused on development of robust estimation and visual tracking methods in a point process framework.
\end{IEEEbiography}
%
%
%

\begin{IEEEbiography}
[{\includegraphics[width=0.9\columnwidth,draft=false]{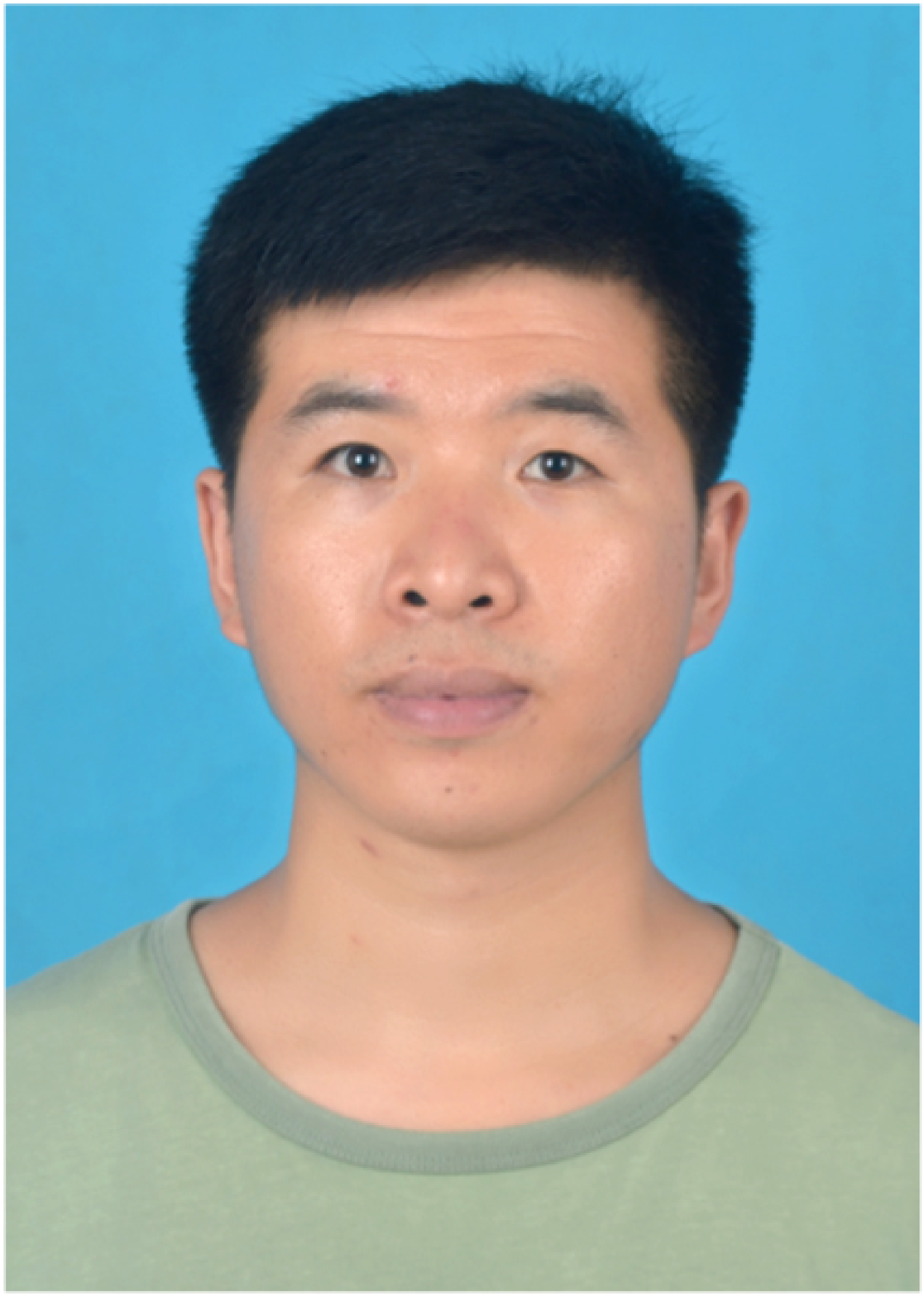}}]{Bailu Wang} received his B.S. degree from the University of Electronic Science and
Technology of China (UESTC) in 2011. He is now working toward his Ph.D. degree
on signal and information processing at UESTC.

From August  2016, he has been a visiting student with the Dipartimento di Ingegneria dell' Informazione (DINFO), Universit$\grave{\mbox{a}}$ degli Studi di Firenze, Italy. His current research interests include
radar and statistical signal processing, and multi-sensor multi-target fusion.
\end{IEEEbiography}
\begin{IEEEbiography}
[{\includegraphics[width=0.9\columnwidth,draft=false]
{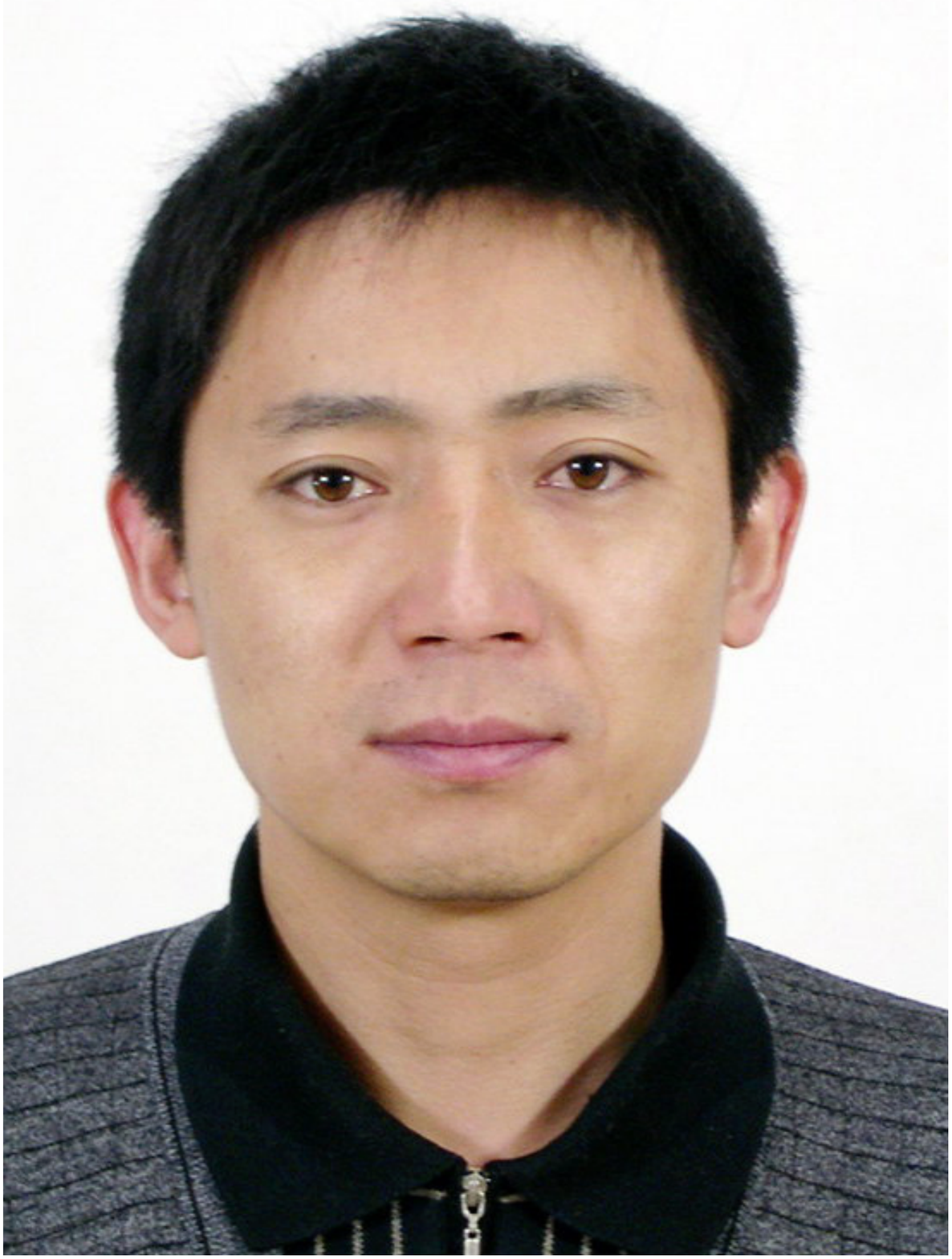}}]{Lingjiang Kong} was born in 1974. He received the B.S., M.S., and Ph.D. degrees from the
University of Electronic Science and Technology of China (UESTC) in 1997, 2000
and 2003, respectively.

From September 2009 to March 2010, he was a visiting researcher with the
University of Florida. 

He is currently a professor with the School of
Electronic Engineering, University of Electronic Science and Technology of
China (UESTC). His research interests include multiple-input multiple-output
(MIMO) radar, through the wall radar, and statistical signal processing.
\end{IEEEbiography}

\end{document}